\documentclass[aps,prd,showkeys,superscriptaddress,twocolumn,nofootinbib,floatfix]{revtex4-2}

\usepackage{subfiles}
\usepackage{amsmath}
\usepackage{amssymb}
\usepackage{amsthm}
\usepackage{mathrsfs}
\usepackage{graphicx}
\usepackage{epstopdf}
\usepackage{fancyhdr}
\usepackage{array}
\usepackage[all]{xy}
\usepackage{eufrak}
\usepackage{euscript}
\usepackage{enumerate}
\usepackage{slashed}
\usepackage{hyperref}
\usepackage{subfigure} % NEEDED IN ORDER TO USE SUBFIG
\usepackage{epstopdf} % COMPILE EPS (AND PDF) FIGURES USING PDFLATEX!!!
\hypersetup{pdftex,colorlinks=true,linkcolor=blue,citecolor=blue,menucolor=black,urlcolor=blue,filecolor=blue}

\begin{document}

\title{Hydrodynamization times of a holographic fluid far from equilibrium}

\author{Romulo Rougemont}
\email{romulo.pereira@uerj.br}
\affiliation{Departamento de F\'{i}sica Te\'{o}rica, Universidade do Estado do Rio de Janeiro, Rua S\~{a}o Francisco Xavier 524, 20550-013, Maracan\~{a}, Rio de Janeiro, Rio de Janeiro, Brazil}

\author{Willians Barreto}
\email{willians.barreto@ufabc.edu.br}
\affiliation{Centro de Ci\^{e}ncias Naturais e Humanas, Universidade Federal do ABC, Av. dos Estados 5001, 09210-580 Santo Andr\'{e}, S\~{a}o Paulo, Brazil}
\affiliation{Centro de F\'{i}sica Fundamental, Universidad de Los Andes, M\'{e}rida 5101, Venezuela}

\author{Jorge Noronha}
\email{jn0508@illinois.edu}
\affiliation{Illinois Center for Advanced Studies of the Universe\\ Department of Physics, University of Illinois at Urbana-Champaign, Urbana, IL 61801, USA}

\begin{abstract}
We investigate several hydrodynamization times for an ensemble of different far-from-equilibrium solutions of the strongly coupled $\mathcal{N}=4$ Supersymmetric Yang-Mills plasma undergoing Bjorken flow. For the ensemble of initial data analyzed in the present work, we find that, with typical tolerances between 3\% to 5\%, the average hydrodynamization time associated with the late time convergence of the pressure anisotropy to the corresponding Borel resummed hydrodynamic attractor is approximately equal to the average hydrodynamization time associated with the Navier-Stokes result, while both are shorter than the average hydrodynamization time associated with second-order hydrodynamics. On the other hand, we find that the entropy density of the different solutions coalesces to second-order hydrodynamics long before entering in the Navier-Stokes regime. A clear hierarchy between the different average hydrodynamization times of the Bjorken expanding fluid is established for the set of analyzed initial data, comprising also some solutions which, whilst satisfying the dominant and the weak energy conditions at the initial time, evolve such as to transiently violate one or both conditions when the fluid is still far from equilibrium. In particular, solutions violating the weak energy condition are generally found to take a longer time to enter in the hydrodynamic regime than the other solutions.
\end{abstract}

%\pacs{Valid PACS appear here}
% PACS, the Physics and Astronomy Classification Scheme.
% Valid PACS numbers may be entered using the \verb+\pacs{#1} command.

%No need for key words in PRD anymore.

%\keywords{Gauge/gravity duality, far-from-equilibrium dynamics, relativistic viscous hydrodynamics, finite temperature, Bjorken flow, hydrodynamization, pressure anisotropy, non-equilibrium entropy.}
% Use showkeys class option if keyword display desired

\maketitle
\tableofcontents

%%%%%%%%%%%%%%%%%%%%%%%%%%%%%%%%%%
\section{Introduction}
\label{sec:intro}

The holographic gauge/gravity correspondence \cite{Maldacena:1997re,Gubser:1998bc,Witten:1998qj,Witten:1998zw} provides a unique way to perform a first-principles analysis of the real time dynamics of far-from-equilibrium strongly coupled quantum fluids. Since the pioneering work by Chesler and Yaffe on the numerical analysis of the homogeneous isotropization dynamics of the strongly coupled $\mathcal{N}=4$ supersymmetric Yang-Mills (SYM) plasma in Ref.\ \cite{Chesler:2008hg}, many other works have been developed addressing  different aspects concerning the far-from-equilibrium dynamics of holographic models, see e.g. Refs. \cite{Chesler:2009cy,Chesler:2010bi,Heller:2011ju,Heller:2012je,vanderSchee:2012qj,Casalderrey-Solana:2013aba,Chesler:2013lia,vanderSchee:2014qwa,Jankowski:2014lna,Fuini:2015hba,Chesler:2015bba,Attems:2016tby,Casalderrey-Solana:2016xfq,Grozdanov:2016zjj,Attems:2017zam,Romatschke:2017vte,Spalinski:2017mel,Critelli:2017euk,Casalderrey-Solana:2017zyh,Critelli:2018osu,Attems:2018gou,Cartwright:2019opv,Kurkela:2019set,Ecker:2021ukv,Ghosh:2021naw}.

The first numerical analysis of Bjorken flow \cite{Bjorken:1982qr} in a holographic SYM plasma was performed in Ref.\ \cite{Chesler:2009cy}, with further developments regarding the numerical formalism being presented in Refs.\ \cite{Heller:2011ju,Heller:2012je,Chesler:2013lia,vanderSchee:2014qwa}. Additionally, a high statistics analysis of the holographic Bjorken flow of the SYM plasma was presented in Ref.\ \cite{Jankowski:2014lna}. It was observed in those early works that the onset of hydrodynamic behavior, in the sense of the effective applicability of late time constitutive relations such as those defined by Navier-Stokes (NS) or the second-order gradient expansion \cite{Baier:2007ix,Bhattacharyya:2008jc},\footnote{In this context, note that we distinguish the \emph{asymptotic} results obtained using the gradient expansion from different dynamical approaches to hydrodynamics pursued by M\"{u}ller-Israel-Stewart \cite{Muller:1967zza,Israel:1976tn,Israel:1979wp}, which include an extended set of dynamical variables.} generically happens in the SYM plasma when the system is still far from equilibrium, as inferred from a sizable pressure anisotropy at the hydrodynamization time. Such a result was initially surprising because it seems to contradict the usual notion that hydrodynamic behavior only emerges when there small deviations from local equilibrium. However, a broader notion of hydrodynamics was proposed in Ref.\ \cite{Heller:2015dha} as the late time emergence of universal behavior for observables such as the pressure anisotropy,  which is dictated by the decay of the nonhydrodynamic quasinormal modes (QNM) of the system \cite{Heller:2015dha,Romatschke:2017vte,Spalinski:2017mel,Florkowski:2017olj}.

It was proposed in  \cite{Heller:2015dha} that the Borel resummation of the divergent asymptotic gradient series \cite{Heller:2013fn,Buchel:2016cbj,Denicol:2016bjh,Heller:2016rtz} defines a hydrodynamic attractor, to which different far-from-equilibrium solutions would coalesce before converging to the corresponding limits associated with finite order truncations of the hydrodynamic gradient expansion, such as Navier-Stokes theory or the second or higher order hydrodynamic constitutive relations. More recently, it was shown in Refs.\ \cite{Casalderrey-Solana:2017zyh,Kurkela:2019set} that this is not a generic feature because, even though one may find solutions that do coalesce earlier to the Borel resummed attractor, when considering more generic initial conditions the Borel attractor does not provide a significantly earlier effective description of hydrodynamics when compared to the Navier-Stokes result, at least in the case of a holographic SYM plasma undergoing Bjorken flow.

Concerning the entropy of the SYM plasma out of equilibrium, it was argued in Ref.\ \cite{Figueras:2009iu} that this quantity should be associated with the area of the apparent horizon of a black hole within the higher dimensional bulk (instead of the event horizon, to which the apparent horizon converges only at asymptotic times, close to local equilibrium). In this context, Refs.\ \cite{Heller:2011ju,Jankowski:2014lna} analyzed the difference between the final and initial entropies as a function of the initial entropy for the different initial conditions considered. However, in those works it was not explicitly presented the calculation of the non-equilibrium entropy density as function of time.

In the present work we focus on the analysis of the hydrodynamization properties of a SYM plasma  undergoing Bjorken flow \cite{Bjorken:1982qr}. More specifically, we investigate how the pressure anisotropy and the non-equilibrium entropy density of the fluid converge to the hydrodynamic regime at late times for an ensemble of initial data originally studied in Ref.\ \cite{Rougemont:2021qyk}.

For the analyzed ensemble of initial conditions we find that on average, under typical relative tolerances between $3\%$ to $5\%$,\footnote{We choose some different values for the tolerance in order to illustrate how the hydrodynamization times of the system can vary depending on such a choice.} the Borel resummed attractor for the pressure anisotropy does not provide a significantly earlier description of the putative far-from-equilibrium hydrodynamic universal behavior than the corresponding Navier-Stokes result. More specifically, by analyzing how each of the different initial conditions approach the Navier-Stokes regime, second-order hydrodynamics, and the Borel resummed attractors within the aforementioned tolerances, we notice that different initial conditions can converge first to different attractors. By averaging over all the initial data considered, we find that the average hydrodynamization time associated with the Borel attractor is approximately equal to the average hydrodynamization time associated with convergence to the NS regime, with both being shorter than the average hydrodynamization time associated with the second-order hydrodynamic truncation of the pressure anisotropy. The Borel resummed attractor only provides a clearly better description of the hydrodynamization process than the NS result if one considers just very small relative tolerances in the long time regime.

On the other hand, we find that the non-equilibrium entropy density of the different solutions always coalesces to the second-order hydrodynamic truncation long before it converges to the corresponding NS regime. Moreover, the average hydrodynamization time for the entropy density associated with  second-order hydrodynamics is also considerably smaller than the different average hydrodynamization times related to the pressure anisotropy. Consequently, the time scales at which the fluid approaches local equilibrium may be rather different depending on which physical observable one considers to probe the evolution of the medium, and we clearly identify a hierarchy between the different average hydrodynamization times of the Bjorken expanding fluid.

The ensemble of initial data studied in the present work was first considered in \cite{Rougemont:2021qyk}, where a class of numerical solutions was found in Bjorken flow that transiently violates the dominant energy condition or even the weak energy condition at early times when the system is still far from equilibrium. Such violations are not present in the initial state and, thus, they are indeed generated by the subsequent evolution of the plasma.

This work is organized as follows. In section \ref{sec:HolBjorken} we review in detail the main steps required for the holographic calculation of dynamics of the SYM plasma undergoing Bjorken flow using the characteristic formulation of general relativity --- although these results are not new, we are not aware of any previous work in the literature that provides such a detailed account of the step-by-step procedure, which can be very useful when trying to reproduce the results discussed here and in some other works in the literature. In section \ref{sec:HolResults} we present our main results, consisting in the analysis of the pressure anisotropy, the non-equilibrium entropy density and their associated hydrodynamization times for the ensemble of initial data originally investigated in \cite{Rougemont:2021qyk}. Our main conclusions are summarized in section \ref{sec:conclusion}. In Appendix \ref{sec:appA} we present further details regarding the numerics of our work and in Appendix \ref{sec:appB} we present a brief discussion on the performance of our numerical code.

We use in this work a mostly plus metric signature and natural units $\hbar = c = k_B = 1$.

%%%%%%%%%%%%%%%%%%%%%%%%%%%%%%%%%%
\section{Holographic SYM plasma undergoing bjorken flow}
\label{sec:HolBjorken}

Bjorken flow \cite{Bjorken:1982qr} corresponds to a rapidly expanding inhomogeneous relativistic fluid that possesses the following set of symmetries: there is boost invariance in the single spatial direction $z$ along which the fluid expands at the speed of light, plus translation and $O(2)$ rotation invariance in the transverse $xy$ plane (also, $Z_2$ invariance is imposed along the spatial rapidity direction \cite{Gubser:2010ze}). This setting is usually taken as a first (and crude) approximation to the expanding quark-gluon plasma formed in high-energy heavy-ion collisions \cite{Arsene:2004fa,Adcox:2004mh,Back:2004je,Adams:2005dq,Aad:2013xma} near mid-rapidity, i.e. close to the collision axis (the transverse expansion to the collision axis is completely neglected in this simple model).

Bjorken symmetry is more easily handled by changing from  Cartesian coordinates $(t,x,y,z)$ to the so-called Milne coordinates $(\tau,\xi,x,y)$, where $\tau$ and $\xi$ are the propertime and the spacetime rapidity, respectively, defined as follows,
\begin{equation}
\tau = \sqrt{t^2-z^2},\qquad
\xi = \ln\left( \frac{t+z}{t-z}\right),
\label{eq1}
\end{equation}
in terms of which the metric of the 4D Minkowski spacetime where the fluid is defined reads,
\begin{align}
ds^2_{\textrm{(4D)}} = -d\tau^2+\tau^2 d\xi^2+dx^2+dy^2.
\label{eq2}
\end{align}

The holographic gauge/gravity modeling of the Bjorken flow of a \textit{relativistic and strongly coupled quantum fluid} can be implemented by considering that the 4D flat spacetime \eqref{eq2}, where the fluid is defined, is (up to a global conformal factor) the boundary of a 5D curved spacetime that is asymptotically anti-de Sitter (AdS$_5$) by the standards of the holographic dictionary \cite{Maldacena:1997re,Gubser:1998bc,Witten:1998qj,Witten:1998zw}, which in turn requires that the classical gravity action for the higher dimensional bulk has a negative cosmological constant. The extra holographic direction is related to a geometrization of the renormalization group flow \cite{deBoer:1999tgo} of the quantum field theory describing the fluid living at the boundary of the higher dimensional bulk spacetime.

There are infinitely many different holographic models that can be constructed under such assumptions, each one of them supposedly describing a different kind of strongly coupled quantum field theory at the boundary. The simplest and better-known top-down holographic construction that can be considered in this regard corresponds to the conformal and strongly coupled SYM plasma, whose dual bulk action is given simply by the 5D Einstein-Hilbert action with a negative cosmological constant,
\begin{align}
S = \frac{1}{2\kappa_5^2} \int_{\mathcal{M}_5}d^5x\sqrt{-g}\left[R-2\Lambda_L\right],
\label{eq3}
\end{align}
where $\kappa_5^2\equiv 8\pi G_5$ is the 5D gravitational Newton's constant, which is holographically related to the number of colors $N_c$ of the SYM theory as $\kappa_5^2=4\pi^2/N_c^2$, $\Lambda_L=-6/L^2$ is the negative cosmological constant associated with the asymptotically AdS$_5$ spacetime, and $L$ is the asymptotic AdS$_5$ radius (which we set to unity here). The bulk action \eqref{eq3} is supplemented by boundary terms which do not contribute to the bulk equations of motion but are necessary for the holographic computation of some observables, such as the Green's functions of the dual boundary quantum field theory. They comprise the Gibbons-Hawking-York action \cite{York:1972sj,Gibbons:1976ue}, needed for the well-posedness of the boundary value problem, and the counterterm action associated with  holographic renormalization \cite{deHaro:2000vlm,Bianchi:2001kw,Skenderis:2002wp,Lindgren:2015lia,Elvang:2016tzz} of the bulk action.%\footnote{The correspondence relates $\mathcal{N}=4$ SYM in 4D to type IIB superstring theory in AdS$_5 \otimes S_5$. However, the dynamics within the 5-sphere plays no role in the analysis done in this work. Therefore, the relevant bulk geometry used in this paper is defined in five dimensions.}

The Ansatz for the 5D bulk metric field compatible with diffeomorphism invariance and Bjorken symmetry can be written using infalling Eddington-Finkelstein (EF) coordinates as follows \cite{Chesler:2009cy,Chesler:2013lia},
\begin{align}
ds^2 =& \,2d\tau\left[dr-A(\tau,r) d\tau \right]+\Sigma(\tau,r)^2\left[e^{-2B(\tau,r)}d\xi^2 \right.\nonumber\\ 
      & \left. +\, e^{B(\tau,r)}(dx^2+dy^2)\right],
\label{lineElement}
\end{align}
where $r$ is the radial holographic direction, in terms of which the boundary lies at $r\to\infty$ (the EF time $\tau$ reduces to the propertime of the gauge theory fluid at the boundary). In the EF coordinates infalling radial null geodesics satisfy $\tau=\textrm{constant}$, while outgoing radial null geodesics satisfy $dr = A(\tau,r)d\tau$. By foliating the bulk spacetime in slices of constant $\tau$, which in the EF coordinates correspond to null hypersurfaces, and then evolving the equations of motion in the EF time, one implements a time evolution of the system according to the so-called \textit{characteristic formulation of general relativity}, which for asymptotically AdS spacetimes in the context of holography is reviewed in detail in Ref.\ \cite{Chesler:2013lia} (for the original formulation involving asymptotically flat spacetimes see Refs.\ \cite{Bondi:1962px,Sachs:1962wk}).

The holographic Bjorken flow for the SYM plasma is therefore formulated on the gravity side of the correspondence in terms of the three bulk metric coefficients $A(\tau,r)$, $B(\tau,r)$, and $\Sigma(\tau,r)$, which are functions of two variables: the holographic direction $r$ and the EF time $\tau$. The line element \eqref{lineElement} still has a residual diffeomorphism invariance under radial shifts of the form $r\mapsto r+\lambda(\tau)$, with $\lambda(\tau)$ being an arbitrary function of time. One major technical difficult of the characteristic formulation of general relativity concerns the possible breakdown of the numerical evolution of the system in regions with strong curvatures due to the formation of caustics \cite{Chesler:2013lia}. In order to integrate the equations of motion of the system in the radial direction one must consider the entire portion of the bulk geometry causally connected to the boundary. When there is an apparent horizon within the bulk\footnote{We are going to discuss this in more detail in section \ref{sec:HolHor}.}, this condition is met if one performs the radial integration from the horizon to the boundary. However, in general, the radial position of the apparent horizon (when it exists) can widely fluctuate between the different time slices. If one adopts a fixed infrared (IR) radial cutoff in the interior of the bulk to start the integration of the equations of motion in the radial direction on every time slice, then some different possibilities may happen. One of them is the following: in some time slices this fixed IR radial position in the interior of the bulk may be behind the apparent horizon, and in this case we are sure that we are covering the region of the bulk causally connected to the boundary. However, it may be that in some of these time slices the chosen fixed IR radial cutoff penetrates too deep into the horizon and eventually reaches a caustic, and in this case the numerical simulation breaks down. 

On the other hand, if one chooses some fixed IR radial position which does not penetrate too far within the bulk, the numerical simulations will probably never find a caustic and proceed without breaking down, but it may be that at some time slices this fixed IR radial cutoff lies beyond the apparent horizon, and in this case we may lose information of part of the bulk geometry causally connected to the boundary, which may lead to physically inaccurate results. A possible way, discussed in detail in Ref.\ \cite{Chesler:2013lia} to deal with this issue, is to use the aforementioned residual diffeomorphism invariance and fix different values for the function $\lambda(\tau)$ on the different time slices by requiring that the radial position of the apparent horizon remains fixed for all time slices. We will discuss an implementation of this scheme in section \ref{sec:HolHor}.

At the boundary of the bulk spacetime the metric coefficients must satisfy boundary conditions such that for $r\to\infty$ one recovers from the 5D line element \eqref{lineElement} the 4D metric \eqref{eq2}, up to the global conformal factor $r^2$ of AdS$_5$. This is accomplished by imposing the following boundary conditions associated with  holographic Bjorken flow,
\begin{align}
A(r\to\infty,\tau)&\sim\frac{r^2}{2},\,\,\, B(r\to\infty,\tau)\sim-\frac{2 \ln(\tau)}{3},\nonumber\\
\Sigma(r\to\infty,\tau)&\sim\tau^{1/3}r.
\label{eq5}
\end{align}
In fact, by substituting  \eqref{eq5} into Eq.\ \eqref{lineElement} one recovers the AdS$_5$ metric near the boundary in the EF coordinates,
\begin{align}
ds^2\biggr|_{r\to\infty} \to ds^2_{(\textrm{AdS}_5)} = 2d\tau dr + r^2 ds^2_{\textrm{(4D)}}.
\label{eq6}
\end{align}

Einstein's equations for the metric field can be worked out to give the following set of coupled $1+1$ partial differential equations (PDEs) for the metric coefficients $A(r,\tau)$, $B(r,\tau)$, and $\Sigma(r,\tau)$\footnote{We remark that the metric coefficient $A(r,\tau)$ defined in Eq.\ \eqref{lineElement} and also in Ref.\ \cite{Chesler:2013lia} corresponds to half of the corresponding function as defined in Ref.\ \cite{Chesler:2009cy}.}
\begin{subequations}
\begin{align}
\Sigma''+\frac{\Sigma B'\,^2}{2} &=0, \label{pde1}\\
(d_+\Sigma)'+\frac{2\Sigma ' d_+\Sigma}{\Sigma} - 2\Sigma &=0, \label{pde2}\\
\Sigma(d_+B)' +\frac{3(B' d_+\Sigma+\Sigma 'd_+B)}{2} &=0, \label{pde3}\\
A''+\frac{4+3B' d_+B - 12(\Sigma ' d_+\Sigma)/\Sigma^2}{2} &=0, \label{pde4}\\
d_+(d_+\Sigma)+\frac{\Sigma (d_+B)^2}{2} -A' d_+\Sigma &=0, \label{pde5}
\end{align}
\end{subequations}
where $'\equiv\partial_r$ is the directional derivative along infalling radial null geodesics (with $\tau=\textrm{constant}$) and $d_{+}\equiv \partial_{\tau}+A(r,\tau)\partial_r$ is the directional derivative along outgoing radial null geodesics (with $dr/d\tau=A(r,\tau)$). Eq.\ \eqref{pde1} is the so-called Hamiltonian constraint. Eq.\ \eqref{pde5} is a constraint that can be used in order to check the accuracy of the numerical solutions obtained by solving the nested equations \eqref{pde1} --- \eqref{pde4}.

In fact, the nested or hierarchical structure observed in Eqs.\ \eqref{pde1} --- \eqref{pde4} is a common feature of the characteristic formulation of general relativity. In view of this structure, one can devise the following general ordered steps in order to solve Einstein's equations in the characteristic formulation:

\begin{enumerate}[i.]
\item One must choose some initial profile for the metric anisotropy $B(r,\tau_0)$ specified over the null hypersurface corresponding to the initial time $\tau_0$;\footnote{Alternatively, one could also choose to specify $\Sigma(r,\tau_0)$ as an initial data and then solve Eq. \eqref{pde1} for $B(r,\tau_0)$; however, in such case one would need to solve a nonlinear equation instead of a linear one.}\\
\item Next, one radially solves the Hamiltonian constraint \eqref{pde1} to obtain $\Sigma(r,\tau_0)$;\\
\item Then, one radially solves Eq.\ \eqref{pde2} to obtain $d_+\Sigma(r,\tau_0)$;\\
\item Next, one radially solves Eq. \eqref{pde3} to obtain $d_+B(r,\tau_0)$;\\
\item Then, one radially solves Eq. \eqref{pde4} to obtain $A(r,\tau_0)$;\\
\item At this point, we already know $B(r,\tau_0)$, $A(r,\tau_0)$, and $d_+B(r,\tau_0)$ and, thus, we can determine the time derivative of the metric anisotropy function evaluated on the initial time slice, $\partial_\tau B(r,\tau_0)$, by using that $d_+B(r,\tau_0)=\partial_\tau B(r,\tau_0)+A(r,\tau_0)\partial_r B(r,\tau_0)$. With $\{B(r,\tau_0),\partial_\tau B(r,\tau_0)\}$ at hand, one can evolve the metric anisotropy to the next time slice $\tau_0+\Delta\tau$ (notice that in the characteristic formulation the relevant PDEs to be solved are always first-order in the time derivatives);\\
\item Steps i --- vi are then repeated until reaching the final time of the numerical simulation.
\end{enumerate}

\subsection{UV expansions and renormalized 1-point functions}
\label{sec:HolRen}

Before going through some details of the numerics involved in the actual implementation of the aforementioned algorithm, let us first review some important results regarding the ultraviolet (UV) near-boundary expansions of the metric coefficients and their relation to the holographically renormalized one-point Green's function of the energy-momentum tensor of the boundary SYM gauge theory, from which we are going to extract the energy density and the longitudinal and transverse pressures of the strongly coupled quantum fluid under consideration.

Given the boundary conditions \eqref{eq5} of holographic Bjorken flow, the UV near-boundary expansions of the bulk metric coefficients assume the following form \cite{Chesler:2009cy,Critelli:2018osu}
\begin{subequations}
\begin{align}
A(r,\tau) &= \frac{(r+\lambda(\tau))^2}{2}-\partial_\tau\lambda(\tau)+\sum_{n=1}^\infty \frac{a_n(\tau)}{r^n},\label{UVA}\\
B(r,\tau) &= -\frac{2\ln(\tau)}{3}+\sum_{n=1}^\infty \frac{b_n(\tau)}{r^n},\label{UVB}\\
\Sigma(r,\tau) &= \tau^{1/3}r+\sum_{n=0}^\infty \frac{s_n(\tau)}{r^n}.\label{UVS}
\end{align}
\end{subequations}
By substituting the above UV expansions (truncated e.g. at eighth-order) back into the equations of motion and solving the resulting algebraic equations order by order in $r$, one can fix the values of the UV coefficients $\{a_n(\tau),b_n(\tau),s_n(\tau)\}$ as functions of $\tau$, $\lambda(\tau)$, and the single dynamical UV coefficient which remains undetermined in such UV analysis, $a_2(\tau)$, and its derivatives. In fact, as we are going to discuss, the value of the UV coefficient $a_2(\tau_0)$ at the initial time $\tau_0$, together with the initial profile for the metric anisotropy, $B(r,\tau_0)$, are the initial data that must be chosen on the gravity side of the gauge/gravity duality to generate different solutions of Einstein's equations corresponding to different evolutions of the far-from-equilibrium SYM plasma\footnote{Actually, since we are going to consider $\lambda(\tau)\neq 0$, also $\lambda(\tau_0)$ needs to be specified as an initial data.}.

Explicitly, the first few UV coefficients so obtained are shown below
\begin{subequations}
\begin{align}
A(r,\tau) &= \frac{(r+\lambda(\tau))^2}{2}-\partial_\tau\lambda(\tau) + \frac{a_2(\tau)}{r^2} + \mathcal{O}(r^{-3}), \label{expA}\\
B(r,\tau) &= - \frac{2 \ln(\tau)}{3} - \frac{2}{3r\tau} + \frac{1+2\tau\lambda(\tau)}{3r^2 \tau^2} \nonumber\\
& -\, \frac{2+6\tau\lambda(\tau)+6\tau^2\lambda^2(\tau)}{9r^3 \tau^3}\nonumber\\
& +\, \frac{6+24\tau\lambda(\tau)+36\tau^2\lambda^2(\tau)+24\tau^3\lambda^3(\tau)}{36r^4 \tau^4} \nonumber\\
& -\, \frac{36\tau^4 a_2(\tau)+27\tau^5 \partial_\tau a_2(\tau)}{36r^4 \tau^4} + \mathcal{O}(r^{-5}),\label{expB}\\
\Sigma(r,\tau) &= \tau^{1/3}r + \frac{1+3\tau\lambda(\tau)}{3\tau^{2/3}} - \frac{1}{9r \tau^{5/3}} \nonumber\\
&\!\!\!\!\!\!\!\!\!\!\!\!\!\!\!\!\!\!\!\!\!\!\!\!\!\! +\, \frac{5+9\tau\lambda(\tau)}{81r^2 \tau^{8/3}} - \frac{10+30\tau\lambda(\tau)+27\tau^2\lambda^2(\tau)}{243r^3 \tau^{11/3}} + \mathcal{O}(r^{-4}),\label{expS}\\
d_+\Sigma(r,\tau) &= \frac{\tau^{1/3}r^2}{2}+\frac{(1+3\tau\lambda(\tau))r}{3\tau^{2/3}} \nonumber\\
& -\, \frac{1-2\tau\lambda(\tau)-3\tau^2\lambda^2(\tau)}{6\tau^{5/3}}+\frac{10}{81r \tau^{8/3}}\nonumber\\
& +\, \frac{-25-30\tau\lambda(\tau)+243\tau^4 a_2(\tau)}{243r^2 \tau^{11/3}} + \mathcal{O}(r^{-3}),\label{expdS}\\
d_+B(r,\tau) &= -\frac{1}{3\tau}+\frac{1}{3r\tau^2}-\frac{1+\tau\lambda(\tau)}{3r^2 \tau^3}\nonumber\\
&\!\!\!\!\!\!\!\!\!\!\!\!\!\!\!\!\!\!\!\!\!\!\!\!\!\!\!\!\!\!\! +\! \frac{2\!+\!4\tau\lambda(\tau)\!+\!2\tau^2\lambda^2(\tau)\! +\!12\tau^4 a_2(\tau)\!+\!9\tau^5 \partial_\tau a_2(\tau)}{6 r^3 \tau^4}\! +\! \mathcal{O}(r^{-4}).\label{expdB}
\end{align}
\end{subequations}

As discussed in detail e.g. in Ref. \cite{Critelli:2017euk}, the holographic renormalization of the model provides formulas relating the one-point function of the boundary energy-momentum tensor with the UV coefficients of the bulk fields, generally written in Fefferman-Graham (FG) coordinates. Then, in order to use these formulas one must first find a relation between the holographic radial direction $r$ in EF coordinates and the holographic radial direction $\rho$ in FG coordinates. This relation reads,
\begin{align}
\int\frac{dr}{\sqrt{2A(r,\tau)}} = \ln(\rho^{-1/2}),
\label{FGcoord}
\end{align}
which may be perturbatively solved close to the boundary using the UV expansion of the metric coefficient $A(r,\tau)$, which gives
\begin{align}
r(\rho)=\frac{1}{\sqrt{\rho}}-\frac{a_2(\tau)\rho^{3/2}}{4}-\frac{\partial_\tau a_2(\tau)\rho^2}{10} + \mathcal{O}(\rho^{5/2}).
\label{rrho}
\end{align}
By substituting the result above into the UV expansions of the bulk fields \eqref{expA} --- \eqref{expdB}, one identifies the relevant UV coefficients in FG coordinates entering the holographic renormalization formula for the one-point function of the energy-momentum tensor of the boundary quantum gauge theory \cite{Critelli:2017euk}. The final results are given by \cite{Critelli:2018osu}\footnote{We remark that our definition of the normalized one-point function of the boundary energy-momentum tensor, $\langle\hat{T}_{\mu\nu}\rangle\equiv\kappa_5^2 \langle T_{\mu\nu}\rangle = (4\pi^2/N_c^2)\langle T_{\mu\nu}\rangle$, corresponds to twice the value of the definition used in Ref.\ \cite{Chesler:2009cy}.}
\begin{subequations}
\begin{align}
\hat{\varepsilon}(\tau) &\equiv \kappa_{5}^{2}\langle T_{\tau\tau} \rangle = -3a_2(\tau),\label{Ttt}\\
\hat{p}_T(\tau) &\equiv \kappa_{5}^{2}\langle T_x^x \rangle = -3a_2(\tau)-\frac{3}{2}\tau \partial_\tau a_2(\tau),\label{Txx}\\
\hat{p}_L(\tau) &\equiv \kappa_{5}^{2}\langle T_\xi^\xi \rangle =  3a_2(\tau)+3\tau \partial_\tau a_2(\tau),\label{Txixi}
\end{align}
\end{subequations}
where $\hat{\varepsilon}(\tau)$, $\hat{p}_T(\tau)$, and $\hat{p}_L(\tau)$ are, respectively, the (normalized) energy density, the transverse pressure, and the longitudinal pressure of the SYM plasma\footnote{Note that from Eqs.\ \eqref{Ttt} --- \eqref{Txixi} the trace anomaly $\hat{I}\equiv g^{\mu\nu}_{\textrm{(4D)}}\langle \hat{T}_{\mu\nu}\rangle=-\hat{\varepsilon}+\hat{p}_L+2\hat{p}_T$ vanishes for the SYM plasma, as expected from  conformal invariance.}. From Eqs.\ \eqref{Ttt} --- \eqref{Txixi} we see that once we determine the time evolution of the dynamical UV coefficient $a_2(\tau)$, we have also the dynamical evolution of these physical observables at the boundary. We also see from Eq.\ \eqref{expA} that $a_2(\tau)$ is a subleading UV coefficient of the metric coefficient $A(r,\tau)$ close to the boundary $r\to\infty$. In order to better extract it from the numerical solution for $A(r,\tau)$, we may define subtracted fields to get rid of the leading terms in the UV expansions of the metric coefficients, which are associated with the boundary conditions \eqref{eq5}, and also get rid of the factor of $r^{-2}$ multiplying $a_2(\tau)$ in Eq.\ \eqref{expA}.

In order to do so, and also with a view on the numerics to discussed next, we first define a new holographic radial variable  (which, as discussed before, goes from some fixed IR radial cutoff, which should lie behind the apparent horizon, to the boundary located at $r\to\infty$),
\begin{equation}
u\equiv\frac{1}{r}.
\label{eq:u}
\end{equation}
Now we define \textit{subtracted fields} as follows, $u^p X_s(u,\tau)\equiv X(u,\tau) - X_{\textrm{UV}}(u,\tau)$, where $X$ generically denotes any of the metric coefficients, $p$ is an integer, and $X_{\textrm{UV}}$ is some UV truncation of $X$. Using this reasoning and looking at Eqs.\ \eqref{expA} --- \eqref{expdB}, we define 
\begin{subequations}
\begin{align}
u^2 A_s(u,\tau) &\equiv A(u,\tau) - \frac{1}{2}\left(\frac{1}{u}+\lambda(\tau)\right)^2+\partial_\tau\lambda(\tau),\label{As}\\
u^4 B_s(u,\tau) &\equiv B(u,\tau) + \frac{2\ln(\tau)}{3} + \frac{2u}{3\tau} \nonumber\\
&\!\!\!\!\!\!\!\!\!\!\!\!\!\!\!\!\!\!\!\!\!\!\!\!\!\!\! -\, \frac{(1+2\tau\lambda(\tau))u^2}{3\tau^2} + \frac{(2+6\tau\lambda(\tau)+6\tau^2\lambda^2(\tau))u^3}{9\tau^3},\label{Bs}\\
u^3 \Sigma_s(u,\tau) &\equiv \Sigma(u,\tau) -\frac{\tau^{1/3}}{u}-\frac{1+3\tau\lambda(\tau)}{3\tau^{2/3}}+\frac{u}{9\tau^{5/3}} \nonumber\\
& -\, \frac{(5+9\tau\lambda(\tau))u^2}{81\tau^{8/3}},\label{Ss}\\
u^2 (d_+\Sigma)_s(u,\tau) &\equiv d_+\Sigma(u,\tau)-\frac{\tau^{1/3}}{2u^2}-\frac{1+3\tau\lambda(\tau)}{3u\tau^{2/3}}\nonumber\\
& +\, \frac{1-2\tau\lambda(\tau)-3\tau^2\lambda^2(\tau)}{6\tau^{5/3}} - \frac{10u}{81\tau^{8/3}},\label{dSs}\\
u^3 (d_+B)_s(u,\tau) &\equiv d_+B(u,\tau)+\frac{1}{3\tau}-\frac{u}{3\tau^2}+\frac{(1+\tau\lambda(\tau))u^2}{3\tau^3}.\label{dBs}
\end{align}
\end{subequations}

Then, the boundary values of the subtracted fields are simply given by
\begin{subequations}
\begin{align}
A_s(u=0,\tau) &= a_2(\tau),\label{AAs}\\
B_s(u=0,\tau) &= -a_2(\tau)-\frac{3\tau\partial_\tau a_2(\tau)}{4}+\frac{1}{6\tau^4}\nonumber\\
& +\, \frac{2\lambda(\tau)}{3\tau^3}+\frac{\lambda^2(\tau)}{\tau^2}+\frac{2\lambda^3(\tau)}{3\tau},\label{ABs}\\
\Sigma_s(u=0,\tau) &= -\frac{10+30\tau\lambda(\tau)+27\tau^2\lambda^2(\tau)}{243\tau^{11/3}},\label{ASs}\\
(d_+\Sigma)_s(u=0,\tau) &= \tau^{1/3} a_2(\tau)-\frac{25+30\tau\lambda(\tau)}{243\tau^{11/3}},\label{AdSs}\\
(d_+B)_s(u=0,\tau) &= 2a_2(\tau)+\frac{3\tau\partial_\tau a_2(\tau)}{2}+\frac{1}{3\tau^4}\nonumber\\
& +\, \frac{2\lambda(\tau)}{3\tau^3} + \frac{\lambda^2(\tau)}{3\tau^2}.\label{AdBs}
\end{align}
\end{subequations}

Therefore, in terms of the subtracted fields, the dynamical UV coefficient $a_2(\tau)$ entering in the holographic formulas \eqref{Ttt} --- \eqref{Txixi} can be simply obtained as the boundary value of the subtracted field $A_s(u,\tau)$. In order to solve the equations of motions for the subtracted fields, one must rewrite Eqs.\ \eqref{pde1} --- \eqref{pde4} in terms of the new radial direction $u$ and also use Eqs.\ \eqref{As} --- \eqref{dBs} to express the original fields in terms of the subtracted ones.

\subsection{The apparent horizon}
\label{sec:HolHor}

As mentioned before, for the radial integration of Einstein's equations of motion one needs to consider the entire region of the bulk geometry causally connected to the boundary. For this sake, we need first to briefly discuss the event and apparent horizons of a black hole within the bulk geometry.

The event horizon of a black hole within the asymptotically AdS$_5$ bulk is the surface where the congruence of null geodesics bifurcates, with some geodesics escaping up to the boundary and some plunging deep into the bulk. Consequently, light rays inside the event horizon can never leave the interior of the black hole, and the portion of the bulk geometry within the black hole event horizon is, therefore, causally disconnected from observers at the boundary. Thus, when there is an event horizon within the bulk, in order to integrate the radial part of the equations of motion without losing physical information about the bulk geometry, one should integrate from the event horizon (or from some radial position inside it, which adds no extra physical information) to the boundary. The radial location of the event horizon is determined by the solution of the outgoing radial null geodesics equation subjected to the condition that at asymptotically large times it is given by a zero of the metric coefficient $A(r,t)$,
\begin{align}
\frac{dr_\textrm{H}(\tau)}{d\tau} = A(r_\textrm{H}(\tau),\tau)\,\,\,\,\,\,\,\,\,\, | \,\,\,\,\,\,\,\,\,\, r_\textrm{H}(\tau\to\infty) = r_\textrm{H}^{\textrm{(eq)}},
\label{EventHor}
\end{align}
where $r_\textrm{H}^{\textrm{(eq)}}$ is the largest simple root of equation $A(r,\tau\to\infty)=0$, corresponding to the radial position of the event horizon in equilibrium. It is clear from Eq.\ \eqref{EventHor} that the event horizon in a far-from-equilibrium setup is a global feature of the bulk geometry which can be only determined by determining first the entire time evolution of the metric coefficient $A(r,\tau)$.

The aforementioned fact makes it technically inconvenient to use the event horizon as the infrared cutoff of the radial domain of integration, since we do not know its position at the beginning of the numerical simulations. One possibility to deal with this issue is to adopt a fixed IR radial cutoff for all time slices, and then check afterwards whether the event horizon was in fact beyond the chosen value radial cutoff on top of each time slice considered. If this was not the case, then the simulations should be run again with a new tentative value for the fixed IR radial cutoff. There is, however, a much easier and convenient way to handle this question, which involves the consideration of the radial position of the apparent horizon on each time slice, instead of the event horizon.

The apparent horizon corresponds to the outermost trapped null surface within the event horizon, which separates a region of the spacetime where the geodesics are directed outward with light rays moving outward and a region where the light rays along the same geodesics move inward. Therefore, within an apparent horizon all light rays move inward. A very nice and clear illustration of such state of affairs is depicted in Fig.\ 2 of Ref.\ \cite{Chesler:2009cy}. The apparent horizon converges to the event horizon at late times, therefore, they coincide in equilibrium. Notice that, since in a far-from-equilibrium setting the apparent horizon lies within the event horizon, by taking the fixed IR radial cutoff to lie at or within the apparent horizon, one automatically guarantees that the radial domain of the bulk geometry causally connected to the boundary is being properly taken into account and no physical information is being lost.

The main technical advantage of considering the apparent horizon instead of the event horizon is that the former is local in time and its radial position can be fully determined at each individual time slice without requiring knowledge about the entire time evolution of the system. In the holographic Bjorken flow of the SYM plasma the radial position of the apparent horizon significantly varies between the different time slices, which could make it unfeasible to be used as a fixed IR radial cutoff to start the integration of the equations of motion. However, one can use the residual diffeomorphism invariance discussed before to conveniently choose the value of the function $\lambda(\tau)$ on top of each time slice so that the radial position of the apparent horizon is held fixed at all times.

For any metric satisfying an Ansatz of the form given in Eq.\ \eqref{lineElement}, the radial position of the apparent horizon, \textit{when taken as a constant in time}, $r_{\textrm{AH}}$, may be determined by finding the value of the radial coordinate which solves the following equation \cite{Chesler:2013lia},
\begin{align}
d_+\Sigma(r_{\textrm{AH}},\tau)=0.
\label{AppHor}
\end{align}
The requirements that $\partial_\tau r_{\textrm{AH}}(\tau)=0$ and that Eq.\ \eqref{AppHor} holds at all times imply that $\partial_\tau d_+\Sigma(r_{\textrm{AH}},\tau)=0$, which in turn implies that $d_+(d_+\Sigma)(r_{\textrm{AH}},\tau)=A\partial_r d_+\Sigma(r_{\textrm{AH}},\tau)$. Using this condition into the constraint equation \eqref{pde5}, and then combining the obtained result with the other components of Einstein's equations, one can show that the aforementioned requirements are realized by the following condition (already written in the radial coordinate $u=1/r$),
\begin{align}
A(u_{\textrm{AH}},\tau) &= -2\Sigma(u_{\textrm{AH}},\tau)\left[3(d_+B(u_{\textrm{AH}},\tau))^2\Sigma(u_{\textrm{AH}},\tau)  \right.\nonumber\\
& \left. +\, 6u_{\textrm{AH}}^2A'(u_{\textrm{AH}},\tau)d_+\Sigma(u_{\textrm{AH}},\tau)\right] / \nonumber\\
& \left[24(u_{\textrm{AH}}^2\Sigma '(u_{\textrm{AH}},\tau)d_+\Sigma(u_{\textrm{AH}},\tau) + \Sigma^2(u_{\textrm{AH}},\tau))\right] \nonumber\\
&= -(d_+B(u_{\textrm{AH}},\tau))^2 / 4,
\label{Astar}
\end{align}
where above $'\equiv\partial_u$ and we used in the last step Eq.\ \eqref{AppHor}, $d_+\Sigma(u_{\textrm{AH}},\tau)=0$. Then, one can use Eq.\ \eqref{As} to obtain,
\begin{align}
\partial_\tau\lambda(\tau) &= u_{\textrm{AH}}^2 A_s(u_{\textrm{AH}},\tau)+\frac{1}{2u_{\textrm{AH}}^2}+\frac{\lambda(\tau)}{u_{\textrm{AH}}}+\frac{\lambda^2(\tau)}{2} \nonumber\\
& -\, A(u_{\textrm{AH}},\tau),
\label{dlambda}
\end{align}
where $A_s(u_{\textrm{AH}},\tau)$ is the numerical solution for the subtracted metric coefficient $A_s(u,\tau)$ evaluated at the apparent horizon and $A(u_{\textrm{AH}},\tau)$ is given by Eq. \eqref{Astar}. Then, once it is chosen a value for $\lambda(\tau_0)$ on the initial time slice $\tau_0$, $\lambda(\tau)$ can be evolved to the next time slice using Eq.\ \eqref{dlambda} such that the radial position of the apparent horizon remains fixed during this time evolution. For this purpose, we set the initial condition $\lambda(\tau_0)=0$ and solve Eq.\ \eqref{AppHor} using Eq.\ \eqref{dSs} and the Newton-Raphson algorithm. In general, the value of $u_{\textrm{AH}}$ does not coincide with any of the \textit{collocation points} (to be discussed in section \ref{sec:HolNum}), and we find such a value with good precision, given by the tolerance (or the number of iterations) of the method. For the initial data considered in the present work, we typically find that the radial position of the event horizon is held fixed within $\sim 10^{-5}\,\%$.

Alternatively, we also considered an approximation to calculate the radial position of the apparent horizon consisting in the following procedure: first we set the function $\lambda(\tau)\equiv 0$ and integrate the equations of motion for a short period of time and then evaluate the radial position of the apparent horizon at the initial time slice on top of numerically interpolated results. Next, we search for the point $u_\star$ within the \textit{radial grid} (to be discussed in section \ref{sec:HolNum}) which is closest to the radial position of the apparent horizon $u_{\textrm{AH}}$ at the initial time $\tau_0$ and take $u_\star$ (instead of $u_{\textrm{AH}}$) to determine an approximation for Eq.\ \eqref{dlambda}. In this case, we take $\lambda(\tau_0)=0$ as the initial condition for evolving $\lambda(\tau)$ in time. Within the aforementioned approximation, the radial position of the apparent horizon for the different initial conditions considered was held fixed within $\sim 10^{-2}\,\%$.

For all the initial data considered in the present work the apparent horizon calculated using both approaches discussed above was always found within the chosen radial domain of integration, $u\in[0,u_{\textrm{IR}}]$, where we took $u_{\textrm{IR}}=1$ as the fixed IR radial cutoff (for the different initial conditions considered in this work, we obtained apparent horizons between $u\sim 0.7$ --- $0.99$). Although the precision in the calculation of the radial position of the apparent horizon for the two approaches discussed above is different, the results for the physical observables analyzed in the present work (namely, the pressure anisotropy, and the non-equilibrium entropy density) are indistinguishable by eye.

In fact, we also considered a third way of evolving the system, where we simply set the function $\lambda(\tau)\equiv 0$ and let the radial position of the apparent horizon to freely fluctuate between the time slices. This approach is more difficult to handle in practice because we had to choose different values of the fixed IR radial cutoff $u_\textrm{IR}$ for different initial conditions. Moreover, it is also more limited in the sense that in order to avoid the breakdown of the numerical simulations due to the caustics in the deep interior of the bulk, the chosen values of $u_\textrm{IR}$, contrary to what happens in the method with fixed apparent horizon, are not guaranteed to cover the radial domain where the apparent horizon $u_\textrm{AH}$ lies within for every time slice --- in fact, at large enough times, depending on the chosen initial conditions, we could not find the apparent horizon within the radial domain of integration $[0,u_{\textrm{IR}}]$ in this method with fluctuating apparent horizon without resorting to numerical extrapolations, which may be physically unreliable. However, for the time intervals where it was possible to run the numerical code using $\lambda(\tau) = 0$, the results obtained for the pressure anisotropy and the non-equilibrium entropy density coincide with the ones obtained with fixed apparent horizon and nontrivial $\lambda(\tau)$. Such an observation is consistent with the fact that $\lambda(\tau)$ may be freely chosen since it is associated with a residual diffeomorphism invariance of the system, as discussed before.
% But $\lambda(tau_0)$ is an IC!!!

\subsection{Numerics and initial data}
\label{sec:HolNum}

In order to ensure the reproducibility of our results, in this section we give the details behind our numerical work. 

We numerically integrate the equations of motion using a discretization of both the radial and time directions. Let us first briefly discuss the discretization of the radial domain of integration of the PDEs. This is implemented here using the \textit{pseudospectral or collocation method} \cite{boyd01}, where the discrete radial points are described by the Chebyshev-Gauss-Lobatto grid,
\begin{equation}
u_k = \frac{u_{\textrm{IR}}}{2}\left[1+\cos\left(\frac{k\pi}{N-1}\right)\right], \ \ \ k=0,\dots , N-1,
\label{CBLgrid}
\end{equation}
where $N$ is the number of grid points (also known as collocation points) and $u_{\textrm{IR}}$ is the fixed infrared radial cutoff in the interior of the bulk, from which one must radially integrate the equations of motion up to the boundary at $u=0$. As discussed previously, for the present work we take $u_{\textrm{IR}}=1$.

The main reasoning involved in the use of the pseudospectral method consists in radially expanding the bulk fields in the basis of Chebyshev polynomials of the first kind on each point of the Chebyshev-Gauss-Lobato radial grid \eqref{CBLgrid}. The expansion coefficients so obtained are called the \textit{spectral coefficients} of the Chebyshev expansion. The expansion is truncated at the same order of the number of collocation points used in the discrete radial grid, therefore, in principle, the larger the number of collocation points the more accurate the numerical results should be. One major numerical advantage of the pseudospectral method over alternative methods, such as finite differences, is that the convergence of the Chebyshev expansion is expected to be exponential (rather than polynomial) in the number of collocation points.

As discussed in detail e.g. in section 5.4 of Ref.\ \cite{Critelli:2017euk}, by discretizing the radial part of the continuum differential equations of motion using the pseudospectral method, one is left with a linear algebra eigenvalue problem essentially consisting in the inversion of a diagonal $(N-1)\times (N-1)$ matrix for each of the bulk fields. These matrices are given by the homogeneous part\footnote{The multiplication of these inverse matrices by the column vectors corresponding to the inhomogeneous part of the equations of motion gives the numerical solutions for the bulk fields.} of the discretized differential equations of motion evaluated at each radial grid point, excluding the point corresponding to the boundary. Indeed, at the boundary grid point one must impose the boundary conditions \eqref{AAs} --- \eqref{AdBs} for the (subtracted) bulk fields. Then, in practice, one simply joins to the $(N-1)$-dimensional eigenvectors obtained as solutions of the aforementioned eigenvalue problem the values of the respective bulk fields determined at the boundary grid point by the corresponding boundary conditions. With this, one constructs the complete $N$-dimensional eigenvectors corresponding to the numerical solutions of the radial part of the Einstein's equations of motion (the components of the $N$-dimensional eigenvectors are the values of the bulk fields on each of the $N$ collocation points of the discrete radial grid).

At the initial time slice $\tau_0$, the value of $a_2(\tau_0)=A_s(u=0,\tau_0)$ must be specified. Therefore, as mentioned before, the initial value of the dynamical UV coefficient $a_2(\tau_0)$ is one of the initial conditions which must be specified on the gravity side of the gauge/gravity duality (the other initial condition being the initial profile of the bulk metric anisotropy, $B_s(u,\tau_0)$)\footnote{We remark that by using $\lambda(\tau)\neq 0$ the initial value $\lambda(\tau_0)$ must also be specified.}.

% OBS (R.R.): PELO O QUE TESTEI, O FILTRO PARECE SER DESNECESSARIO PARA SOLUCOES DE BJOKEN FLOW NA SYM PURA!!!

In order to evolve in time the set of initial data $\{a_2(\tau),B_s(u,\tau);\lambda(\tau)\}$, one also needs to determine the values of the time derivatives $\partial_\tau a_2(\tau)$, $\partial_\tau B_s(u,\tau)$, and $\partial_\tau\lambda(\tau)$. The latter is calculated by using Eq.\ \eqref{dlambda}. Moreover, having knowledge of the values of $a_2(\tau_0)$, $B_s(u=0,\tau_0)$, and $\lambda(\tau_0)$, which on the initial time slice are simply the freely chosen initial conditions, one can determine the value of $\partial_\tau a_2(\tau)$ at the initial time $\tau_0$ using Eq.\ \eqref{ABs}.

Finally, we also need to determine  $\partial_\tau B_s(u,\tau)$. This can be done by using Eq.\ \eqref{dBs} to relate the numerical field $(d_+ B)_s(u,\tau)$ to $d_+ B(u,\tau)=\partial_\tau B(u,\tau)-u^2 A(u,\tau) \partial_u B(u,\tau)$, and then expressing in this relation $A(u,\tau)$ and $B(u,\tau)$ in terms of the corresponding subtracted fields as given by Eqs.\ \eqref{As} and \eqref{Bs}. The resulting equation is solved for $\partial_\tau B_s(u,\tau)$ giving
\begin{align}
\partial_\tau B_s(u,\tau) &= \frac{(d_+B)_s(u,\tau)}{u} -\frac{2}{3\tau^4 u} - \frac{2A_s(u,\tau)}{3\tau} \nonumber\\
& +\, \frac{2uA_s(u,\tau)}{3\tau^2} - \frac{2u^2A_s(u,\tau)}{3\tau^3} \nonumber\\
& +\, 4u^3A_s(u,\tau)B_s(u,\tau) + \frac{2B_s(u,\tau)}{u} \nonumber\\
& +\, \frac{B_s'(u,\tau)}{2} + u^4A_s(u,\tau)B_s'(u,\tau) \nonumber\\
& +\, \left(4B_s(u,\tau)-\frac{2}{u\tau^3}+\frac{4u A_s(u,\tau)}{3\tau} \right.\nonumber\\
& \left. -\, \frac{2u^2 A_s(u,\tau)}{\tau^2}+uB_s'(u,\tau)\right)\lambda(\tau)\nonumber\\
& +\! \left(\! - \frac{1}{3\tau^3} \!-\! \frac{7}{3\tau^2 u} \!+\! 2uB_s(u,\tau) \!-\! \frac{2u^2A_s(u,\tau)}{\tau} \right.\nonumber\\
& \left. +\, \frac{u^2B_s'(u,\tau)}{2} \right)\lambda^2(\tau) - \left( \frac{1}{\tau^2} + \frac{4}{3\tau u} \right)\lambda^3(\tau) \nonumber\\
& -\, \frac{\lambda^4(\tau)}{\tau} + \left( \frac{2}{3\tau^3} - 4uB_s(u\tau) - u^2B_s'(u,\tau) \right.\nonumber\\
& \left. +\, \frac{2\lambda(\tau)}{\tau^2} + \frac{2\lambda^2(\tau)}{\tau} \right)\partial_\tau\lambda(\tau) ,
\label{dtBs}
\end{align}
where $B_s'(u,\tau)\equiv \partial_u B_s(u,\tau)$ can be obtained at any constant time slice by simply applying the pseudospectral finite differentiation matrix \cite{Critelli:2017euk,boyd01} to the numerical solution $B_s(u,\tau)$ (which is expressed as a vector field with $N$ components, as discussed before).

For the time evolution of the gravitational system we employ here the fourth-order Adams-Bashforth (AB) integration method. This method requires earlier initialization by other methods, so in order to evolve the system from the initial time slice to the next one we use the Euler method (also known as first-order AB),
\begin{align}
X(\tau+\Delta\tau) = X(\tau) + \Delta\tau\partial_\tau X(\tau),
\end{align}
the next time evolution is done with second-order AB,
\begin{align}
X(\tau+\Delta\tau) = X(\tau) + \frac{\Delta\tau}{2} [3\, \partial_\tau X(\tau) - \partial_\tau X(\tau-\Delta\tau)],
\end{align}
the subsequent time evolution is done with third-order AB,
\begin{align}
X(\tau+\Delta\tau) &= X(\tau) + \frac{\Delta\tau}{12} [23\, \partial_\tau X(\tau) - 16\, \partial_\tau X(\tau-\Delta\tau) \nonumber\\
& +\, 5\, \partial_\tau X(\tau-2\Delta\tau)],
\end{align}
and then, finally, all the next steps corresponding to the subsequent time slices are done using fourth-order AB,
\begin{align}
X(\tau+\Delta\tau) &= X(\tau) + \frac{\Delta\tau}{24} [55\, \partial_\tau X(\tau) - 59\, \partial_\tau X(\tau-\Delta\tau) \nonumber\\ & +\, 37\, \partial_\tau X(\tau-2\Delta\tau) - 9\, \partial_\tau X(\tau-3\Delta\tau)],
\end{align}
where $\Delta\tau$ is the time step size and $X(\tau)$ denotes either $a_2(\tau)$, $B_s(u,\tau)$, or $\lambda(\tau)$. In the present work we used $N = 33$ collocation points and $\Delta\tau=12\times 10^{-5}$. Further numerical details and an error analysis can be found in Appendix \ref{sec:appA}.

The form of the initial conditions used in the present work are similar to the ones chosen in Refs.\ \cite{Critelli:2018osu,wilkaodamassa},
\begin{subequations}
\begin{align}
%& \,\,\,\,\,\,\,\;\! a_2(\tau_0) = -20/3,\label{a20}\\
& B_s(u,\tau_0) = \Omega_1 \cos(\gamma_1 u) + \Omega_2 \tan(\gamma_2 u) + \Omega_3 \sin(\gamma_3 u) \nonumber\\ 
& + \sum_{i=0}^{5}\beta_i u^i + \frac{\alpha}{u^4} \left[-\frac{2}{3} \ln\left(1+ \frac{u}{\tau _0}\right) + \frac{2 u^3}{9 \tau_0^3} - \frac{u^2}{3 \tau _0^2}+\frac{2 u}{3 \tau _0}\right],\label{Bs0}\\
& \,\,\,\,\,\,\,\,\, \lambda(\tau_0) = 0\label{lambda0},
\end{align}
\end{subequations}
with the chosen values for $a_2(\tau_0)$, which determine the initial energy density of the fluid through Eq. \eqref{Ttt}, displayed in table \ref{tabICs}. We choose $\tau_0=0.2$ as the initial time of our numerical simulations. By choosing different values for the set of parameters $\{\Omega_i,\gamma_i,\beta_i,\alpha\}$ in the initial metric anisotropy \eqref{Bs0}, as depicted in table \ref{tabICs}, one generates very different solutions for the physical observables of the SYM plasma.

\begin{widetext}
\begin{table*}[ht]
\centering
\begin{tabular}{|c||c|c|c|c|c|c|c|c|c|c|c|c|c||c|}
\hline
IC$\#$ & $\Omega_1$ & $\gamma_1$ & $\Omega_2$ & $\gamma_2$ & $\Omega_3$ & $\gamma_3$ & $\beta_0$ & $\beta_1$ & $\beta_2$ & $\beta_3$ & $\beta_4$ & $\beta_5$ & $\alpha$ & $a_2(\tau_0)$ \\
\hline
\hline
1 & 0 & 0 & 0 & 0 & 0 & 0 & 0.5 & -0.5 & 0.4 & 0.2 & -0.3 & 0.1 & 1 & -20/3 \\
\hline
2 & 0 & 0 & 0 & 0 & 0 & 0 & 0.2 & 0.1 & -0.1 & 0.1 & 0.2 & 0.5 & 1.02 & -20/3 \\
\hline
3 & 0 & 0 & 0 & 0 & 0 & 0 & 0.1 & -0.5 & 0.5 & 0 & 0 & 0 & 1 & -20/3 \\
\hline
4 & 0 & 0 & 0 & 0 & 0 & 0 & 0.1 & 0.2 & -0.5 & 0 & 0 & 0 & 1 & -20/3 \\
\hline
5 & 0 & 0 & 0 & 0 & 0 & 0 & -0.1 & -0.4 & 0 & 0 & 0 & 0 & 1 & -20/3 \\
\hline
6 & 0 & 0 & 0 & 0 & 0 & 0 & -0.2 & -0.5 & 0.3 & 0.1 & -0.2 & 0.4 & 1 & -20/3 \\
\hline
7 & 0 & 0 & 0 & 0 & 0 & 0 & 0.1 & -0.4 & 0.3 & 0 & -0.1 & 0 & 1 & -20/3 \\
\hline
8 & 0 & 0 & 0 & 0 & 0 & 0 & 0 & 0.2 & 0 & 0.4 & 0 & 0.1 & 1 & -20/3 \\
\hline
9 & 0 & 0 & 0 & 0 & 0 & 0 & 0.1 & -0.2 & 0.3 & 0 & -0.4 & 0.2 & 1.03 & -20/3 \\
\hline
10 & 0 & 0 & 0 & 0 & 0 & 0 & 0.1 & -0.4 & 0.3 & 0 & -0.1 & 0 & 1.01 & -20/3 \\
\hline
11 & 1 & 1 & 0 & 0 & 0 & 0 & 0 & 0 & 0 & 0 & 0 & 0 & 1 & -20/3 \\
\hline
12 & 0 & 0 & 1 & 1 & 0 & 0 & 0 & 0 & 0 & 0 & 0 & 0 & 1 & -20/3 \\
\hline
13 & 0 & 0 & 0 & 0 & 0 & 0 & 0.1 & -0.4 & 0.4 & 0 & -0.1 & 0 & 1 & -20/3 \\
\hline
14 & 0 & 0 & 0 & 0 & 0 & 0 & -0.2 & -0.5 & 0.3 & 0.1 & -0.2 & 0.3 & 1.01 & -20/3 \\
\hline
15 & 0 & 0 & 0 & 0 & 0 & 0 & -0.2 & -0.3 & 0 & 0 & 0 & 0 & 1 & -20/3 \\
\hline
16 & 0 & 0 & 0 & 0 & 0 & 0 & -0.2 & -0.5 & 0 & 0 & 0 & 0 & 1 & -20/3 \\
\hline
17 & 0 & 0 & 0 & 0 & 0 & 0 & -0.1 & -0.3 & 0 & 0 & 0 & 0 & 1 & -20/3 \\
\hline
18 & 0 & 0 & 0 & 0 & 0 & 0 & -0.1 & -0.2 & 0 & 0 & 0 & 0 & 1 & -20/3 \\
\hline
19 & 0 & 0 & 0 & 0 & 0 & 0 & -0.5 & 0.2 & 0 & 0 & 0 & 0 & 1 & -20/3 \\
\hline
20 & 0 & 0 & 0 & 0 & 0 & 0 & -0.2 & -0.4 & 0 & 0 & 0 & 0 & 1 & -20/3 \\
\hline
21 & 0 & 0 & 0 & 0 & 0 & 0 & -0.2 & -0.6 & 0 & 0 & 0 & 0 & 1 & -20/3 \\
\hline
22 & 0 & 0 & 0 & 0 & 0 & 0 & -0.3 & -0.5 & 0 & 0 & 0 & 0 & 1 & -20/3 \\
\hline
23 & 0 & 0 & 0 & 0 & 1 & 8 & 0 & 0 & 0 & 0 & 0 & 0 & 1 & -20/3 \\
\hline
24 & 1 & 8 & 0 & 0 & 0 & 0 & -0.2 & -0.5 & 0 & 0 & 0 & 0 & 1 & -7.75 \\
\hline
25 & 0.5 & 8 & 0 & 0 & 0 & 0 & -0.2 & -0.5 & 0 & 0 & 0 & 0 & 1 & -7.1 \\
\hline
\end{tabular}
\caption{Initial conditions (ICs) used in the present work, which were originally considered in \cite{Rougemont:2021qyk}. ICs $\# 16$, $\# 21$, and $\# 22$ (the magenta, red, and salmon curves in Figs. \ref{fig:result1} and \ref{fig:result2}, respectively) generate solutions for the pressure anisotropy transiently violating the dominant energy condition at early times, while ICs $\# 23$ (which was originally proposed in \cite{wilkaodamassa}), $\# 24$, and $\# 25$ (the blue, orange, and purple curves, respectively) also transiently violate the weak energy condition when the fluid is far from equilibrium.}
\label{tabICs}
\end{table*}
\end{widetext}

\subsection{Energy conditions}
\label{sec:ECs}

Energy conditions \cite{HawkingEllisBook,WaldBookGR1984} are usually postulated in general relativity to constrain the form of the energy-momentum tensor of matter used in Einstein's equations based on classical expectations related to the positiveness of energy, even though some quantum effects are known to violate such energy conditions \cite{Visser:1999de}.

The \textit{weak energy condition} (WEC) states that $\langle \hat{T}_{\mu\nu}\rangle t^\mu t^\nu\ge 0$ for any timelike vector $t^\mu$. It was shown in \cite{Janik:2005zt} that this implies the following set of inequalities for the Bjorken flow of a conformal field theory (as e.g. the SYM plasma),
\begin{align}
\hat{\varepsilon}(\tau)\ge 0,\qquad \partial_\tau\hat{\varepsilon}(\tau)\le 0,\qquad \tau\partial_\tau\ln(\hat{\varepsilon}(\tau))\ge -4.
\label{WEC}
\end{align}
In face of Eq.\ \eqref{deltaP}, the two last inequalities above imply that $-4\le \Delta\hat{p}/\hat{\varepsilon}\le 2$. Therefore, one can see that the WEC leads to direct constraints on the magnitude of the pressure anisotropy. Also, its violation does not necessarily imply that the local energy density is negative --- that is just one of the conditions in \eqref{WEC}.

We also remark that the \textit{strong energy condition} (SEC), which states that $\langle\hat{T}_{\mu\nu}\rangle t^\mu t^\nu\ge -\langle \hat{T}_\mu^\mu\rangle /2$, is trivially equivalent to the WEC for a conformal fluid, since in this case $\langle\hat{T}_\mu^\mu\rangle=0$.

The \textit{dominant energy condition} (DEC) states that for any future directed timelike vector $t^\mu$ (i.e. $t^\tau > 0$), the vector $X^\mu \equiv -\langle\hat{T}^{\mu\nu}\rangle t_\nu$ must also be a future directed timelike or null vector. This condition is important to establish causal propagation of matter \cite{WaldBookGR1984}. Let us now work in detail the DEC for Bjorken expanding SYM fluid closely following the reasoning discussed in Ref.\ \cite{Janik:2005zt}, originally used to derive the WEC.
The energy-momentum tensor for a conformal fluid undergoing Bjorken flow may be written as follows
\begin{align}
\hat{T}_{\mu\nu} &= \textrm{diag}\left(\hat{\varepsilon},\hat{p}_T,\hat{p}_T,\tau^2\hat{p}_L\right)\nonumber\\
&= \textrm{diag}\left(\hat{\varepsilon},\hat{\varepsilon}+\tau\partial_\tau\hat{\varepsilon}/2,\hat{\varepsilon}+\tau\partial_\tau\hat{\varepsilon}/2,-\tau^2\hat{\varepsilon}-\tau^3\partial_\tau\hat{\varepsilon}\right),
\label{eq:TCFT}
\end{align}
where in the last line we made use of Eqs.\ \eqref{Ttt} --- \eqref{Txixi}.
On the other hand, taking $s,v,w\in\mathbb{R}$, the most general timelike or null 4-vector at the flat boundary is given by
\begin{align}
t^\mu = \left(\sqrt{s^2+2w^2+\tau^2v^2},w,w,v\right) \Rightarrow t_\mu t^\mu = -s^2 \le 0.
\label{eq:4tmu}
\end{align}
The condition that $X^\tau > 0$ then implies that $\hat{\varepsilon} > 0$, while $X_\mu X^\mu \le 0$ leads to a more restricted condition for $\Delta p/\varepsilon$ than in the WEC, namely, $-1\le \Delta\hat{p}/\hat{\varepsilon}\le 2$. Therefore, the dominant energy condition leads to more stringent constraints on the pressure anisotropy of the fluid. Given that the pressure anisotropy (or, equivalently, the shear-stress tensor) is only nonzero out of equilibrium, the discussion concerning the investigation of energy conditions, and their possible violations, can be useful to systematically characterize the far-from-equilibrium dynamics of rapidly expanding systems.

We remark that it is known that in SYM theory the weak energy condition can be violated in holographic shockwave collisions, as shown in \cite{Arnold:2014jva}. Our results displayed in Fig.\ \ref{fig:result1} (a) show that the DEC and also the WEC can be violated even in the much simpler holographic Bjorken flow for the SYM plasma, as originally discussed in \cite{Rougemont:2021qyk}. Therefore, this suggests that the violation of energy conditions in strongly coupled holographic fluids far from equilibrium is a common feature of such systems. This should be contrasted with other approaches commonly used to investigate the quark-gluon plasma in heavy-ion collisions, such as relativistic kinetic theory \cite{degroot}, where the positiveness of the distribution function ensures that these energy conditions cannot be violated \cite{Arnold:2014jva}.

\subsection{Holographic non-equilibrium entropy}
\label{sec:HolS}

Now we discuss the calculation of the holographic non-equilibrium entropy density. The Bekenstein-Hawking relation \cite{Bekenstein:1973ur,Hawking:1974sw} associates the thermodynamical entropy of a black hole in equilibrium with the area of its event horizon. However, in out-of-equilibrium settings, it was argued in Ref.\ \cite{Figueras:2009iu} that the holographic non-equilibrium entropy should be associated with the area of the apparent horizon instead of the area of the event horizon\footnote{Ref.\ \cite{Figueras:2009iu} provided a clear example to justify this argument: in the case of conformal soliton flow \cite{Friess:2006kw}, which corresponds to an ideal fluid, entropy production must be identically zero at all times. While the entropy calculated through the area of the apparent horizon in this case is in fact constant, the area of the event horizon diverges showing that it is an inadequate measure of the non-equilibrium entropy.}. In fact, the holographic non-equilibrium entropy has been considered in many other works \cite{Chesler:2009cy,Heller:2011ju,Heller:2012je,Jankowski:2014lna,vanderSchee:2014qwa,Grozdanov:2016zjj,Buchel:2016cbj,Muller:2020ziz,Engelhardt:2017aux} as being related to the area of the apparent horizon. As mentioned before, the apparent horizon lies behind the event horizon and converges to the latter at late times and, thus, for sufficiently long times the areas of both horizons coincide giving the same result for the entropy in equilibrium.

In order to obtain the radial position of the apparent horizon, $u_{\textrm{AH}}(\tau)$, at each time slice of the numerically generated background geometries, we look for the largest root of the transcendental equation \eqref{AppHor}, which in terms of the numerically known subtracted field $(d_+\Sigma)_s(u,\tau)$ reads,
\begin{align}
(d_+\Sigma)_s(u_{\textrm{AH}},\tau) &= -\, \frac{10}{81u_{\textrm{AH}}\tau^{8/3}} - \frac{\tau^{1/3}}{2u_{\textrm{AH}}^4} - \frac{1+3\tau\lambda(\tau)}{3u_{\textrm{AH}}^3\tau^{2/3}} \nonumber\\
& +\, \frac{1-2\tau\lambda(\tau)-3\tau^2\lambda^2(\tau)}{6u_{\textrm{AH}}^2\tau^{5/3}}.
\label{dShor}
\end{align}

The area of the apparent horizon in  holographic Bjorken flow is given by
\begin{align}
A_{\textrm{AH}}(\tau) = \int d^3x \sqrt{-g}\biggr|_{u=u_{\textrm{AH}}} &= \int dx dy d\xi \sqrt{-g}\biggr|_{u=u_{\textrm{AH}}} \nonumber\\
&= \sqrt{-g}\biggr|_{u=u_{\textrm{AH}}} \mathcal{A} \nonumber\\
&= |\Sigma(u_{\textrm{AH}},\tau)|^3 \mathcal{A},
\label{AreaHor}
\end{align}
where $\mathcal{V}(\tau) = \tau \mathcal{A} = \tau \int dx dy d\xi$ is the expanding spatial volume of the fluid in Milne coordinates (note that the spacetime rapidity $\xi$ is dimensionless).

The Bekenstein-Hawking relation for the non-equilibrium holographic entropy reads
\begin{align}
S(\tau) = \frac{A_{\textrm{AH}}(\tau)}{4G_5} = \frac{2\pi |\Sigma(u_{\textrm{AH}},\tau)|^3 \mathcal{A}}{\kappa_5^2},
\label{BHrel}
\end{align}
while the (normalized) entropy density is given by
\begin{align}
\hat{s}(\tau) \equiv \kappa_5^2\, s(\tau) = \kappa_5^2\, \frac{S(\tau)}{\mathcal{V}(\tau)} = \frac{2\pi|\Sigma(u_{\textrm{AH}},\tau)|^3}{\tau},
\label{shat}
\end{align}
where $\Sigma(u_{\textrm{AH}},\tau)$ can be found in terms of the numerically known subtracted field $\Sigma_s(u,\tau)$ using \eqref{Ss},
\begin{align}
\Sigma(u_{\textrm{AH}},\tau) &= u_{\textrm{AH}}^3\Sigma_s(u_{\textrm{AH}},\tau)+\frac{\tau^{1/3}}{u_{\textrm{AH}}}-\frac{u_{\textrm{AH}}}{9\tau^{5/3}} \nonumber\\
& +\, \frac{1+3\tau\lambda(\tau)}{3\tau^{2/3}} + \frac{(5+9\tau\lambda(\tau)) u_{\textrm{AH}}^2}{81\tau^{8/3}}.
\label{lolol}
\end{align}

In the rest frame of fluid, the flow velocity 4-vector is $u^\mu=(1,0,0,0)$, while the (normalized) entropy 4-current density $\hat{s}^\mu(\tau)$ is simply given by
\begin{align}
\hat{s}^\mu(\tau) = \hat{s}(\tau) u^\mu = (\hat{s}(\tau),0,0,0).
\label{smu}
\end{align}
The entropy production is given by the 4-divergence of the entropy 4-current density,
\begin{align}
\nabla_\mu \hat{s}^\mu(\tau) &= \frac{1}{\sqrt{-g_{\textrm{(4D)}}}}\, \partial_\mu\left[\sqrt{-g_{\textrm{(4D)}}}\, \hat{s}(\tau) \delta^{\mu\tau} \right] \nonumber\\
&= \frac{\hat{s}(\tau)}{\tau} + \frac{d\hat{s}(\tau)}{d\tau} = \frac{2\pi}{\tau\mathcal{A}} \frac{d A_{\textrm{AH}}(\tau)}{d\tau}.
\label{sprod}
\end{align}

The covariant form of the second law of thermodynamics is expressed as the requirement that the entropy production \eqref{sprod} is non-negative or, equivalently, that the area of the apparent horizon is non-decreasing,
\begin{align}
\nabla_\mu \hat{s}^\mu(\tau) \ge 0 \qquad\Rightarrow\qquad \frac{d A_{\textrm{AH}}(\tau)}{d\tau} \ge 0,
\label{2ndlaw}
\end{align}
where, in Bjorken flow, the equality should be saturated when $\tau\to\infty$.

%%%%%%%%%%%%%%%%%%%%%%%%%%%%%%%%%%
\section{Holographic results and hydrodynamization}
\label{sec:HolResults}

\begin{figure*}%[h]
\center
\subfigure[]{\includegraphics[width=0.47\textwidth]{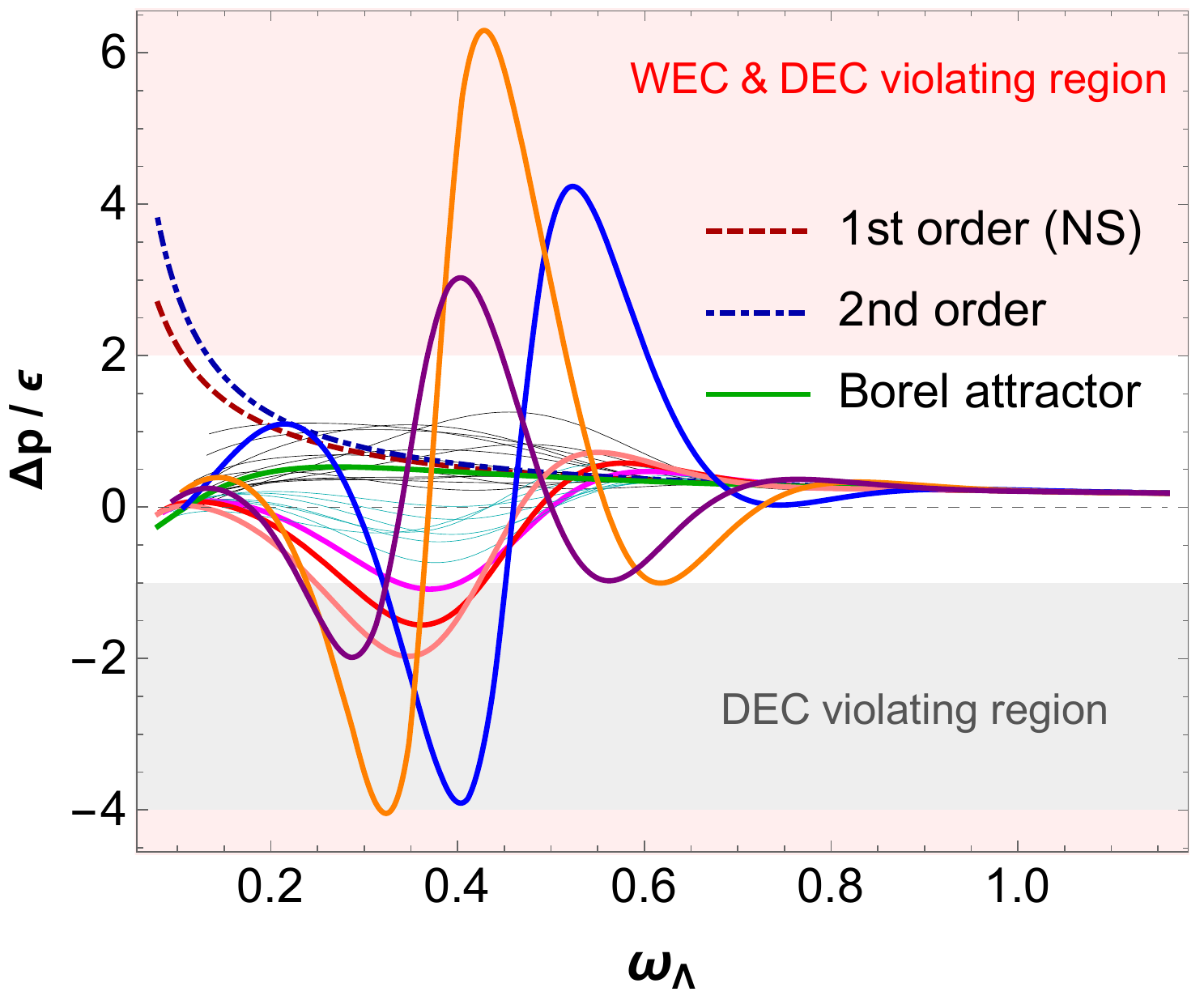}}
\qquad
\subfigure[]{\includegraphics[width=0.47\textwidth]{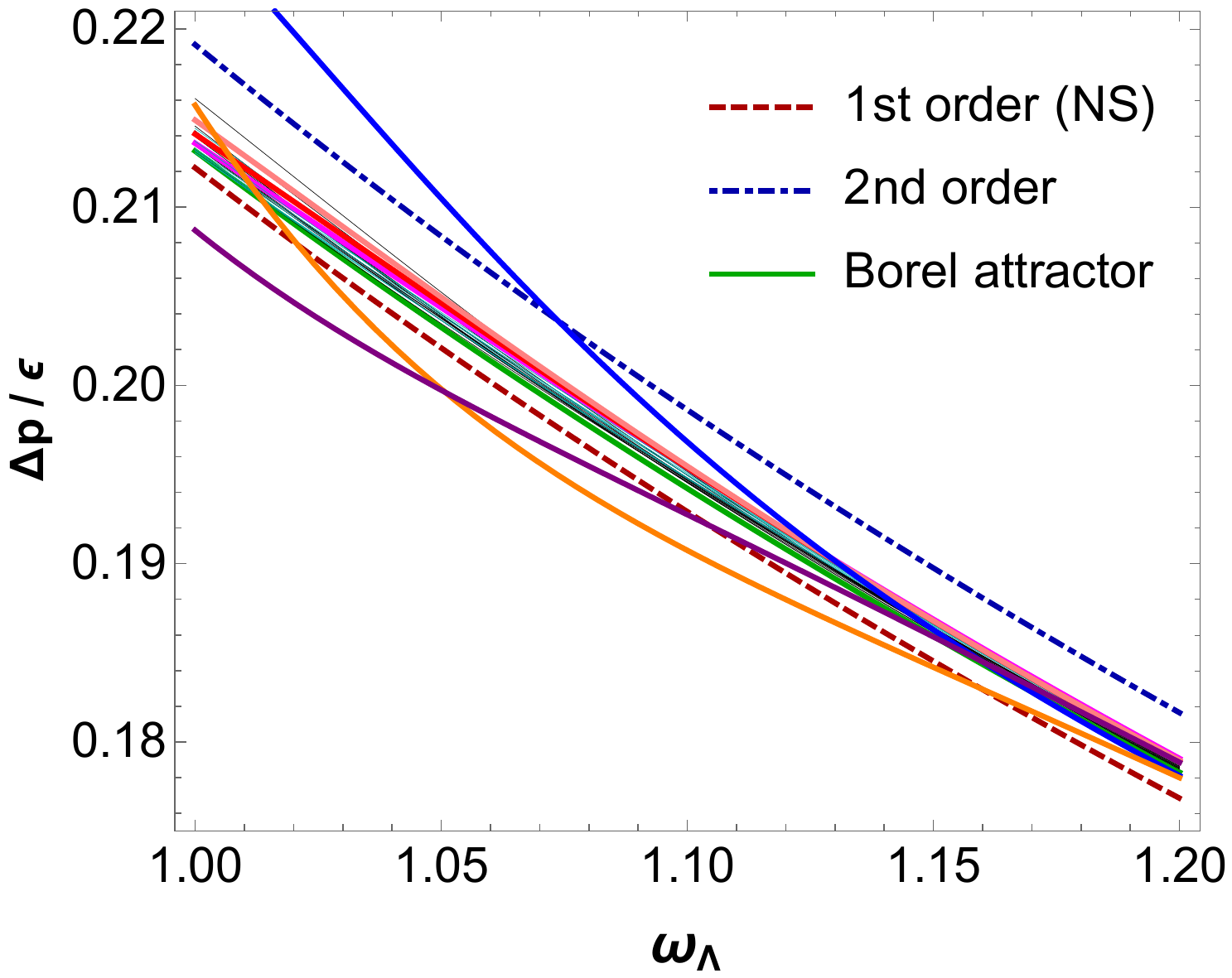}}
\qquad
\subfigure[]{\includegraphics[width=0.47\textwidth]{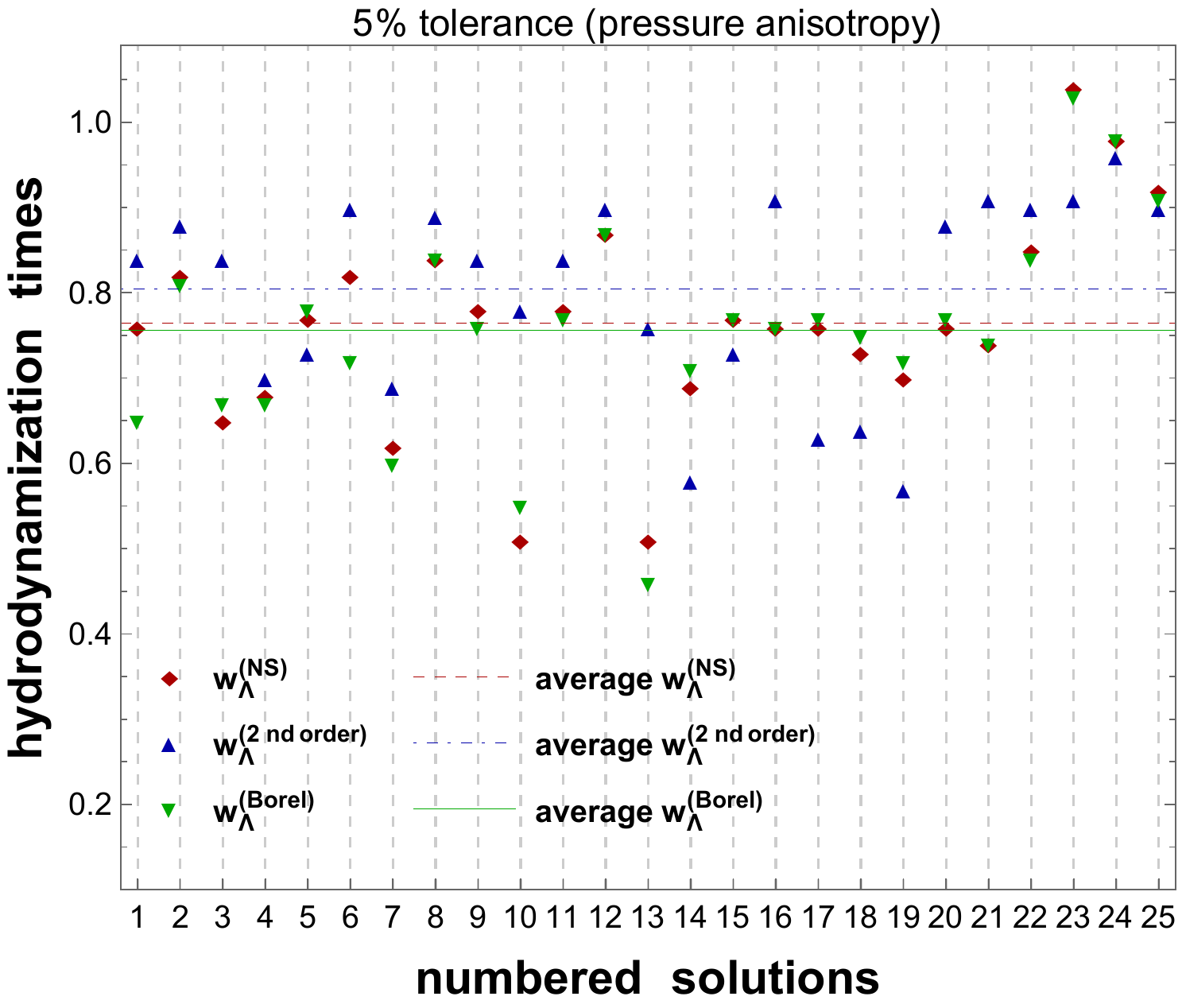}}
\qquad
\subfigure[]{\includegraphics[width=0.47\textwidth]{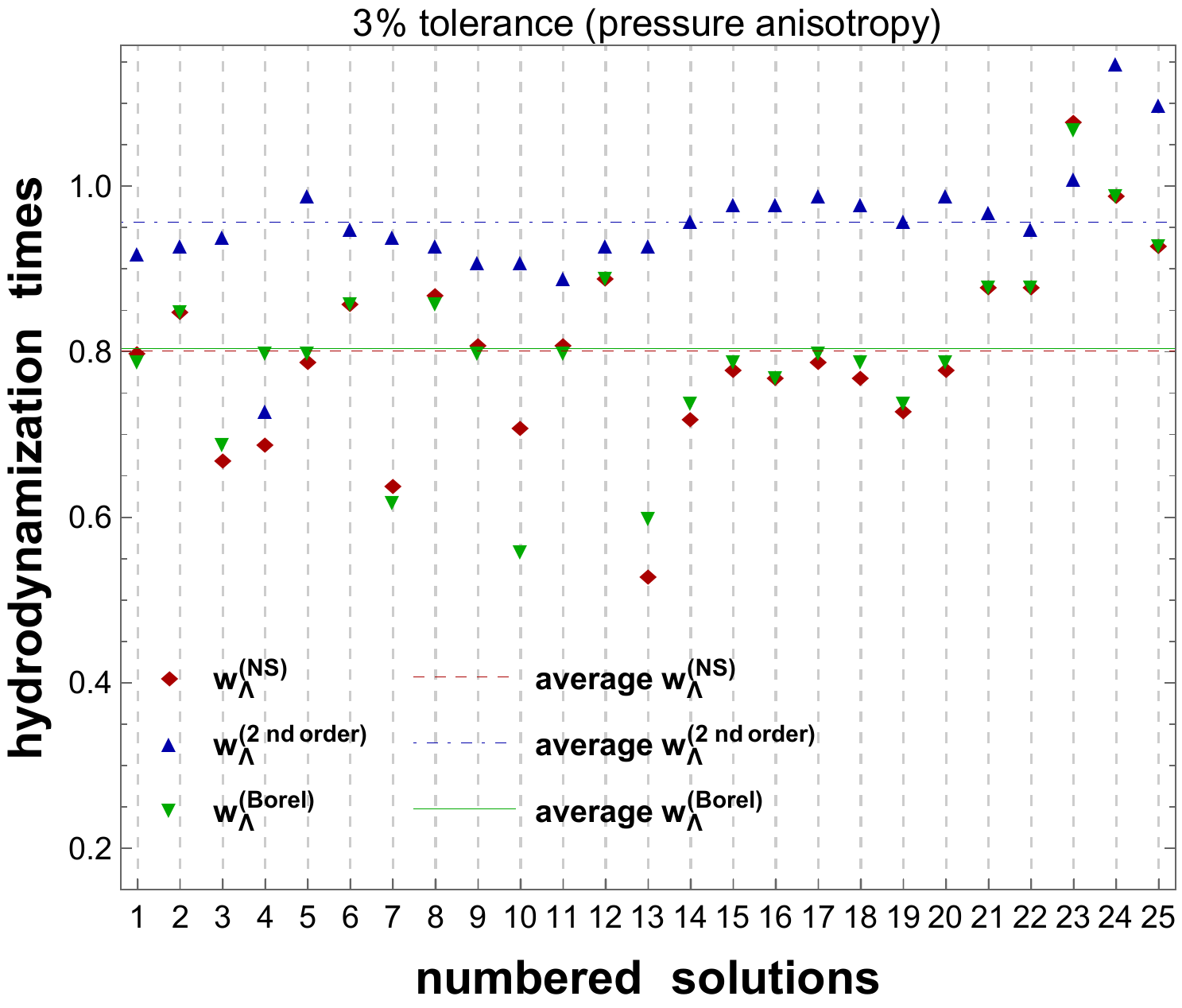}}
\caption{(a) Pressure anisotropy for the ensemble of far-from-equilibrium solutions and the corresponding late time hydrodynamic attractors (the 6 full colored thick curves are used to highlight the solutions transiently violating the energy conditions). (b) Zoom of the late time region for the pressure anisotropy. (c) Individual hydrodynamization times for the pressure anisotropy of the different solutions and the corresponding average times (taking into account all the solutions) with $5\%$ tolerance and (d) with $3\%$ tolerance.}
\label{fig:result1}
\end{figure*}

\begin{figure*}%[h]
\center
\subfigure[]{\includegraphics[width=0.47\textwidth]{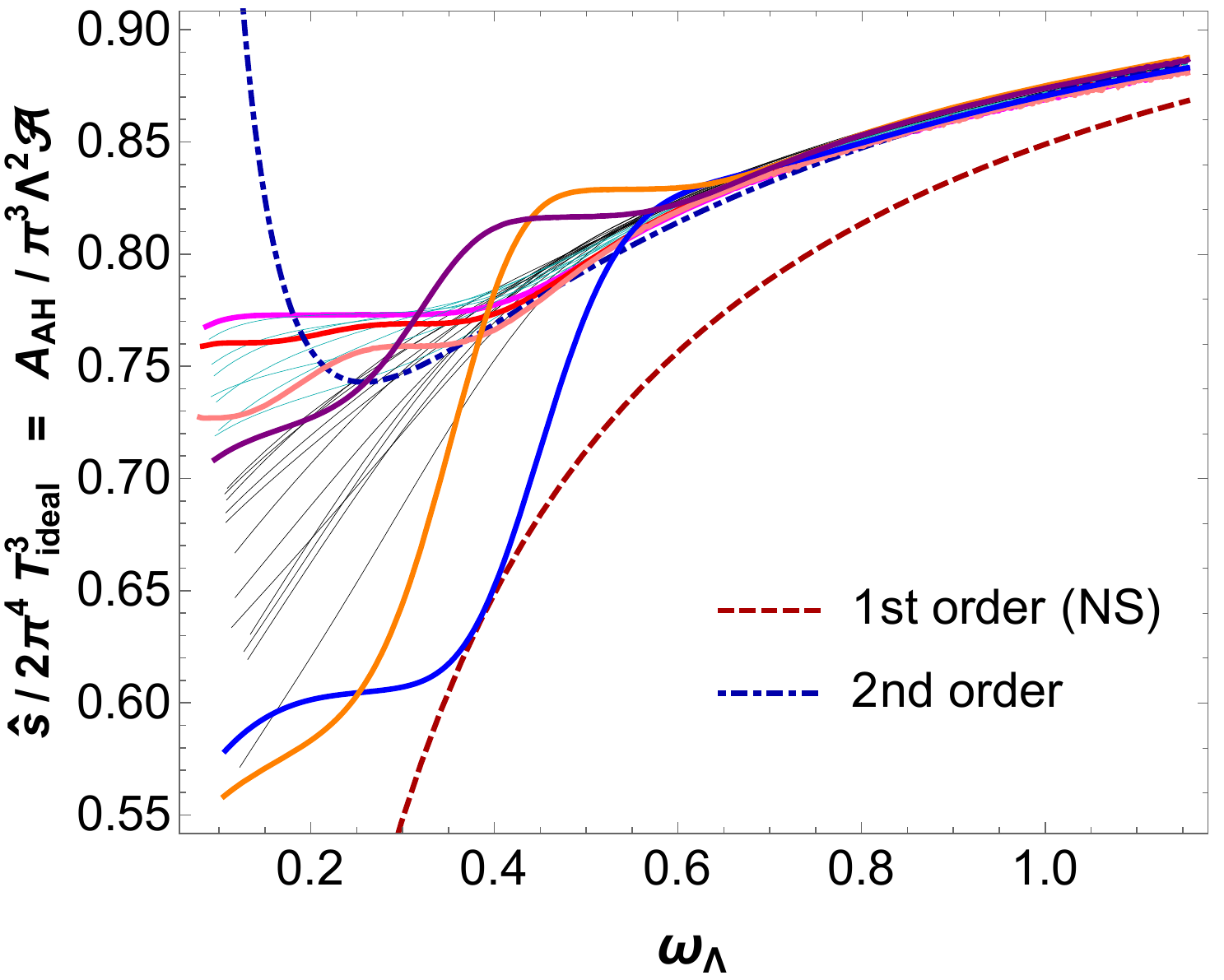}}
\qquad
\subfigure[]{\includegraphics[width=0.47\textwidth]{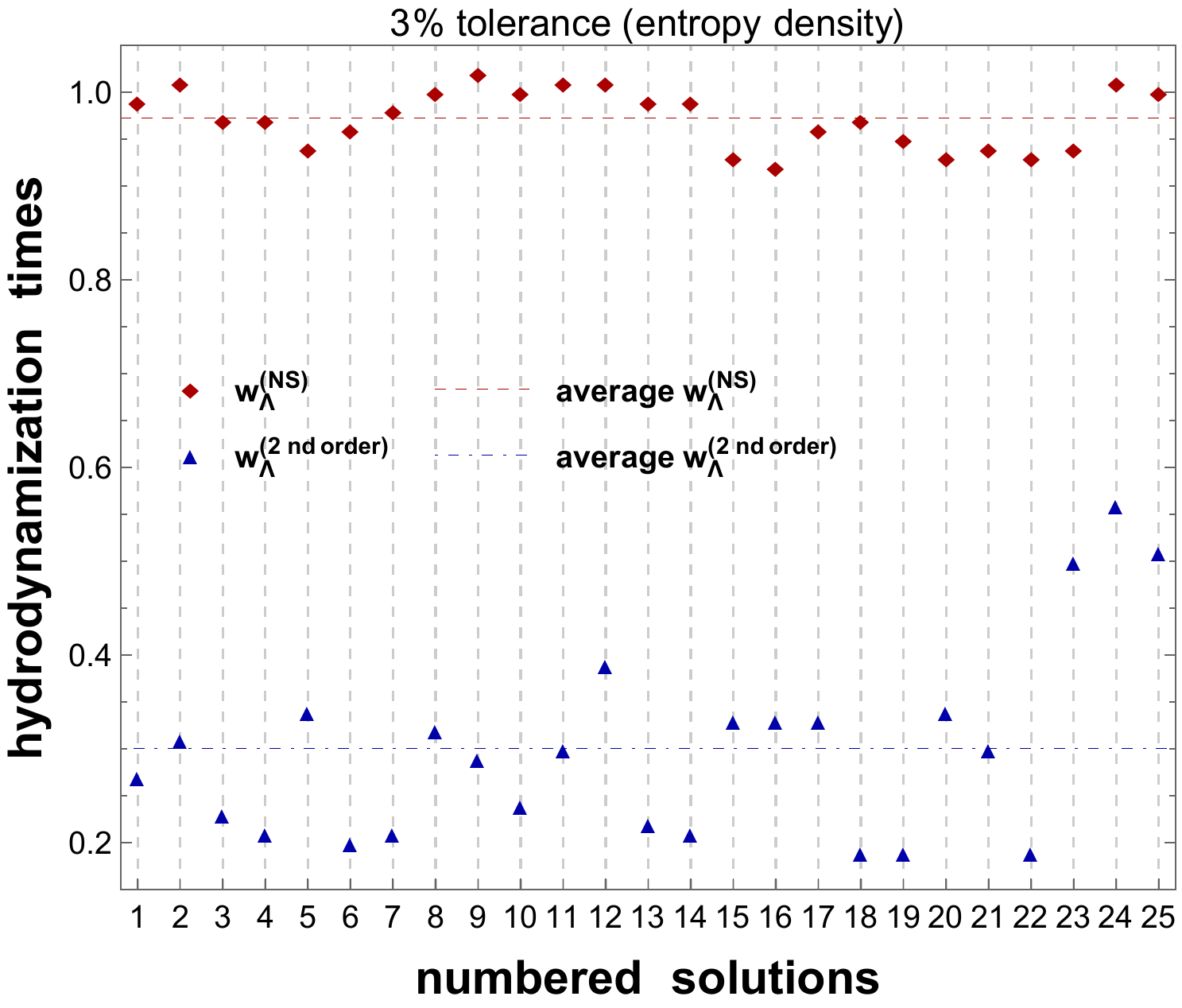}}
\caption{(a) Holographic non-equilibrium entropy density for the ensemble of far-from-equilibrium solutions and the corresponding late time hydrodynamic attractors (the 6 full colored thick curves are used to highlight the solutions transiently violating the energy conditions). (b) Individual hydrodynamization times for the non-equilibrium entropy density of the different solutions and the corresponding average times (taking into account all the solutions) with $3\%$ tolerance.}
\label{fig:result2}
\end{figure*}

In the present work we are interested in calculating the holographic results for the propertime evolution of the pressure anisotropy and the non-equilibrium entropy density of the strongly coupled quantum SYM plasma undergoing Bjorken flow. In a conformal setup, the ratio of the pressure anisotropy over the energy density is simply given by
\begin{equation}
\frac{\Delta \hat{p}}{\hat{\varepsilon}}\equiv \frac{\hat{p}_T-\hat{p}_L}{\hat{\varepsilon}} = 2+\frac{3}{2}\,\tau\, \partial_\tau\ln(\hat{\varepsilon}).
\label{deltaP}
\end{equation}

For the pressure anisotropy of the SYM plasma, the corresponding analytical hydrodynamic expressions for the NS regime, the second-order gradient expansion \cite{Baier:2007ix,Romatschke:2017vte}, and the Borel resummation \cite{Spalinski:2017mel,Florkowski:2017olj} of the divergent gradient expansion \cite{Heller:2013fn} are given by, respectively,
\begin{subequations}
\begin{align}
\left[\frac{\Delta \hat{p}}{\hat{\varepsilon}}\right]_{\textrm{NS}} &= \frac{2}{3\pi\omega_\Lambda}, \label{NSP}\\
\left[\frac{\Delta \hat{p}}{\hat{\varepsilon}}\right]_{\textrm{2nd order}} &= \frac{2}{3\pi\omega_\Lambda} + \frac{2\,(1-\ln(2))}{9\pi^2\omega_\Lambda^2}, \label{2ndP}\\
\left[\frac{\Delta \hat{p}}{\hat{\varepsilon}}\right]_{\textrm{Borel}} &= \frac{-276+2530\,\omega_\Lambda}{3\,(120-570\,\omega_\Lambda+3975\,\omega^2_\Lambda)}, \label{BorelP}
\end{align}
\end{subequations}
where $\omega_\Lambda(\tau)\equiv \tau\, T_{\textrm{eff}}(\tau)$ is an effective dimensionless time measure, with $T_{\textrm{eff}}(\tau)$ being an effective temperature defined out of equilibrium. 

We recall that temperature is actually a thermodynamical concept which, strictly speaking, is only unambiguously defined in equilibrium. However, one usually defines in holographic calculations an out of equilibrium effective temperature as follows. For the SYM plasma in equilibrium, conformal invariance dictates that the relation between the energy density and the temperature of the fluid is given by \cite{Baier:2007ix},
\begin{align}
\varepsilon_{\textrm{eq}} = \frac{3\pi^2N_c^2}{8}\, T^4 = \frac{3\pi^4}{2\kappa_5^2}\, T^4.
\label{Teq}
\end{align}
Therefore, in equilibrium, the temperature can be written in terms of the normalized energy density, $\hat{\varepsilon} \equiv \kappa_5^2 \varepsilon$, as follows,
\begin{align}
T = \frac{(2/3)^{1/4}}{\pi}\, \hat{\varepsilon}_{\textrm{eq}}^{1/4}.
\label{Teq2}
\end{align}
Notice that Eq.\ \eqref{Teq2} only holds in equilibrium. However, nothing prevents one to simply \textit{define} an object, which we shall call the ``out of equilibrium effective temperature", in analogy with Eq.\ \eqref{Teq2},
\begin{align}
T_{\textrm{eff}}(\tau) \equiv \frac{(2/3)^{1/4}}{\pi}\, \hat{\varepsilon}^{1/4}(\tau),
\label{Teff}
\end{align}
which is nothing more than a constant multiplied by the fourth root of the time-dependent (normalized) energy density of the medium. More precisely, in holographic calculations one usually considers in Eq.\ \eqref{Teff} not the full numerical result for the energy density, but takes instead some finite order hydrodynamic truncation of the energy density\footnote{At late times the full numerical energy density is expected to converge to its analytical hydrodynamic expansion.}. Here we use the third-order hydrodynamic truncation for the energy density of the SYM plasma \cite{Booth:2009ct} to define the effective temperature as in Ref.\ \cite{Florkowski:2017olj},
\begin{align}
T_{\textrm{3rd order}}(\tau) &= \frac{\Lambda}{(\Lambda\tau)^{1/3}} \left[ 1 - \frac{1}{6\pi(\Lambda\tau)^{2/3}} + \frac{-1+\ln(2)}{36\pi^2(\Lambda\tau)^{4/3}} \right.\nonumber\\
&\left. +\, \frac{-21 + 2\pi^2 + 51\ln(2) - 24\ln^2(2)}{1944\pi^3 (\Lambda\tau)^2} \right],
\label{eq:time}
\end{align}
where $\Lambda$ is an energy scale which depends on the chosen initial conditions. From Eq.\ \eqref{eq:time} we also see that the ideal hydrodynamic effective temperature achieved at late times is given by \cite{Chesler:2009cy}
\begin{align}
T_{\textrm{ideal}}(\tau) = \frac{\Lambda}{(\Lambda\tau)^{1/3}}.
\label{eq:tideal}
\end{align}
In order to fix the value of the energy scale $\Lambda$ for each initial condition, as done in Ref.\ \cite{Critelli:2018osu}, we consider here the late time NS result for the energy density \cite{Janik:2005zt,Chesler:2009cy,Critelli:2018osu}\footnote{We recall again that our definition of the normalized energy density, $\hat{\varepsilon}\equiv\kappa_5^2\varepsilon= (4\pi^2/N_c^2)\,\varepsilon$, corresponds to twice the value of the definition used in Ref.\ \cite{Chesler:2009cy}.}
\begin{align}
\hat{\varepsilon}_{\textrm{NS}}(\tau) = \frac{3\pi^4 \Lambda^4}{2(\Lambda\tau)^{4/3}} \left[1 - \frac{2}{3\pi(\Lambda\tau)^{2/3}}\right].
\label{asymE}
\end{align}
Here we fit to the above analytical expression the late time result for the full numerical energy density in order to extract the value of $\Lambda$ for each initial condition considered.

The late time expansions for the area of the apparent horizon of the gravity dual of the Bjorken expanding SYM plasma \cite{Kinoshita:2008dq,Nakamura:2006ih,Figueras:2009iu,Chesler:2009cy}, give the following results for the NS and second-order hydrodynamic truncation of the holographic entropy density \eqref{shat} divided by the cube of the asymptotic ideal effective temperature \eqref{eq:tideal}
\begin{subequations}
\begin{align}
\frac{\hat{s}_{\textrm{NS}}(\tau)}{T^3_{\textrm{ideal}}(\tau)} &= 2\pi^4 \left[1-\frac{1}{2\pi(\Lambda\tau)^{2/3}}\right], \label{NSS}\\
\frac{\hat{s}_{\textrm{2nd order}}(\tau)}{T^3_{\textrm{ideal}}(\tau)} &= 2\pi^4 \left[1-\frac{1}{2\pi(\Lambda\tau)^{2/3}} + \frac{2+\pi+\ln(2)}{24\pi^2 (\Lambda\tau)^{4/3}}\right], \label{2ndS}
\end{align}
\end{subequations}
from which we also see that the asymptotic ideal hydrodynamic limit obtained when $\tau\to\infty$ for this dimensionless ratio is given by $2\pi^4$, which is just the thermodynamic equilibrium value for the normalized entropy density of the SYM plasma (see e.g. \cite{Finazzo:2014cna}).

The numerical results for the non-equilibrium entropy density of the fluid will be presented in terms of the following dimensionless ratio (which becomes unity in equilibrium),
\begin{align}
\frac{\hat{s}(\tau)}{2\pi^4 T_{\textrm{ideal}}^3(\tau)} = \frac{A_{\textrm{AH}}(\tau)}{\pi^3 \Lambda^2 \mathcal{A}} = \frac{|\Sigma(u_{\textrm{AH}},\tau)|^3}{\pi^3\Lambda^2}.
\label{entnum}
\end{align}
It is important to remark that in the hydrodynamic regime of Bjorken flow the \textit{entropy density} falls as $s(\tau)\sim\tau^{-1}$, while for the second law of thermodynamics given by Eq.\ \eqref{2ndlaw} to be satisfied, the requirement is that the entropy itself, $S(\tau)$, or equivalently, the area of the apparent horizon $A_{\textrm{AH}}(\tau)$ is non-decreasing in time. Consequently, the dimensionless ratio in Eq.\ \eqref{entnum} must be non-decreasing for solutions satisfying the second law of thermodynamics. This is the case for all the numerical solutions analyzed in the present work.

Now that we collected the relevant analytical hydrodynamic results for the holographic SYM plasma in Eqs.\ \eqref{NSP}, \eqref{2ndP}, \eqref{BorelP}, \eqref{NSS}, and \eqref{2ndS}, we need to organize our conventions for the dimensionless time measure in order to properly compare these results with the outcomes of our numerical gauge/gravity simulations.

We note that Eqs.\ \eqref{NSP}, \eqref{2ndP}, and \eqref{BorelP} are written as a function of the dimensionless time measure, which we take here to be given by $\omega_\Lambda\equiv\tau\,T_{\textrm{3rd order}}(\tau)$. Then, we interpolate our numerical results for all the physical observables in terms of this dimensionless time measure. The comparison between the full evolution of the pressure anisotropy for the initial conditions listed in Table \ref{tabICs} and the analytical hydrodynamics results of Eqs.\ \eqref{NSP}, \eqref{2ndP}, and \eqref{BorelP} is shown in Fig.\ \ref{fig:result1}. We see that the full numerical pressure anisotropy of the gauge/gravity solutions, although highly dependent on the chosen initial condition at early times, indeed converges at late times to the corresponding hydrodynamic results for all the initial conditions. And, as it is also well-known from previous results in the literature \cite{Chesler:2009cy,Heller:2011ju,Heller:2012je,Jankowski:2014lna,Romatschke:2017vte,Spalinski:2017mel,Florkowski:2017olj,Casalderrey-Solana:2017zyh,Kurkela:2019set}, we see from Fig.\ \ref{fig:result1} that the holographic SYM plasma hydrodynamizes, i.e. it acquires an effective hydrodynamic description while still having a sizable pressure anisotropy being, thus, far from thermodynamic equilibrium.

In order to investigate in a more quantitative way the onset of hydrodynamic behavior in the far-from-equilibrium numerical solutions, for each initial condition we define the corresponding dimensionless hydrodynamization time measure associated with some specific hydrodynamic attractor as the first value of $\omega_\Lambda$ for which the condition bellow is satisfied and such that it keeps being valid until the final time of the simulations,
\begin{align}
|X(\omega_\Lambda)-X_{\textrm{attractor}}(\omega_\Lambda)| \le \textrm{tol}\, |X_{\textrm{attractor}}(\omega_\Lambda)|,
\label{hydrotime}
\end{align}
where $X(\omega_\Lambda)$ denotes any physical observable of the SYM plasma at the boundary, $X_{\textrm{attractor}}(\omega_\Lambda)$ denotes some corresponding hydrodynamic expression (which, for simplicity, we call as an ``attractor"), and $\textrm{tol}$ denotes some specified relative tolerance. In Fig.\ \ref{fig:result1} one can see the different hydrodynamization times for the pressure anisotropy of each initial condition in table \ref{tabICs} with $5\%$ and $3\%$ relative tolerances, and also the corresponding average hydrodynamization times taking into account all the initial conditions. One concludes that within the specified relative tolerances, the pressure anisotropy for different initial data can converge first either to the hydrodynamic attractor corresponding to the NS regime \eqref{NSP}, or to the second-order hydrodynamic truncation \eqref{2ndP}, or to the Borel resummed result \eqref{BorelP}, depending on the chosen initial data. For the ensemble of initial conditions considered here, the average hydrodynamization time associated with the Borel resummed attractor is approximately equal to the corresponding NS result, while both are clearly smaller than the average hydrodynamization time of the second-order hydrodynamic truncation for the pressure anisotropy. One can also notice from Fig.\ \ref{fig:result1} (b) that the Borel resummed attractor only provides a clearly better description of hydrodynamization than the NS result if one considers just very small relative tolerances in the \emph{long time} regime of the system.

Concerning the holographic non-equilibrium entropy density, we notice that the hydrodynamic results given in Eqs.\ \eqref{NSS} and \eqref{2ndS} are expressed in terms of $\omega_0\equiv \tau\,T_{\textrm{ideal}}(\tau)=(\Lambda\tau)^{2/3}$, while our interpolations for the full numerical results were done in terms of $\omega_\Lambda = \tau\,T_{\textrm{3rd order}}(\tau)$, as discussed before. In order to obtain $\omega_0(\omega_\Lambda)$, one just needs to invert the relation $\omega_\Lambda(\omega_0)$ obtained from Eq.\ \eqref{eq:time}. This involves considering a third order algebraic equation, whose roots can be analytically obtained. Only one of the three roots is real, and this simple real root gives the desired relation $\omega_0(\omega_\Lambda)$, which can be plugged into Eqs.\ \eqref{NSS} and \eqref{2ndS} to express them as functions of $\omega_\Lambda$. By doing so, we compared in Fig.\ \ref{fig:result2} the full numerical results for the entropy density of the different initial conditions and the corresponding analytical hydrodynamic expansions\footnote{As far as we know, a Borel resummed attractor for the area of the apparent horizon/non-equilibrium entropy density has not been derived for the SYM plasma undergoing Bjorken flow. This is the reason why in Fig.\ \ref{fig:result2} we only plot the corresponding results for NS and  second-order hydrodynamics.}.

One can see that hydrodynamization process, as seen by the holographic non-equilibrium entropy density, is rather different than that seen by the pressure anisotropy. In fact, although the values of the entropy density of the different initial conditions are widely spread at early times, they all coalesce to the second-order hydrodynamics long before they converge to the corresponding NS regime, in striking contrast to what happens with the pressure anisotropy. Moreover, at the level of $3\%$ relative tolerance, the average second-order hydrodynamization time of the entropy density is considerably shorter than the different average hydrodinamization times of the pressure anisotropy.

We recall that the set of initial data analyzed here, as originally discussed in \cite{Rougemont:2021qyk}, leads to the solutions shown in Fig. \ref{fig:result1} (a) for the pressure anisotropy. This demonstrates that in spite of satisfying all the energy conditions at the initial time, some solutions evolve in such a way that the DEC and even the WEC can be transiently violated in Bjorken flow at early times when the system is still far from equilibrium. Moreover, as also discussed in \cite{Rougemont:2021qyk}, by comparing Figs. \ref{fig:result1} (a) and \ref{fig:result2} (a), one notices that

\begin{enumerate}[i.]
\item When there is (multiple or single) transient plateau formation for $\hat{s}/2\pi^4T^3_{\textrm{ideal}}$ far-from-equilibrium, the normalized entropy density only trespass its last (or single) plateau around the 2nd order hydrodynamization time, which happens if and only if a local minimum is observed for the normalized pressure anisotropy with $\Delta p/\epsilon \le -1$ \emph{after} such a plateau has been formed; in such cases, a single plateau for the normalized entropy density (see the magenta, orange, and purple curves) \emph{later} implies a local minimum with $\Delta p/\epsilon = -1$ (boundary to DEC violation), while the presence of multiple plateaus (see the red and salmon curves) \emph{later} implies a local minimum with $\Delta p/\epsilon < -1$ (DEC violation);

\item On the other hand, there are solutions (see the blue curves) violating DEC and also WEC which display no transient plateau for $\hat{s}/2\pi^4T^3_{\textrm{ideal}}$, and therefore such solutions always have nonvanishing entropy production when the medium is far-from-equilibrium --- in particular, we found no special features for the normalized entropy density associated with the region violating WEC and DEC with $\Delta p/\epsilon > 2$.
\end{enumerate}

In particular, from Figs. \ref{fig:result1} and \ref{fig:result2} it is clear that the WEC violating solutions (namely, IC's $\# 23$, $\# 24$, and $\# 25$, which correspond, respectively, to the blue, orange, and purple curves) generally take a longer time to enter in the hydrodynamic regime than the other solutions.

\begin{figure*}%[h]
\center
\subfigure[]{\includegraphics[width=0.47\textwidth]{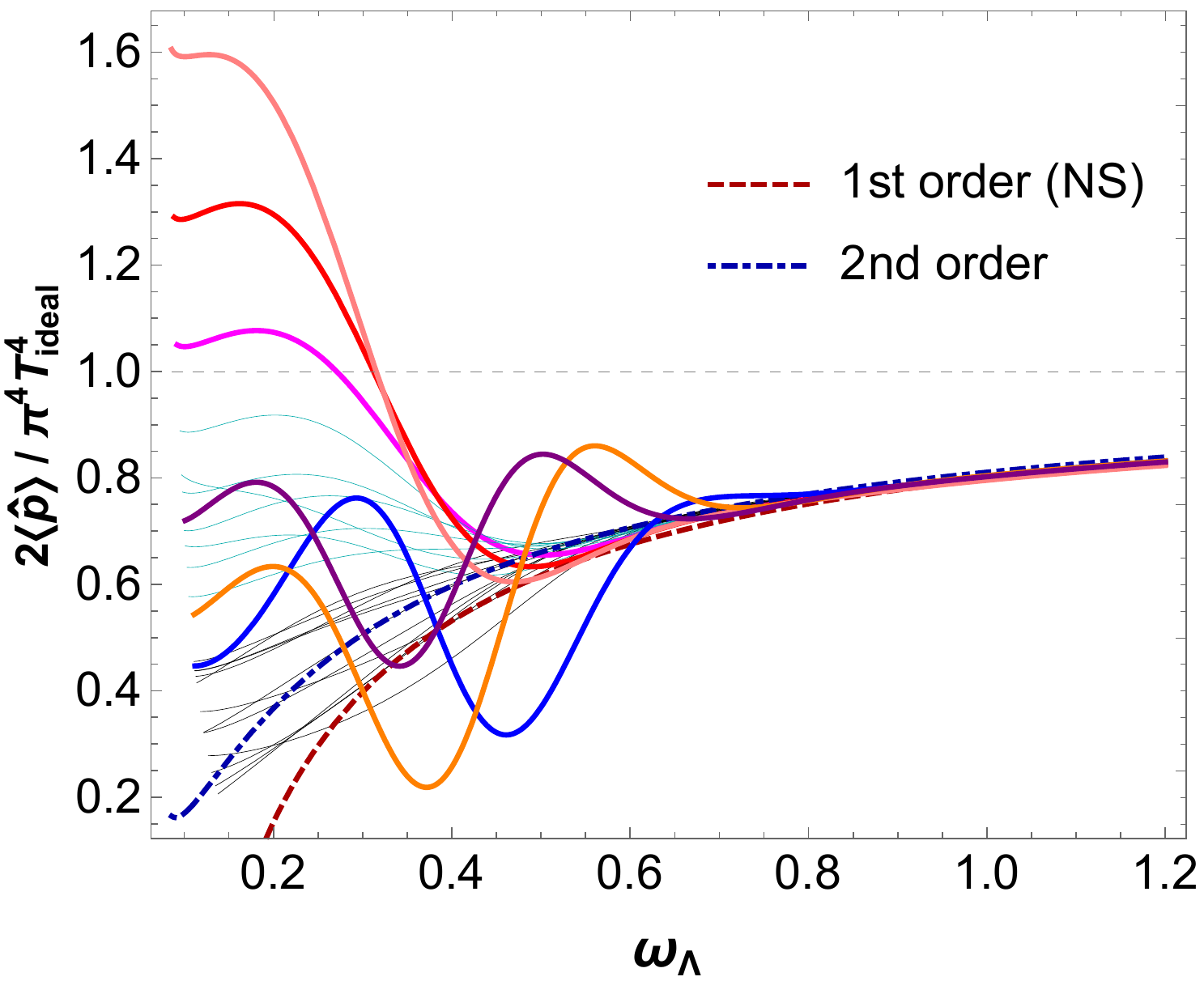}}
\qquad
\subfigure[]{\includegraphics[width=0.47\textwidth]{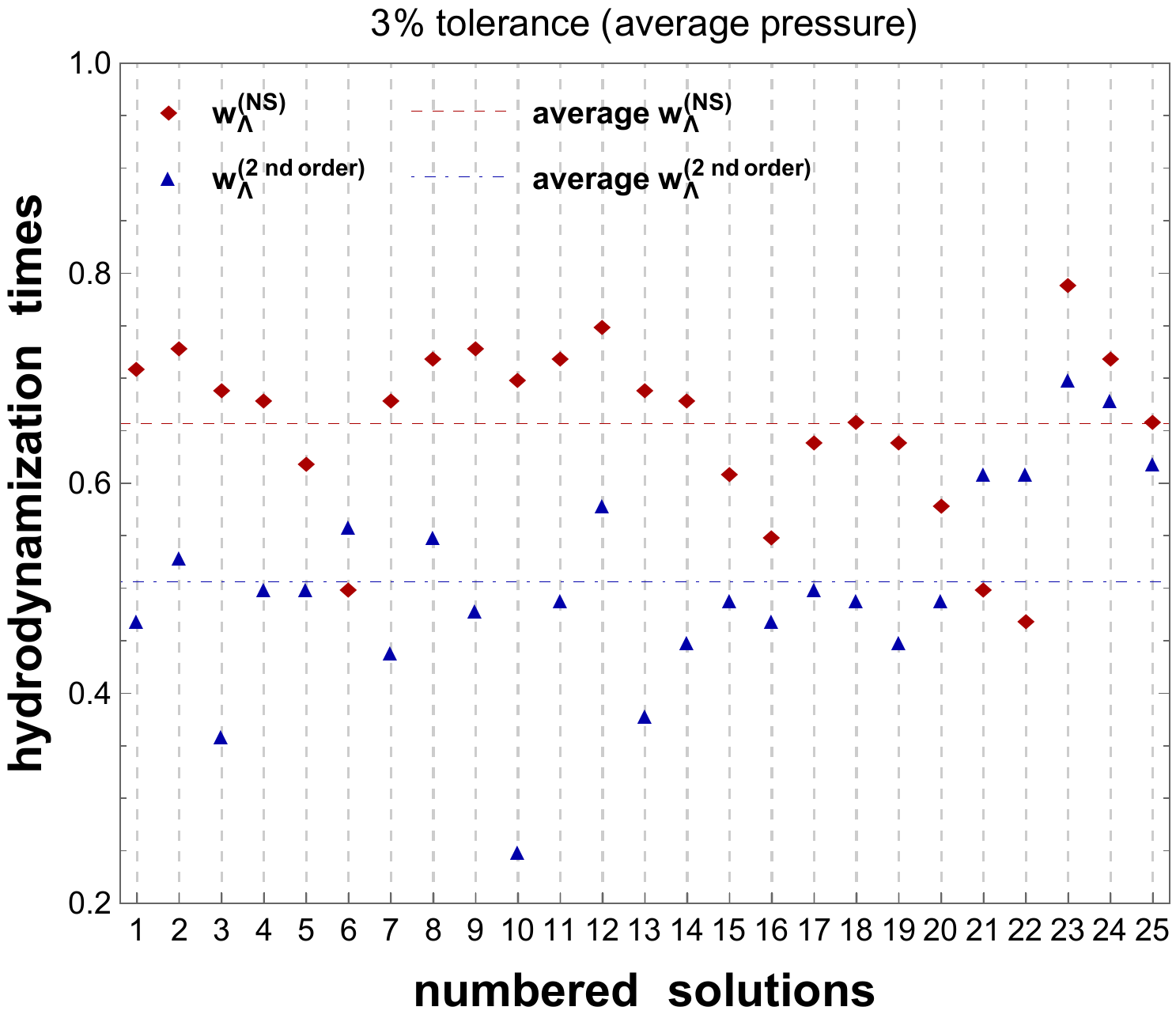}}
\caption{(a) Average pressure for the ensemble of far-from-equilibrium solutions and the corresponding late time hydrodynamic attractors (the 6 full colored thick curves are used to highlight the solutions transiently violating the energy conditions). (b) Individual hydrodynamization times for the non-equilibrium entropy density of the different solutions and the corresponding average times (taking into account all the solutions) with $3\%$ tolerance.}
\label{fig:result3}
\end{figure*}

It is also interesting to notice that, e.g. for IC $\# 16$ (corresponding to the magenta curves in Figs. \ref{fig:result1} and \ref{fig:result2}), at and around the initial time the pressure anisotropy is close to zero, and also the entropy production is close to zero, since the normalized entropy density \eqref{entnum} remains almost constant during a certain period of time at early times. These two points together may give the false impression that the system is close to equilibrium. However, this is certainly not the case. Indeed, because the approximately constant value of the normalized entropy density during this early period of time is far from its equilibrium value, the system is out of equilibrium, since the entropy needs to converge to its equilibrium value as $\tau\to\infty$. In order to pursue such a convergence, the system is driven out of an apparent and false ``near-equilibrium state'' at the initial time, and the pressure anisotropy becomes larger before going to the corresponding hydrodynamic regime, while the entropy production and the normalized entropy density increase with time.

We close this section by considering the behavior of another interesting physical observable, namely, the average pressure,
\begin{align}
\langle\hat{p}\rangle \equiv \frac{\hat{p}_L+2\hat{p}_T}{3},
\label{avpress}
\end{align}
which tends to $\pi^4 T_{\textrm{ideal}}^4/2$ for asymptotically large times in the SYM plasma. Indeed, its hydrodynamic gradient expansion up to second order reads \cite{Chesler:2009cy},\footnote{We recall once again that our definition of the normalized energy-momentum tensor of the boundary theory corresponds to twice the value of the definition used in \cite{Chesler:2009cy}. Moreover, we point out that there are two kinds of typos in Eqs. (24) and (25) of \cite{Chesler:2009cy}: first, the correct numerator in the coefficient $C_2$ is $1+2\ln(2)$, which can be checked by comparing the energy density in Eq. (24a) with the result in Eq. (4.21) of \cite{Baier:2007ix} (doing the identification $\Lambda = 4^{3/8}/3^{3/8}\pi^{3/2}$); second, there are missing factors of 3 which should multiply the coefficients $C_1$ and $C_2$ in Eqs. (24b) and (24c), as noticed in \cite{wilkaodamassa} --- in fact, it is immediate to check that without considering these multiplicative factors of 3 the trace anomaly of the hydrodynamic expansion in Eqs. (24a-c) of \cite{Chesler:2009cy} does not vanish, while the SYM plasma is a CFT and, as such, has zero trace anomaly.}
\begin{align}
\frac{2 \langle\hat{p}\rangle_\textrm{2nd order}}{\pi^4 T_{\textrm{ideal}}^4} = 1 - \frac{2}{3\pi\omega_0}+\frac{1+2\ln(2)}{18\pi^2\omega_0^2},
\label{avHpress}
\end{align}
where, as before, $\omega_0\equiv \tau\,T_{\textrm{ideal}}(\tau)=(\Lambda\tau)^{2/3}$, and one must invert $\omega_\Lambda(\omega_0)$ to obtain $\omega_0(\omega_\Lambda)$ in order to plot all the observables in terms of the effective dimensionless time $\omega_\Lambda$.

Since the SYM plasma is a CFT, it follows immediately that, $2\langle\hat{p}\rangle/\pi^4 T_\textrm{ideal}^4=2\hat{\epsilon}/3\pi^4 T_\textrm{ideal}^4$, so we are equivalently considering here the behavior of the normalized energy density of the fluid. The results for the average pressure of the ensemble of solutions considered in the present work and their associated hydrodynamization times are shown in Fig. \ref{fig:result3}.

%%%%%%%%%%%%%%%%%%%%%%%%%%%%%%%%%%
\section{Conclusions}
\label{sec:conclusion}

In the present work we analyzed in a quantitative way the different hydrodynamization times of the pressure anisotropy and of the non-equilibrium entropy density for a given ensemble of far-from-equilibrium solutions describing the dynamics of the strongly coupled SYM plasma undergoing Bjorken flow. Some of these solutions evolve in time such that a transient violation of energy conditions is developed at early times when the system is still far from equilibrium, even though there is no violation in the initial data.

The main new observation done in the present work concerns the differences on how the pressure anisotropy and the holographic non-equilibrium entropy density converge to their respective hydrodynamic regimes. While the pressure anisotropy can converge first to different hydrodynamic attractors (namely, NS, second order hydrodynamics or Borel resummation) depending on the chosen initial data (for the considered relative tolerances of 3\% to 5\%), the \emph{average} hydrodynamization times found for the ensemble of initial data analyzed here gives,
\begin{align}
\bar{\omega}_{\Lambda\,\,(\textrm{pressure})}^{\textrm{(NS)}}\sim\bar{\omega}_{\Lambda\,\,(\textrm{pressure})}^{\textrm{(Borel)}} < \bar{\omega}_{\Lambda\,\,(\textrm{pressure})}^{\textrm{(2nd order)}}.\nonumber
\end{align}
The Borel resummed attractor can only provide a clearly better description of hydrodynamization than the NS constitutive relation if one restricts the analysis to very small relative tolerances in the long time regime of the fluid.

On the other hand, concerning the non-equilibrium entropy density (determined by the apparent horizon), all the individual solutions converge much earlier to the second-order hydrodynamics regime than to the NS result. Moreover, such a convergence is attained much earlier than any of the characteristic hydrodynamization time scales of the pressure anisotropy.

By considering also the average hydrodynamization time scales of the average pressure (or, equivalently, the energy density) of the medium, we found for the ensemble of solutions considered in the present paper the hierarchy of time scales indicated in table \ref{tabTs}. It is interesting to notice that different hydrodynamic attractors may earlier apply to some observables, while only becoming valid much later for other ones. In fact, 2nd order hydrodynamics describes the behavior of the entropy and energy densities earlier than the corresponding NS results, while the opposite happens for the pressure anisotropy.

\begin{widetext}
\begin{table*}[ht]
\centering
\begin{tabular}{|c||c|c|}
\hline
Ordering & observable / attractor & $\bar{\omega}_{\Lambda} = \tau\, T_{\textrm{eff}}(\tau)$ \\
\hline
\hline
1 & entropy / 2nd order hydro & 0.3004 \\
\hline
2 & energy / 2nd order hydro & 0.5060 \\
\hline
3 & energy / Navier-Stokes & 0.6568 \\
\hline
4 & pressure anisotropy / Navier-Stokes & 0.8008 \\
\hline
5 & pressure anisotropy / Borel resummation & 0.8036 \\
\hline
6 & pressure anisotropy / 2nd order hydro & 0.9568 \\
\hline
7 & entropy / Navier-Stokes & 0.9728 \\
\hline
\end{tabular}
\caption{Average hydrodynamization time scales of the SYM plasma undergoing Bjorken flow for the ensemble of initial data considered in the present work with 3\% relative tolerances, taking into account different physical observables and hydrodynamic attractors.}
\label{tabTs}
\end{table*}
\end{widetext}

%In face of the results obtained for the ensemble of initial data analyzed here, we found the following hierarchy for the average hydrodynamization time scales of the SYM plasma undergoing Bjorken flow:
%\begin{widetext}
%\begin{align}
%\bar{\omega}_{\Lambda\,\,(\textrm{entropy})}^{\textrm{(2nd order)}}\ll\bar{\omega}_{\Lambda\,\,(\textrm{pressure})}^{\textrm{(NS)}}\sim\bar{\omega}_{\Lambda\,\,(\textrm{pressure})}^{\textrm{(Borel)}} < \bar{\omega}_{\Lambda\,\,(\textrm{pressure})}^{\textrm{(2nd order)}} \lesssim \bar{\omega}_{\Lambda\,\,(\textrm{entropy})}^{\textrm{(NS)}}.\nonumber
%\end{align}
%\end{widetext}

Moreover, we also found that the solutions violating the weak energy condition generally take a longer time to enter in the hydrodynamic regime than the other solutions.

It would be interesting to work out the same quantitative analysis of the different hydrodynamization times of the fluid in other holographic models, especially in phenomenologically realistic constructions for the quark-gluon plasma such as e.g. \cite{Finazzo:2016mhm,Critelli:2017oub,Grefa:2021qvt}, and check whether the above hierarchy of average time scales is a general feature of strongly coupled holographic fluids.

%It would also be interesting to see whether other holographic models display an initial entropy bound and energy condition violations in Bjorken flow, such as originally reported in Ref. \cite{Rougemont:2021qyk} in the case of $\mathcal{N}=4$ SYM.

\begin{figure*}%[h]
\center
\subfigure[]{\includegraphics[width=0.47\textwidth]{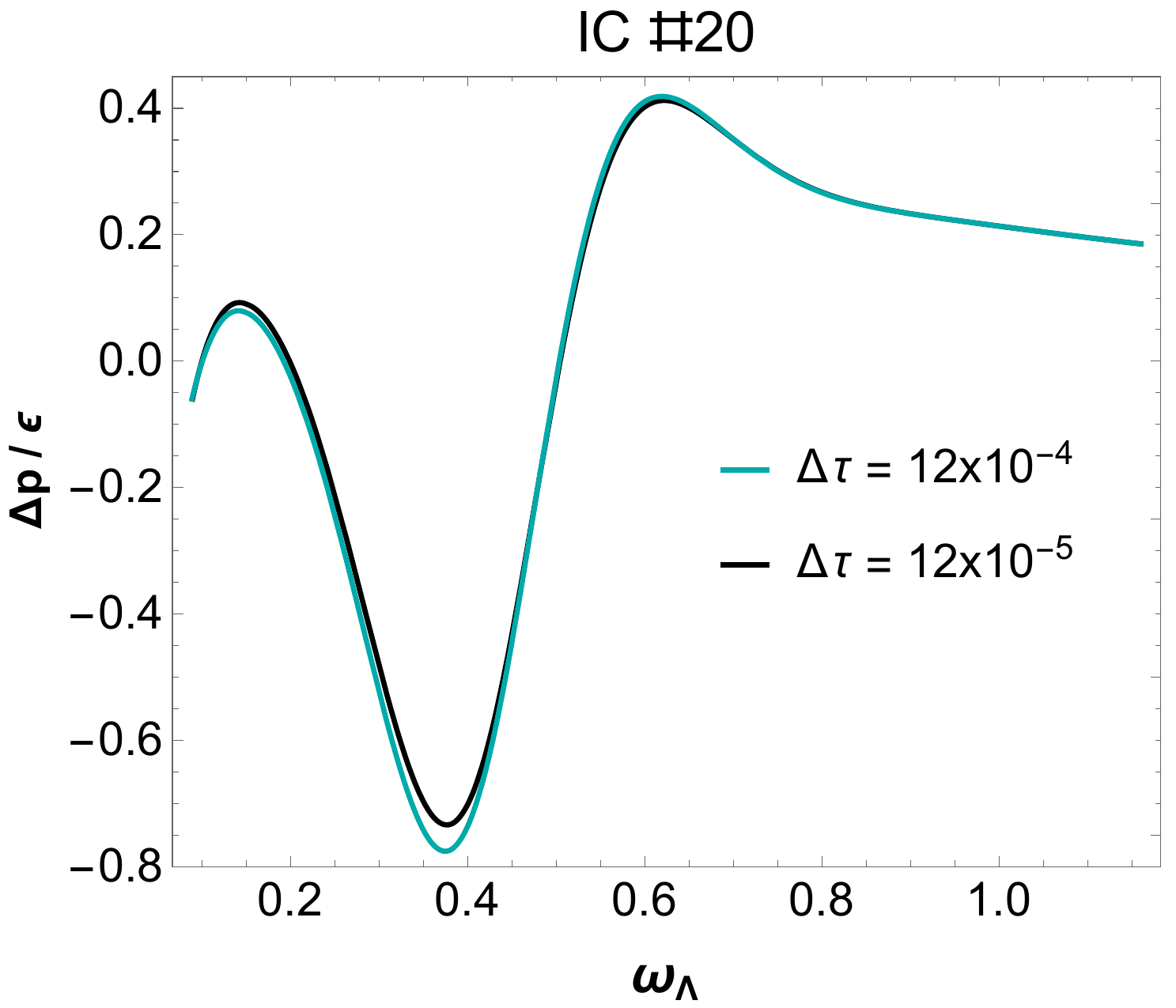}}
\qquad
\subfigure[]{\includegraphics[width=0.47\textwidth]{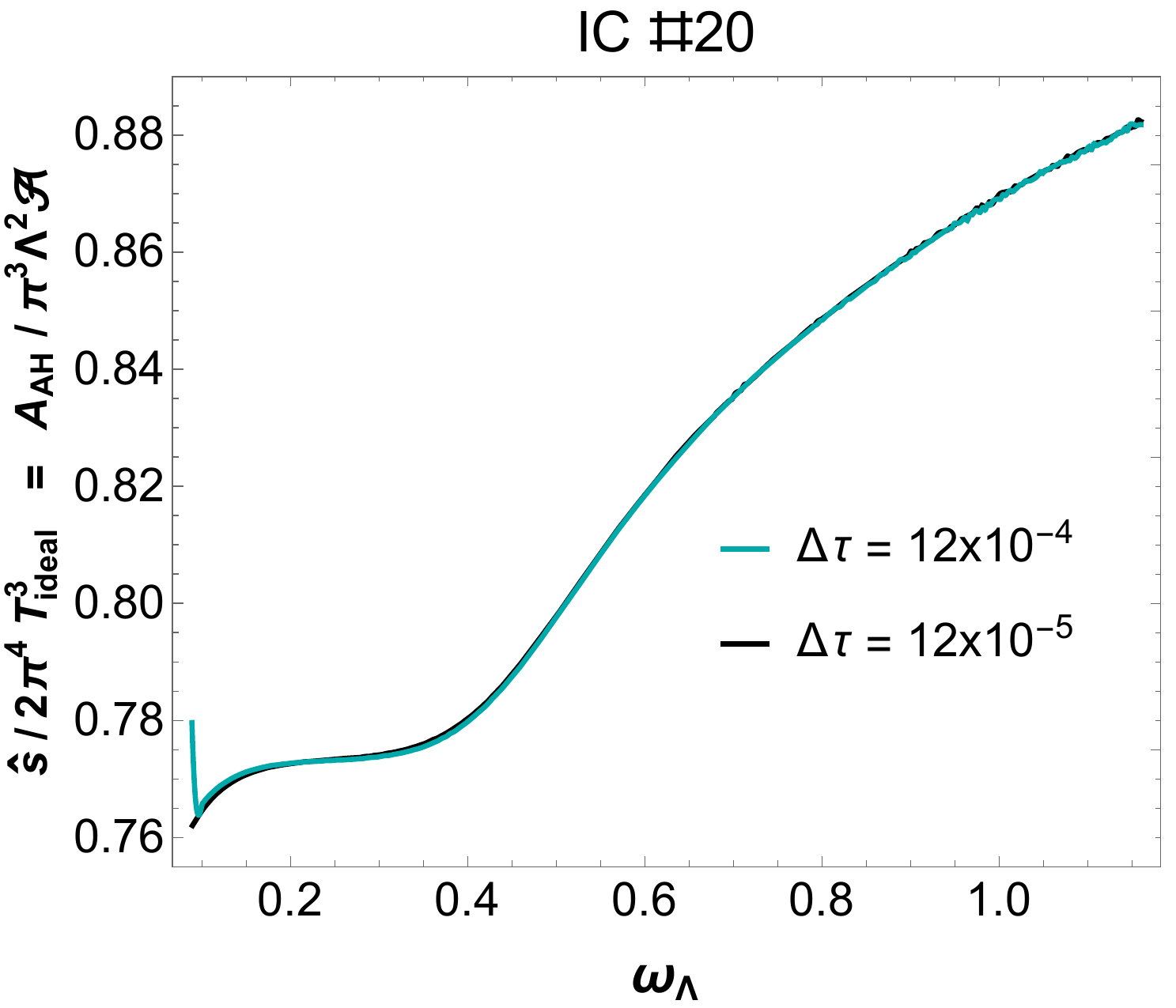}}
\caption{Results obtained with different values of the time step $\Delta\tau$ for (a) the pressure anisotropy and (b) the non-equilibrium entropy density regarding IC $\# 20$ in table \ref{tabICs}.}
\label{fig:test1}
\end{figure*}

%\begin{figure}%[h]
%\includegraphics[width=0.5\textwidth]{error_gauss.pdf}
%\caption{Exponential convergence for a Gaussian initial metric anisotropy \eqref{eq:GA}: observe the expected saturation for any set of parameters close to the working precision of 16 digits.}
%\label{fig:error_gauss}
%\end{figure}

%\begin{figure}%[h]
%\includegraphics[width=0.37\textwidth]{error_ics.pdf}
%\caption{Partial exponential convergence (observed for a not-large number of collocation points) for ICs $\# 15$ (line) and $\# 23$ (circles), given by Eq.\ \eqref{Bs0} and table \ref{tabICs}. The convergence for the whole set of IC's is indistinguishable graphically.}
%\label{fig:error_ics}
%\end{figure}

\begin{figure}%[h]
%\center
\includegraphics[width=0.40\textwidth]{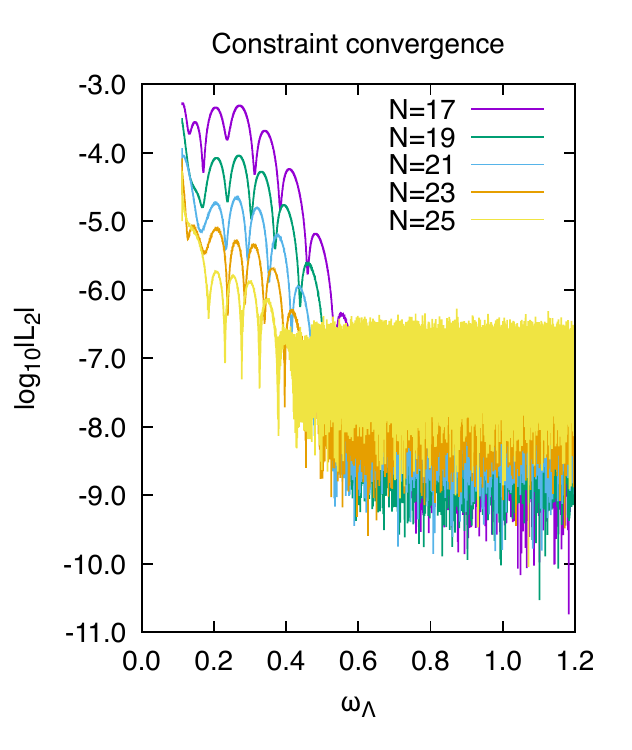}
\caption{Time evolution of the constraint error (Eq.\ \eqref{pde5}) for different truncation orders $N$, evaluated globally with the norm \eqref{eq:L2} for the IC $\# 23$ in table \ref{tabICs}. Convergence is clearly displayed for the expected truncation orders. The situation is similar for the whole set of initial conditions.}
\label{fig:cerror}
\end{figure}

\begin{figure}
\includegraphics[width=0.4\textwidth]{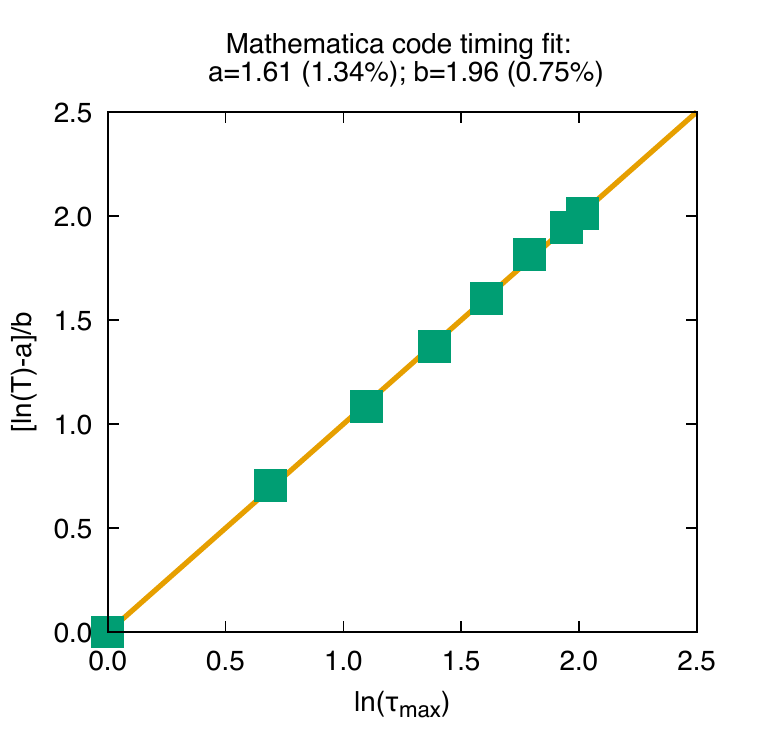}
\caption{Timing fit (with the percent error for the fit parameters informed in the brackets) for the Mathematica code performance:
$\ln(T)=a+b\ln(\tau_{\textrm{max}})$. For $\tau_{\textrm{max}}=7.5$ the Mathematica code takes $T=4h20m32s$.}
\label{fig:timing_mathematica}
\end{figure} 

\begin{figure}
\includegraphics[width=0.4\textwidth]{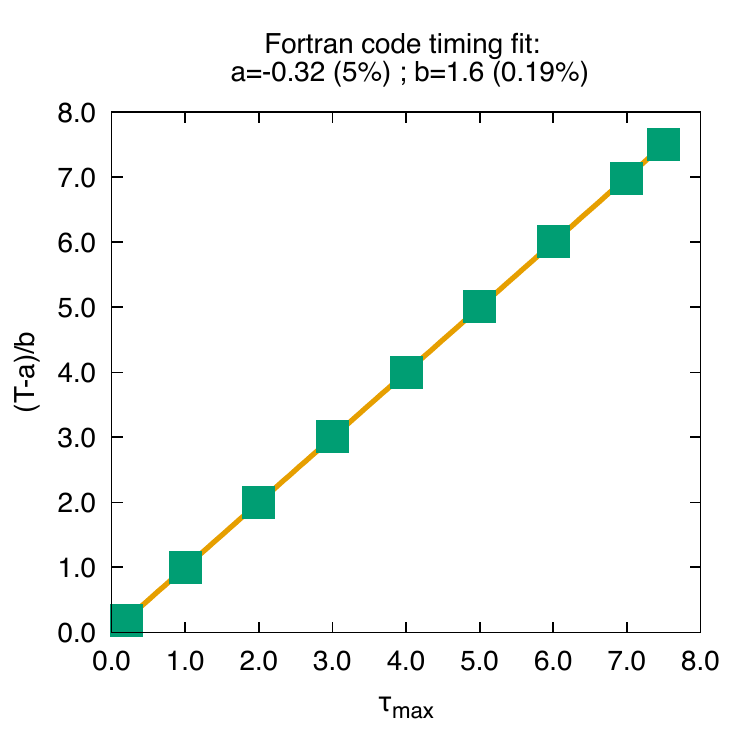}
\caption{Timing fit (with the percent error for the fit parameters informed in the brackets) for the Fortran code performance: $T=a+b\tau_{\textrm{max}}$. For $\tau_{\textrm{max}}=7.5$ the Fortran code takes $T=11.7s$.}
\label{fig:timing_fortran}
\end{figure}

%%%%%%%%%%%%%%%%%%%%%%%%%%%%%%%%%%
\begin{acknowledgments}
The authors thank R.~Critelli for the collaboration during the first stages of this work. R.R. acknowledges financial support by Universidade do Estado do Rio de Janeiro (UERJ) and Funda\c{c}\~{a}o Carlos Chagas Filho de Amparo \`{a} Pesquisa do Estado do Rio de Janeiro (FAPERJ). J.N. is partially supported by the U.S. Department of Energy, Office of Science, Office for Nuclear Physics under Award No.\ DE-SC0021301. J.N. also thanks Funda\c{c}\~{a}o de Amparo \`{a}  Pesquisa do Estado de S\~ao Paulo (FAPESP), grant number 2017/05685-2.
\end{acknowledgments}

%%%%%%%%%%%%%%%%%%%%%%%%%%%%%%%%%%
\appendix

\section{Numerical error analysis}
\label{sec:appA}

In this Appendix we present further details on the numerical procedure developed in the present work, with a focus on error analysis.

In Fig.\ \ref{fig:test1} we show the results for the pressure anisotropy and for the entropy density regarding a selected initial data (namely, IC $\# 20$ in table \ref{tabICs}), as evolved in time by considering two different time steps $\Delta\tau$. For most of the evolution both results closely agree graphically, however, one notices a slight reduction on the negative peak of the pressure anisotropy associated with a reduction of the time step from $\Delta\tau=12\times 10^{-4}$ to $\Delta\tau=12\times 10^{-5}$. Furthermore, and more importantly, one notices a spurious violation of the second law of thermodynamics at early times for this initial data when considering $\Delta\tau=12\times 10^{-4}$ (which is manifest in the early decrease of the area of the apparent horizon) --- such a violation is actually a numerical artifact which is eliminated by considering a smaller time step of $\Delta\tau=12\times 10^{-5}$, as shown in the figure. The same issue related to spurious numerical violations of the second law of thermodynamics due to an inadequate large time step of $\Delta\tau=12\times 10^{-4}$ is also observed for ICs $\# 15$, $\# 16$, $\# 21$, $\# 22$,  $\# 23$, $\# 24$, and $\# 25$ in table \ref{tabICs}, with this spurious numerical artifact being removed by considering a smaller time step of $\Delta\tau=12\times 10^{-5}$ (or lower).

We also define an error measure by means of a root mean square (RMS) norm $L_2$,
\begin{equation}
L_2=\left\{\frac{1}{2}\int_0^{u_{\textrm{IR}}}|...|^2 du\right\}^{1/2},
\label{eq:L2}
\end{equation}
where $|...|$ is an expected computational zero.

We apply the error measure \eqref{eq:L2} to analyze the time evolution of the error in the constraint Eq.\ \eqref{pde5} for the chosen truncation scheme. In this case $|...|= |$LHS of Eq.\ \eqref{pde5}$|$. Fig.\ \ref{fig:cerror} shows the result for the initial condition $\# 23$. This result is quite representative when it comes to the whole set of initial data considered in this work.

\section{Timing and performance}
\label{sec:appB}

We developed a prototype code with Wolfram's Mathematica (version 12) and also a serial Fortran code (from scratch, which uses open source libraries and compiler) calibrated with the Mathematica's prototype. For IC $\# 20$ in table \ref{tabICs} with $N=33$ and $\Delta\tau=12 \times 10^{-5}$ running from $\tau_0=0.2$ up to $\tau_{\textrm{max}}=7.5$, and without considering the post-processing, the performance is displayed in Figs.\ \ref{fig:timing_mathematica} and \ref{fig:timing_fortran}. We ran both codes on an Intel core i7-9700k@8x4.9 GHz with 64 Gb of memory, under Ubuntu 18.04 bionic. The timing $T$ for the Mathematica code goes as $\sim \tau_{\textrm{max}}^2$ using the full processor (800\% CPU) and between $0.7\%$ and $1.6\%$ of the total available memory. We remark that in the simulations performed to evolve the set of initial data considered in the present work, we used $\tau_{\textrm{max}}=7.5$. The evolution of the system for a single initial data takes $T=4h20m32s$ with the Mathematica code.

On the other hand, the timing $T$ for the Fortran code goes as $\sim \tau_{\textrm{max}}$. For $\tau_{\textrm{max}}=7.5$ it takes $T=11.7s$ (100\% CPU and a negligible use of memory). For this particular setting the Fortran code largely outperforms the Mathematica code by reducing the computation time by a factor of $\sim 1,336$.

%We intend to make a generalized version of our Fortran code , which is currently being developed to deal with a more general holographic model comprising also a chemical potential. In this more general code the results for the neutral SYM plasma considered here will be simply obtained as a particular case at zero chemical potential.

%\newpage

%%%%%%%%%%%%%%%%%%%%%%%%%%%%%%%%%%
\bibliographystyle{apsrev4-2}
\bibliography{Bibliography}

\begin{thebibliography}{80}%
\makeatletter
\providecommand \@ifxundefined [1]{%
 \@ifx{#1\undefined}
}%
\providecommand \@ifnum [1]{%
 \ifnum #1\expandafter \@firstoftwo
 \else \expandafter \@secondoftwo
 \fi
}%
\providecommand \@ifx [1]{%
 \ifx #1\expandafter \@firstoftwo
 \else \expandafter \@secondoftwo
 \fi
}%
\providecommand \natexlab [1]{#1}%
\providecommand \enquote  [1]{``#1''}%
\providecommand \bibnamefont  [1]{#1}%
\providecommand \bibfnamefont [1]{#1}%
\providecommand \citenamefont [1]{#1}%
\providecommand \href@noop [0]{\@secondoftwo}%
\providecommand \href [0]{\begingroup \@sanitize@url \@href}%
\providecommand \@href[1]{\@@startlink{#1}\@@href}%
\providecommand \@@href[1]{\endgroup#1\@@endlink}%
\providecommand \@sanitize@url [0]{\catcode `\\12\catcode `\$12\catcode
  `\&12\catcode `\#12\catcode `\^12\catcode `\_12\catcode `\%12\relax}%
\providecommand \@@startlink[1]{}%
\providecommand \@@endlink[0]{}%
\providecommand \url  [0]{\begingroup\@sanitize@url \@url }%
\providecommand \@url [1]{\endgroup\@href {#1}{\urlprefix }}%
\providecommand \urlprefix  [0]{URL }%
\providecommand \Eprint [0]{\href }%
\providecommand \doibase [0]{https://doi.org/}%
\providecommand \selectlanguage [0]{\@gobble}%
\providecommand \bibinfo  [0]{\@secondoftwo}%
\providecommand \bibfield  [0]{\@secondoftwo}%
\providecommand \translation [1]{[#1]}%
\providecommand \BibitemOpen [0]{}%
\providecommand \bibitemStop [0]{}%
\providecommand \bibitemNoStop [0]{.\EOS\space}%
\providecommand \EOS [0]{\spacefactor3000\relax}%
\providecommand \BibitemShut  [1]{\csname bibitem#1\endcsname}%
\let\auto@bib@innerbib\@empty
%</preamble>
\bibitem [{\citenamefont {Maldacena}(1998)}]{Maldacena:1997re}%
  \BibitemOpen
  \bibfield  {author} {\bibinfo {author} {\bibfnamefont {J.~M.}\ \bibnamefont
  {Maldacena}},\ }\href {https://doi.org/10.1023/A:1026654312961} {\bibfield
  {journal} {\bibinfo  {journal} {Adv. Theor. Math. Phys.}\ }\textbf {\bibinfo
  {volume} {2}},\ \bibinfo {pages} {231} (\bibinfo {year} {1998})},\ \Eprint
  {https://arxiv.org/abs/hep-th/9711200} {arXiv:hep-th/9711200} \BibitemShut
  {NoStop}%
\bibitem [{\citenamefont {Gubser}\ \emph {et~al.}(1998)\citenamefont {Gubser},
  \citenamefont {Klebanov},\ and\ \citenamefont {Polyakov}}]{Gubser:1998bc}%
  \BibitemOpen
  \bibfield  {author} {\bibinfo {author} {\bibfnamefont {S.~S.}\ \bibnamefont
  {Gubser}}, \bibinfo {author} {\bibfnamefont {I.~R.}\ \bibnamefont
  {Klebanov}},\ and\ \bibinfo {author} {\bibfnamefont {A.~M.}\ \bibnamefont
  {Polyakov}},\ }\href {https://doi.org/10.1016/S0370-2693(98)00377-3}
  {\bibfield  {journal} {\bibinfo  {journal} {Phys. Lett. B}\ }\textbf
  {\bibinfo {volume} {428}},\ \bibinfo {pages} {105} (\bibinfo {year}
  {1998})},\ \Eprint {https://arxiv.org/abs/hep-th/9802109}
  {arXiv:hep-th/9802109} \BibitemShut {NoStop}%
\bibitem [{\citenamefont {Witten}(1998{\natexlab{a}})}]{Witten:1998qj}%
  \BibitemOpen
  \bibfield  {author} {\bibinfo {author} {\bibfnamefont {E.}~\bibnamefont
  {Witten}},\ }\href {https://doi.org/10.4310/ATMP.1998.v2.n2.a2} {\bibfield
  {journal} {\bibinfo  {journal} {Adv. Theor. Math. Phys.}\ }\textbf {\bibinfo
  {volume} {2}},\ \bibinfo {pages} {253} (\bibinfo {year}
  {1998}{\natexlab{a}})},\ \Eprint {https://arxiv.org/abs/hep-th/9802150}
  {arXiv:hep-th/9802150} \BibitemShut {NoStop}%
\bibitem [{\citenamefont {Witten}(1998{\natexlab{b}})}]{Witten:1998zw}%
  \BibitemOpen
  \bibfield  {author} {\bibinfo {author} {\bibfnamefont {E.}~\bibnamefont
  {Witten}},\ }\href {https://doi.org/10.4310/ATMP.1998.v2.n3.a3} {\bibfield
  {journal} {\bibinfo  {journal} {Adv. Theor. Math. Phys.}\ }\textbf {\bibinfo
  {volume} {2}},\ \bibinfo {pages} {505} (\bibinfo {year}
  {1998}{\natexlab{b}})},\ \Eprint {https://arxiv.org/abs/hep-th/9803131}
  {arXiv:hep-th/9803131} \BibitemShut {NoStop}%
\bibitem [{\citenamefont {Chesler}\ and\ \citenamefont
  {Yaffe}(2009)}]{Chesler:2008hg}%
  \BibitemOpen
  \bibfield  {author} {\bibinfo {author} {\bibfnamefont {P.~M.}\ \bibnamefont
  {Chesler}}\ and\ \bibinfo {author} {\bibfnamefont {L.~G.}\ \bibnamefont
  {Yaffe}},\ }\href {https://doi.org/10.1103/PhysRevLett.102.211601} {\bibfield
   {journal} {\bibinfo  {journal} {Phys. Rev. Lett.}\ }\textbf {\bibinfo
  {volume} {102}},\ \bibinfo {pages} {211601} (\bibinfo {year} {2009})},\
  \Eprint {https://arxiv.org/abs/0812.2053} {arXiv:0812.2053 [hep-th]}
  \BibitemShut {NoStop}%
\bibitem [{\citenamefont {Chesler}\ and\ \citenamefont
  {Yaffe}(2010)}]{Chesler:2009cy}%
  \BibitemOpen
  \bibfield  {author} {\bibinfo {author} {\bibfnamefont {P.~M.}\ \bibnamefont
  {Chesler}}\ and\ \bibinfo {author} {\bibfnamefont {L.~G.}\ \bibnamefont
  {Yaffe}},\ }\href {https://doi.org/10.1103/PhysRevD.82.026006} {\bibfield
  {journal} {\bibinfo  {journal} {Phys. Rev. D}\ }\textbf {\bibinfo {volume}
  {82}},\ \bibinfo {pages} {026006} (\bibinfo {year} {2010})},\ \Eprint
  {https://arxiv.org/abs/0906.4426} {arXiv:0906.4426 [hep-th]} \BibitemShut
  {NoStop}%
\bibitem [{\citenamefont {Chesler}\ and\ \citenamefont
  {Yaffe}(2011)}]{Chesler:2010bi}%
  \BibitemOpen
  \bibfield  {author} {\bibinfo {author} {\bibfnamefont {P.~M.}\ \bibnamefont
  {Chesler}}\ and\ \bibinfo {author} {\bibfnamefont {L.~G.}\ \bibnamefont
  {Yaffe}},\ }\href {https://doi.org/10.1103/PhysRevLett.106.021601} {\bibfield
   {journal} {\bibinfo  {journal} {Phys. Rev. Lett.}\ }\textbf {\bibinfo
  {volume} {106}},\ \bibinfo {pages} {021601} (\bibinfo {year} {2011})},\
  \Eprint {https://arxiv.org/abs/1011.3562} {arXiv:1011.3562 [hep-th]}
  \BibitemShut {NoStop}%
\bibitem [{\citenamefont {Heller}\ \emph
  {et~al.}(2012{\natexlab{a}})\citenamefont {Heller}, \citenamefont {Janik},\
  and\ \citenamefont {Witaszczyk}}]{Heller:2011ju}%
  \BibitemOpen
  \bibfield  {author} {\bibinfo {author} {\bibfnamefont {M.~P.}\ \bibnamefont
  {Heller}}, \bibinfo {author} {\bibfnamefont {R.~A.}\ \bibnamefont {Janik}},\
  and\ \bibinfo {author} {\bibfnamefont {P.}~\bibnamefont {Witaszczyk}},\
  }\href {https://doi.org/10.1103/PhysRevLett.108.201602} {\bibfield  {journal}
  {\bibinfo  {journal} {Phys. Rev. Lett.}\ }\textbf {\bibinfo {volume} {108}},\
  \bibinfo {pages} {201602} (\bibinfo {year} {2012}{\natexlab{a}})},\ \Eprint
  {https://arxiv.org/abs/1103.3452} {arXiv:1103.3452 [hep-th]} \BibitemShut
  {NoStop}%
\bibitem [{\citenamefont {Heller}\ \emph
  {et~al.}(2012{\natexlab{b}})\citenamefont {Heller}, \citenamefont {Janik},\
  and\ \citenamefont {Witaszczyk}}]{Heller:2012je}%
  \BibitemOpen
  \bibfield  {author} {\bibinfo {author} {\bibfnamefont {M.~P.}\ \bibnamefont
  {Heller}}, \bibinfo {author} {\bibfnamefont {R.~A.}\ \bibnamefont {Janik}},\
  and\ \bibinfo {author} {\bibfnamefont {P.}~\bibnamefont {Witaszczyk}},\
  }\href {https://doi.org/10.1103/PhysRevD.85.126002} {\bibfield  {journal}
  {\bibinfo  {journal} {Phys. Rev. D}\ }\textbf {\bibinfo {volume} {85}},\
  \bibinfo {pages} {126002} (\bibinfo {year} {2012}{\natexlab{b}})},\ \Eprint
  {https://arxiv.org/abs/1203.0755} {arXiv:1203.0755 [hep-th]} \BibitemShut
  {NoStop}%
\bibitem [{\citenamefont {van~der Schee}(2013)}]{vanderSchee:2012qj}%
  \BibitemOpen
  \bibfield  {author} {\bibinfo {author} {\bibfnamefont {W.}~\bibnamefont
  {van~der Schee}},\ }\href {https://doi.org/10.1103/PhysRevD.87.061901}
  {\bibfield  {journal} {\bibinfo  {journal} {Phys. Rev. D}\ }\textbf {\bibinfo
  {volume} {87}},\ \bibinfo {pages} {061901} (\bibinfo {year} {2013})},\
  \Eprint {https://arxiv.org/abs/1211.2218} {arXiv:1211.2218 [hep-th]}
  \BibitemShut {NoStop}%
\bibitem [{\citenamefont {Casalderrey-Solana}\ \emph
  {et~al.}(2013)\citenamefont {Casalderrey-Solana}, \citenamefont {Heller},
  \citenamefont {Mateos},\ and\ \citenamefont {van~der
  Schee}}]{Casalderrey-Solana:2013aba}%
  \BibitemOpen
  \bibfield  {author} {\bibinfo {author} {\bibfnamefont {J.}~\bibnamefont
  {Casalderrey-Solana}}, \bibinfo {author} {\bibfnamefont {M.~P.}\ \bibnamefont
  {Heller}}, \bibinfo {author} {\bibfnamefont {D.}~\bibnamefont {Mateos}},\
  and\ \bibinfo {author} {\bibfnamefont {W.}~\bibnamefont {van~der Schee}},\
  }\href {https://doi.org/10.1103/PhysRevLett.111.181601} {\bibfield  {journal}
  {\bibinfo  {journal} {Phys. Rev. Lett.}\ }\textbf {\bibinfo {volume} {111}},\
  \bibinfo {pages} {181601} (\bibinfo {year} {2013})},\ \Eprint
  {https://arxiv.org/abs/1305.4919} {arXiv:1305.4919 [hep-th]} \BibitemShut
  {NoStop}%
\bibitem [{\citenamefont {Chesler}\ and\ \citenamefont
  {Yaffe}(2014)}]{Chesler:2013lia}%
  \BibitemOpen
  \bibfield  {author} {\bibinfo {author} {\bibfnamefont {P.~M.}\ \bibnamefont
  {Chesler}}\ and\ \bibinfo {author} {\bibfnamefont {L.~G.}\ \bibnamefont
  {Yaffe}},\ }\href {https://doi.org/10.1007/JHEP07(2014)086} {\bibfield
  {journal} {\bibinfo  {journal} {JHEP}\ }\textbf {\bibinfo {volume} {07}},\
  \bibinfo {pages} {086}},\ \Eprint {https://arxiv.org/abs/1309.1439}
  {arXiv:1309.1439 [hep-th]} \BibitemShut {NoStop}%
\bibitem [{\citenamefont {van~der Schee}(2014)}]{vanderSchee:2014qwa}%
  \BibitemOpen
  \bibfield  {author} {\bibinfo {author} {\bibfnamefont {W.}~\bibnamefont
  {van~der Schee}},\ }\emph {\bibinfo {title} {{Gravitational collisions and
  the quark-gluon plasma}}},\ \href@noop {} {Ph.D. thesis},\ \bibinfo  {school}
  {Utrecht U.} (\bibinfo {year} {2014}),\ \Eprint
  {https://arxiv.org/abs/1407.1849} {arXiv:1407.1849 [hep-th]} \BibitemShut
  {NoStop}%
\bibitem [{\citenamefont {Jankowski}\ \emph {et~al.}(2014)\citenamefont
  {Jankowski}, \citenamefont {Plewa},\ and\ \citenamefont
  {Spalinski}}]{Jankowski:2014lna}%
  \BibitemOpen
  \bibfield  {author} {\bibinfo {author} {\bibfnamefont {J.}~\bibnamefont
  {Jankowski}}, \bibinfo {author} {\bibfnamefont {G.}~\bibnamefont {Plewa}},\
  and\ \bibinfo {author} {\bibfnamefont {M.}~\bibnamefont {Spalinski}},\ }\href
  {https://doi.org/10.1007/JHEP12(2014)105} {\bibfield  {journal} {\bibinfo
  {journal} {JHEP}\ }\textbf {\bibinfo {volume} {12}},\ \bibinfo {pages}
  {105}},\ \Eprint {https://arxiv.org/abs/1411.1969} {arXiv:1411.1969 [hep-th]}
  \BibitemShut {NoStop}%
\bibitem [{\citenamefont {Fuini}\ and\ \citenamefont
  {Yaffe}(2015)}]{Fuini:2015hba}%
  \BibitemOpen
  \bibfield  {author} {\bibinfo {author} {\bibfnamefont {J.~F.}\ \bibnamefont
  {Fuini}}\ and\ \bibinfo {author} {\bibfnamefont {L.~G.}\ \bibnamefont
  {Yaffe}},\ }\href {https://doi.org/10.1007/JHEP07(2015)116} {\bibfield
  {journal} {\bibinfo  {journal} {JHEP}\ }\textbf {\bibinfo {volume} {07}},\
  \bibinfo {pages} {116}},\ \Eprint {https://arxiv.org/abs/1503.07148}
  {arXiv:1503.07148 [hep-th]} \BibitemShut {NoStop}%
\bibitem [{\citenamefont {Chesler}(2015)}]{Chesler:2015bba}%
  \BibitemOpen
  \bibfield  {author} {\bibinfo {author} {\bibfnamefont {P.~M.}\ \bibnamefont
  {Chesler}},\ }\href {https://doi.org/10.1103/PhysRevLett.115.241602}
  {\bibfield  {journal} {\bibinfo  {journal} {Phys. Rev. Lett.}\ }\textbf
  {\bibinfo {volume} {115}},\ \bibinfo {pages} {241602} (\bibinfo {year}
  {2015})},\ \Eprint {https://arxiv.org/abs/1506.02209} {arXiv:1506.02209
  [hep-th]} \BibitemShut {NoStop}%
\bibitem [{\citenamefont {Attems}\ \emph
  {et~al.}(2017{\natexlab{a}})\citenamefont {Attems}, \citenamefont
  {Casalderrey-Solana}, \citenamefont {Mateos}, \citenamefont
  {Santos-Oliv\'an}, \citenamefont {Sopuerta}, \citenamefont {Triana},\ and\
  \citenamefont {Zilh\~ao}}]{Attems:2016tby}%
  \BibitemOpen
  \bibfield  {author} {\bibinfo {author} {\bibfnamefont {M.}~\bibnamefont
  {Attems}}, \bibinfo {author} {\bibfnamefont {J.}~\bibnamefont
  {Casalderrey-Solana}}, \bibinfo {author} {\bibfnamefont {D.}~\bibnamefont
  {Mateos}}, \bibinfo {author} {\bibfnamefont {D.}~\bibnamefont
  {Santos-Oliv\'an}}, \bibinfo {author} {\bibfnamefont {C.~F.}\ \bibnamefont
  {Sopuerta}}, \bibinfo {author} {\bibfnamefont {M.}~\bibnamefont {Triana}},\
  and\ \bibinfo {author} {\bibfnamefont {M.}~\bibnamefont {Zilh\~ao}},\ }\href
  {https://doi.org/10.1007/JHEP01(2017)026} {\bibfield  {journal} {\bibinfo
  {journal} {JHEP}\ }\textbf {\bibinfo {volume} {01}},\ \bibinfo {pages}
  {026}},\ \Eprint {https://arxiv.org/abs/1604.06439} {arXiv:1604.06439
  [hep-th]} \BibitemShut {NoStop}%
\bibitem [{\citenamefont {Casalderrey-Solana}\ \emph
  {et~al.}(2016)\citenamefont {Casalderrey-Solana}, \citenamefont {Mateos},
  \citenamefont {van~der Schee},\ and\ \citenamefont
  {Triana}}]{Casalderrey-Solana:2016xfq}%
  \BibitemOpen
  \bibfield  {author} {\bibinfo {author} {\bibfnamefont {J.}~\bibnamefont
  {Casalderrey-Solana}}, \bibinfo {author} {\bibfnamefont {D.}~\bibnamefont
  {Mateos}}, \bibinfo {author} {\bibfnamefont {W.}~\bibnamefont {van~der
  Schee}},\ and\ \bibinfo {author} {\bibfnamefont {M.}~\bibnamefont {Triana}},\
  }\href {https://doi.org/10.1007/JHEP09(2016)108} {\bibfield  {journal}
  {\bibinfo  {journal} {JHEP}\ }\textbf {\bibinfo {volume} {09}},\ \bibinfo
  {pages} {108}},\ \Eprint {https://arxiv.org/abs/1607.05273} {arXiv:1607.05273
  [hep-th]} \BibitemShut {NoStop}%
\bibitem [{\citenamefont {Grozdanov}\ and\ \citenamefont {van~der
  Schee}(2017)}]{Grozdanov:2016zjj}%
  \BibitemOpen
  \bibfield  {author} {\bibinfo {author} {\bibfnamefont {S.}~\bibnamefont
  {Grozdanov}}\ and\ \bibinfo {author} {\bibfnamefont {W.}~\bibnamefont
  {van~der Schee}},\ }\href {https://doi.org/10.1103/PhysRevLett.119.011601}
  {\bibfield  {journal} {\bibinfo  {journal} {Phys. Rev. Lett.}\ }\textbf
  {\bibinfo {volume} {119}},\ \bibinfo {pages} {011601} (\bibinfo {year}
  {2017})},\ \Eprint {https://arxiv.org/abs/1610.08976} {arXiv:1610.08976
  [hep-th]} \BibitemShut {NoStop}%
\bibitem [{\citenamefont {Attems}\ \emph
  {et~al.}(2017{\natexlab{b}})\citenamefont {Attems}, \citenamefont
  {Casalderrey-Solana}, \citenamefont {Mateos}, \citenamefont
  {Santos-Oliv\'an}, \citenamefont {Sopuerta}, \citenamefont {Triana},\ and\
  \citenamefont {Zilh\~ao}}]{Attems:2017zam}%
  \BibitemOpen
  \bibfield  {author} {\bibinfo {author} {\bibfnamefont {M.}~\bibnamefont
  {Attems}}, \bibinfo {author} {\bibfnamefont {J.}~\bibnamefont
  {Casalderrey-Solana}}, \bibinfo {author} {\bibfnamefont {D.}~\bibnamefont
  {Mateos}}, \bibinfo {author} {\bibfnamefont {D.}~\bibnamefont
  {Santos-Oliv\'an}}, \bibinfo {author} {\bibfnamefont {C.~F.}\ \bibnamefont
  {Sopuerta}}, \bibinfo {author} {\bibfnamefont {M.}~\bibnamefont {Triana}},\
  and\ \bibinfo {author} {\bibfnamefont {M.}~\bibnamefont {Zilh\~ao}},\ }\href
  {https://doi.org/10.1007/JHEP06(2017)154} {\bibfield  {journal} {\bibinfo
  {journal} {JHEP}\ }\textbf {\bibinfo {volume} {06}},\ \bibinfo {pages}
  {154}},\ \Eprint {https://arxiv.org/abs/1703.09681} {arXiv:1703.09681
  [hep-th]} \BibitemShut {NoStop}%
\bibitem [{\citenamefont {Romatschke}(2018)}]{Romatschke:2017vte}%
  \BibitemOpen
  \bibfield  {author} {\bibinfo {author} {\bibfnamefont {P.}~\bibnamefont
  {Romatschke}},\ }\href {https://doi.org/10.1103/PhysRevLett.120.012301}
  {\bibfield  {journal} {\bibinfo  {journal} {Phys. Rev. Lett.}\ }\textbf
  {\bibinfo {volume} {120}},\ \bibinfo {pages} {012301} (\bibinfo {year}
  {2018})},\ \Eprint {https://arxiv.org/abs/1704.08699} {arXiv:1704.08699
  [hep-th]} \BibitemShut {NoStop}%
\bibitem [{\citenamefont {Spali\'nski}(2018)}]{Spalinski:2017mel}%
  \BibitemOpen
  \bibfield  {author} {\bibinfo {author} {\bibfnamefont {M.}~\bibnamefont
  {Spali\'nski}},\ }\href {https://doi.org/10.1016/j.physletb.2017.11.059}
  {\bibfield  {journal} {\bibinfo  {journal} {Phys. Lett. B}\ }\textbf
  {\bibinfo {volume} {776}},\ \bibinfo {pages} {468} (\bibinfo {year}
  {2018})},\ \Eprint {https://arxiv.org/abs/1708.01921} {arXiv:1708.01921
  [hep-th]} \BibitemShut {NoStop}%
\bibitem [{\citenamefont {Critelli}\ \emph
  {et~al.}(2017{\natexlab{a}})\citenamefont {Critelli}, \citenamefont
  {Rougemont},\ and\ \citenamefont {Noronha}}]{Critelli:2017euk}%
  \BibitemOpen
  \bibfield  {author} {\bibinfo {author} {\bibfnamefont {R.}~\bibnamefont
  {Critelli}}, \bibinfo {author} {\bibfnamefont {R.}~\bibnamefont
  {Rougemont}},\ and\ \bibinfo {author} {\bibfnamefont {J.}~\bibnamefont
  {Noronha}},\ }\href {https://doi.org/10.1007/JHEP12(2017)029} {\bibfield
  {journal} {\bibinfo  {journal} {JHEP}\ }\textbf {\bibinfo {volume} {12}},\
  \bibinfo {pages} {029}},\ \Eprint {https://arxiv.org/abs/1709.03131}
  {arXiv:1709.03131 [hep-th]} \BibitemShut {NoStop}%
\bibitem [{\citenamefont {Casalderrey-Solana}\ \emph
  {et~al.}(2018)\citenamefont {Casalderrey-Solana}, \citenamefont {Gushterov},\
  and\ \citenamefont {Meiring}}]{Casalderrey-Solana:2017zyh}%
  \BibitemOpen
  \bibfield  {author} {\bibinfo {author} {\bibfnamefont {J.}~\bibnamefont
  {Casalderrey-Solana}}, \bibinfo {author} {\bibfnamefont {N.~I.}\ \bibnamefont
  {Gushterov}},\ and\ \bibinfo {author} {\bibfnamefont {B.}~\bibnamefont
  {Meiring}},\ }\href {https://doi.org/10.1007/JHEP04(2018)042} {\bibfield
  {journal} {\bibinfo  {journal} {JHEP}\ }\textbf {\bibinfo {volume} {04}},\
  \bibinfo {pages} {042}},\ \Eprint {https://arxiv.org/abs/1712.02772}
  {arXiv:1712.02772 [hep-th]} \BibitemShut {NoStop}%
\bibitem [{\citenamefont {Critelli}\ \emph {et~al.}(2019)\citenamefont
  {Critelli}, \citenamefont {Rougemont},\ and\ \citenamefont
  {Noronha}}]{Critelli:2018osu}%
  \BibitemOpen
  \bibfield  {author} {\bibinfo {author} {\bibfnamefont {R.}~\bibnamefont
  {Critelli}}, \bibinfo {author} {\bibfnamefont {R.}~\bibnamefont
  {Rougemont}},\ and\ \bibinfo {author} {\bibfnamefont {J.}~\bibnamefont
  {Noronha}},\ }\href {https://doi.org/10.1103/PhysRevD.99.066004} {\bibfield
  {journal} {\bibinfo  {journal} {Phys. Rev. D}\ }\textbf {\bibinfo {volume}
  {99}},\ \bibinfo {pages} {066004} (\bibinfo {year} {2019})},\ \Eprint
  {https://arxiv.org/abs/1805.00882} {arXiv:1805.00882 [hep-th]} \BibitemShut
  {NoStop}%
\bibitem [{\citenamefont {Attems}\ \emph {et~al.}(2018)\citenamefont {Attems},
  \citenamefont {Bea}, \citenamefont {Casalderrey-Solana}, \citenamefont
  {Mateos}, \citenamefont {Triana},\ and\ \citenamefont
  {Zilh\~ao}}]{Attems:2018gou}%
  \BibitemOpen
  \bibfield  {author} {\bibinfo {author} {\bibfnamefont {M.}~\bibnamefont
  {Attems}}, \bibinfo {author} {\bibfnamefont {Y.}~\bibnamefont {Bea}},
  \bibinfo {author} {\bibfnamefont {J.}~\bibnamefont {Casalderrey-Solana}},
  \bibinfo {author} {\bibfnamefont {D.}~\bibnamefont {Mateos}}, \bibinfo
  {author} {\bibfnamefont {M.}~\bibnamefont {Triana}},\ and\ \bibinfo {author}
  {\bibfnamefont {M.}~\bibnamefont {Zilh\~ao}},\ }\href
  {https://doi.org/10.1103/PhysRevLett.121.261601} {\bibfield  {journal}
  {\bibinfo  {journal} {Phys. Rev. Lett.}\ }\textbf {\bibinfo {volume} {121}},\
  \bibinfo {pages} {261601} (\bibinfo {year} {2018})},\ \Eprint
  {https://arxiv.org/abs/1807.05175} {arXiv:1807.05175 [hep-th]} \BibitemShut
  {NoStop}%
\bibitem [{\citenamefont {Cartwright}\ and\ \citenamefont
  {Kaminski}(2019)}]{Cartwright:2019opv}%
  \BibitemOpen
  \bibfield  {author} {\bibinfo {author} {\bibfnamefont {C.}~\bibnamefont
  {Cartwright}}\ and\ \bibinfo {author} {\bibfnamefont {M.}~\bibnamefont
  {Kaminski}},\ }\href {https://doi.org/10.1007/JHEP09(2019)072} {\bibfield
  {journal} {\bibinfo  {journal} {JHEP}\ }\textbf {\bibinfo {volume} {09}},\
  \bibinfo {pages} {072}},\ \Eprint {https://arxiv.org/abs/1904.11507}
  {arXiv:1904.11507 [hep-th]} \BibitemShut {NoStop}%
\bibitem [{\citenamefont {Kurkela}\ \emph {et~al.}(2020)\citenamefont
  {Kurkela}, \citenamefont {van~der Schee}, \citenamefont {Wiedemann},\ and\
  \citenamefont {Wu}}]{Kurkela:2019set}%
  \BibitemOpen
  \bibfield  {author} {\bibinfo {author} {\bibfnamefont {A.}~\bibnamefont
  {Kurkela}}, \bibinfo {author} {\bibfnamefont {W.}~\bibnamefont {van~der
  Schee}}, \bibinfo {author} {\bibfnamefont {U.~A.}\ \bibnamefont
  {Wiedemann}},\ and\ \bibinfo {author} {\bibfnamefont {B.}~\bibnamefont
  {Wu}},\ }\href {https://doi.org/10.1103/PhysRevLett.124.102301} {\bibfield
  {journal} {\bibinfo  {journal} {Phys. Rev. Lett.}\ }\textbf {\bibinfo
  {volume} {124}},\ \bibinfo {pages} {102301} (\bibinfo {year} {2020})},\
  \Eprint {https://arxiv.org/abs/1907.08101} {arXiv:1907.08101 [hep-ph]}
  \BibitemShut {NoStop}%
\bibitem [{\citenamefont {Ecker}\ \emph {et~al.}(2021)\citenamefont {Ecker},
  \citenamefont {Erdmenger},\ and\ \citenamefont {van~der
  Schee}}]{Ecker:2021ukv}%
  \BibitemOpen
  \bibfield  {author} {\bibinfo {author} {\bibfnamefont {C.}~\bibnamefont
  {Ecker}}, \bibinfo {author} {\bibfnamefont {J.}~\bibnamefont {Erdmenger}},\
  and\ \bibinfo {author} {\bibfnamefont {W.}~\bibnamefont {van~der Schee}},\
  }\href {https://doi.org/10.21468/SciPostPhys.11.3.047} {\bibfield  {journal}
  {\bibinfo  {journal} {SciPost Phys.}\ }\textbf {\bibinfo {volume} {11}},\
  \bibinfo {pages} {047} (\bibinfo {year} {2021})},\ \Eprint
  {https://arxiv.org/abs/2103.10435} {arXiv:2103.10435 [hep-th]} \BibitemShut
  {NoStop}%
\bibitem [{\citenamefont {Ghosh}\ \emph {et~al.}(2021)\citenamefont {Ghosh},
  \citenamefont {Grieninger}, \citenamefont {Landsteiner},\ and\ \citenamefont
  {Morales-Tejera}}]{Ghosh:2021naw}%
  \BibitemOpen
  \bibfield  {author} {\bibinfo {author} {\bibfnamefont {J.~K.}\ \bibnamefont
  {Ghosh}}, \bibinfo {author} {\bibfnamefont {S.}~\bibnamefont {Grieninger}},
  \bibinfo {author} {\bibfnamefont {K.}~\bibnamefont {Landsteiner}},\ and\
  \bibinfo {author} {\bibfnamefont {S.}~\bibnamefont {Morales-Tejera}},\ }\href
  {https://doi.org/10.1103/PhysRevD.104.046009} {\bibfield  {journal} {\bibinfo
   {journal} {Phys. Rev. D}\ }\textbf {\bibinfo {volume} {104}},\ \bibinfo
  {pages} {046009} (\bibinfo {year} {2021})},\ \Eprint
  {https://arxiv.org/abs/2105.05855} {arXiv:2105.05855 [hep-ph]} \BibitemShut
  {NoStop}%
\bibitem [{\citenamefont {Bjorken}(1983)}]{Bjorken:1982qr}%
  \BibitemOpen
  \bibfield  {author} {\bibinfo {author} {\bibfnamefont {J.~D.}\ \bibnamefont
  {Bjorken}},\ }\href {https://doi.org/10.1103/PhysRevD.27.140} {\bibfield
  {journal} {\bibinfo  {journal} {Phys. Rev. D}\ }\textbf {\bibinfo {volume}
  {27}},\ \bibinfo {pages} {140} (\bibinfo {year} {1983})}\BibitemShut
  {NoStop}%
\bibitem [{\citenamefont {Baier}\ \emph {et~al.}(2008)\citenamefont {Baier},
  \citenamefont {Romatschke}, \citenamefont {Son}, \citenamefont {Starinets},\
  and\ \citenamefont {Stephanov}}]{Baier:2007ix}%
  \BibitemOpen
  \bibfield  {author} {\bibinfo {author} {\bibfnamefont {R.}~\bibnamefont
  {Baier}}, \bibinfo {author} {\bibfnamefont {P.}~\bibnamefont {Romatschke}},
  \bibinfo {author} {\bibfnamefont {D.~T.}\ \bibnamefont {Son}}, \bibinfo
  {author} {\bibfnamefont {A.~O.}\ \bibnamefont {Starinets}},\ and\ \bibinfo
  {author} {\bibfnamefont {M.~A.}\ \bibnamefont {Stephanov}},\ }\href
  {https://doi.org/10.1088/1126-6708/2008/04/100} {\bibfield  {journal}
  {\bibinfo  {journal} {JHEP}\ }\textbf {\bibinfo {volume} {04}},\ \bibinfo
  {pages} {100}},\ \Eprint {https://arxiv.org/abs/0712.2451} {arXiv:0712.2451
  [hep-th]} \BibitemShut {NoStop}%
\bibitem [{\citenamefont {Bhattacharyya}\ \emph {et~al.}(2008)\citenamefont
  {Bhattacharyya}, \citenamefont {Hubeny}, \citenamefont {Minwalla},\ and\
  \citenamefont {Rangamani}}]{Bhattacharyya:2008jc}%
  \BibitemOpen
  \bibfield  {author} {\bibinfo {author} {\bibfnamefont {S.}~\bibnamefont
  {Bhattacharyya}}, \bibinfo {author} {\bibfnamefont {V.~E.}\ \bibnamefont
  {Hubeny}}, \bibinfo {author} {\bibfnamefont {S.}~\bibnamefont {Minwalla}},\
  and\ \bibinfo {author} {\bibfnamefont {M.}~\bibnamefont {Rangamani}},\ }\href
  {https://doi.org/10.1088/1126-6708/2008/02/045} {\bibfield  {journal}
  {\bibinfo  {journal} {JHEP}\ }\textbf {\bibinfo {volume} {02}},\ \bibinfo
  {pages} {045}},\ \Eprint {https://arxiv.org/abs/0712.2456} {arXiv:0712.2456
  [hep-th]} \BibitemShut {NoStop}%
\bibitem [{\citenamefont {Muller}(1967)}]{Muller:1967zza}%
  \BibitemOpen
  \bibfield  {author} {\bibinfo {author} {\bibfnamefont {I.}~\bibnamefont
  {Muller}},\ }\href {https://doi.org/10.1007/BF01326412} {\bibfield  {journal}
  {\bibinfo  {journal} {Z. Phys.}\ }\textbf {\bibinfo {volume} {198}},\
  \bibinfo {pages} {329} (\bibinfo {year} {1967})}\BibitemShut {NoStop}%
\bibitem [{\citenamefont {Israel}(1976)}]{Israel:1976tn}%
  \BibitemOpen
  \bibfield  {author} {\bibinfo {author} {\bibfnamefont {W.}~\bibnamefont
  {Israel}},\ }\href {https://doi.org/10.1016/0003-4916(76)90064-6} {\bibfield
  {journal} {\bibinfo  {journal} {Annals Phys.}\ }\textbf {\bibinfo {volume}
  {100}},\ \bibinfo {pages} {310} (\bibinfo {year} {1976})}\BibitemShut
  {NoStop}%
\bibitem [{\citenamefont {Israel}\ and\ \citenamefont
  {Stewart}(1979)}]{Israel:1979wp}%
  \BibitemOpen
  \bibfield  {author} {\bibinfo {author} {\bibfnamefont {W.}~\bibnamefont
  {Israel}}\ and\ \bibinfo {author} {\bibfnamefont {J.~M.}\ \bibnamefont
  {Stewart}},\ }\href {https://doi.org/10.1016/0003-4916(79)90130-1} {\bibfield
   {journal} {\bibinfo  {journal} {Annals Phys.}\ }\textbf {\bibinfo {volume}
  {118}},\ \bibinfo {pages} {341} (\bibinfo {year} {1979})}\BibitemShut
  {NoStop}%
\bibitem [{\citenamefont {Heller}\ and\ \citenamefont
  {Spalinski}(2015)}]{Heller:2015dha}%
  \BibitemOpen
  \bibfield  {author} {\bibinfo {author} {\bibfnamefont {M.~P.}\ \bibnamefont
  {Heller}}\ and\ \bibinfo {author} {\bibfnamefont {M.}~\bibnamefont
  {Spalinski}},\ }\href {https://doi.org/10.1103/PhysRevLett.115.072501}
  {\bibfield  {journal} {\bibinfo  {journal} {Phys. Rev. Lett.}\ }\textbf
  {\bibinfo {volume} {115}},\ \bibinfo {pages} {072501} (\bibinfo {year}
  {2015})},\ \Eprint {https://arxiv.org/abs/1503.07514} {arXiv:1503.07514
  [hep-th]} \BibitemShut {NoStop}%
\bibitem [{\citenamefont {Florkowski}\ \emph {et~al.}(2018)\citenamefont
  {Florkowski}, \citenamefont {Heller},\ and\ \citenamefont
  {Spalinski}}]{Florkowski:2017olj}%
  \BibitemOpen
  \bibfield  {author} {\bibinfo {author} {\bibfnamefont {W.}~\bibnamefont
  {Florkowski}}, \bibinfo {author} {\bibfnamefont {M.~P.}\ \bibnamefont
  {Heller}},\ and\ \bibinfo {author} {\bibfnamefont {M.}~\bibnamefont
  {Spalinski}},\ }\href {https://doi.org/10.1088/1361-6633/aaa091} {\bibfield
  {journal} {\bibinfo  {journal} {Rept. Prog. Phys.}\ }\textbf {\bibinfo
  {volume} {81}},\ \bibinfo {pages} {046001} (\bibinfo {year} {2018})},\
  \Eprint {https://arxiv.org/abs/1707.02282} {arXiv:1707.02282 [hep-ph]}
  \BibitemShut {NoStop}%
\bibitem [{\citenamefont {Heller}\ \emph {et~al.}(2013)\citenamefont {Heller},
  \citenamefont {Janik},\ and\ \citenamefont {Witaszczyk}}]{Heller:2013fn}%
  \BibitemOpen
  \bibfield  {author} {\bibinfo {author} {\bibfnamefont {M.~P.}\ \bibnamefont
  {Heller}}, \bibinfo {author} {\bibfnamefont {R.~A.}\ \bibnamefont {Janik}},\
  and\ \bibinfo {author} {\bibfnamefont {P.}~\bibnamefont {Witaszczyk}},\
  }\href {https://doi.org/10.1103/PhysRevLett.110.211602} {\bibfield  {journal}
  {\bibinfo  {journal} {Phys. Rev. Lett.}\ }\textbf {\bibinfo {volume} {110}},\
  \bibinfo {pages} {211602} (\bibinfo {year} {2013})},\ \Eprint
  {https://arxiv.org/abs/1302.0697} {arXiv:1302.0697 [hep-th]} \BibitemShut
  {NoStop}%
\bibitem [{\citenamefont {Buchel}\ \emph {et~al.}(2016)\citenamefont {Buchel},
  \citenamefont {Heller},\ and\ \citenamefont {Noronha}}]{Buchel:2016cbj}%
  \BibitemOpen
  \bibfield  {author} {\bibinfo {author} {\bibfnamefont {A.}~\bibnamefont
  {Buchel}}, \bibinfo {author} {\bibfnamefont {M.~P.}\ \bibnamefont {Heller}},\
  and\ \bibinfo {author} {\bibfnamefont {J.}~\bibnamefont {Noronha}},\ }\href
  {https://doi.org/10.1103/PhysRevD.94.106011} {\bibfield  {journal} {\bibinfo
  {journal} {Phys. Rev. D}\ }\textbf {\bibinfo {volume} {94}},\ \bibinfo
  {pages} {106011} (\bibinfo {year} {2016})},\ \Eprint
  {https://arxiv.org/abs/1603.05344} {arXiv:1603.05344 [hep-th]} \BibitemShut
  {NoStop}%
\bibitem [{\citenamefont {Denicol}\ and\ \citenamefont
  {Noronha}(2016)}]{Denicol:2016bjh}%
  \BibitemOpen
  \bibfield  {author} {\bibinfo {author} {\bibfnamefont {G.~S.}\ \bibnamefont
  {Denicol}}\ and\ \bibinfo {author} {\bibfnamefont {J.}~\bibnamefont
  {Noronha}},\ }\href@noop {} {\  (\bibinfo {year} {2016})},\ \Eprint
  {https://arxiv.org/abs/1608.07869} {arXiv:1608.07869 [nucl-th]} \BibitemShut
  {NoStop}%
\bibitem [{\citenamefont {Heller}\ \emph {et~al.}(2018)\citenamefont {Heller},
  \citenamefont {Kurkela}, \citenamefont {Spali\'nski},\ and\ \citenamefont
  {Svensson}}]{Heller:2016rtz}%
  \BibitemOpen
  \bibfield  {author} {\bibinfo {author} {\bibfnamefont {M.~P.}\ \bibnamefont
  {Heller}}, \bibinfo {author} {\bibfnamefont {A.}~\bibnamefont {Kurkela}},
  \bibinfo {author} {\bibfnamefont {M.}~\bibnamefont {Spali\'nski}},\ and\
  \bibinfo {author} {\bibfnamefont {V.}~\bibnamefont {Svensson}},\ }\href
  {https://doi.org/10.1103/PhysRevD.97.091503} {\bibfield  {journal} {\bibinfo
  {journal} {Phys. Rev. D}\ }\textbf {\bibinfo {volume} {97}},\ \bibinfo
  {pages} {091503} (\bibinfo {year} {2018})},\ \Eprint
  {https://arxiv.org/abs/1609.04803} {arXiv:1609.04803 [nucl-th]} \BibitemShut
  {NoStop}%
\bibitem [{\citenamefont {Figueras}\ \emph {et~al.}(2009)\citenamefont
  {Figueras}, \citenamefont {Hubeny}, \citenamefont {Rangamani},\ and\
  \citenamefont {Ross}}]{Figueras:2009iu}%
  \BibitemOpen
  \bibfield  {author} {\bibinfo {author} {\bibfnamefont {P.}~\bibnamefont
  {Figueras}}, \bibinfo {author} {\bibfnamefont {V.~E.}\ \bibnamefont
  {Hubeny}}, \bibinfo {author} {\bibfnamefont {M.}~\bibnamefont {Rangamani}},\
  and\ \bibinfo {author} {\bibfnamefont {S.~F.}\ \bibnamefont {Ross}},\ }\href
  {https://doi.org/10.1088/1126-6708/2009/04/137} {\bibfield  {journal}
  {\bibinfo  {journal} {JHEP}\ }\textbf {\bibinfo {volume} {04}},\ \bibinfo
  {pages} {137}},\ \Eprint {https://arxiv.org/abs/0902.4696} {arXiv:0902.4696
  [hep-th]} \BibitemShut {NoStop}%
\bibitem [{\citenamefont {Rougemont}\ \emph {et~al.}(2021)\citenamefont
  {Rougemont}, \citenamefont {Noronha}, \citenamefont {Barreto}, \citenamefont
  {Denicol},\ and\ \citenamefont {Dore}}]{Rougemont:2021qyk}%
  \BibitemOpen
  \bibfield  {author} {\bibinfo {author} {\bibfnamefont {R.}~\bibnamefont
  {Rougemont}}, \bibinfo {author} {\bibfnamefont {J.}~\bibnamefont {Noronha}},
  \bibinfo {author} {\bibfnamefont {W.}~\bibnamefont {Barreto}}, \bibinfo
  {author} {\bibfnamefont {G.~S.}\ \bibnamefont {Denicol}},\ and\ \bibinfo
  {author} {\bibfnamefont {T.}~\bibnamefont {Dore}},\ }\href
  {https://doi.org/10.1103/PhysRevD.104.126012} {\bibfield  {journal} {\bibinfo
   {journal} {Phys. Rev. D}\ }\textbf {\bibinfo {volume} {104}},\ \bibinfo
  {pages} {126012} (\bibinfo {year} {2021})},\ \Eprint
  {https://arxiv.org/abs/2105.02378} {arXiv:2105.02378 [nucl-th]} \BibitemShut
  {NoStop}%
\bibitem [{\citenamefont {Gubser}(2010)}]{Gubser:2010ze}%
  \BibitemOpen
  \bibfield  {author} {\bibinfo {author} {\bibfnamefont {S.~S.}\ \bibnamefont
  {Gubser}},\ }\href {https://doi.org/10.1103/PhysRevD.82.085027} {\bibfield
  {journal} {\bibinfo  {journal} {Phys. Rev. D}\ }\textbf {\bibinfo {volume}
  {82}},\ \bibinfo {pages} {085027} (\bibinfo {year} {2010})},\ \Eprint
  {https://arxiv.org/abs/1006.0006} {arXiv:1006.0006 [hep-th]} \BibitemShut
  {NoStop}%
\bibitem [{\citenamefont {Arsene}\ \emph {et~al.}(2005)\citenamefont {Arsene}
  \emph {et~al.}}]{Arsene:2004fa}%
  \BibitemOpen
  \bibfield  {author} {\bibinfo {author} {\bibfnamefont {I.}~\bibnamefont
  {Arsene}} \emph {et~al.} (\bibinfo {collaboration} {BRAHMS}),\ }\href
  {https://doi.org/10.1016/j.nuclphysa.2005.02.130} {\bibfield  {journal}
  {\bibinfo  {journal} {Nucl. Phys. A}\ }\textbf {\bibinfo {volume} {757}},\
  \bibinfo {pages} {1} (\bibinfo {year} {2005})},\ \Eprint
  {https://arxiv.org/abs/nucl-ex/0410020} {arXiv:nucl-ex/0410020} \BibitemShut
  {NoStop}%
\bibitem [{\citenamefont {Adcox}\ \emph {et~al.}(2005)\citenamefont {Adcox}
  \emph {et~al.}}]{Adcox:2004mh}%
  \BibitemOpen
  \bibfield  {author} {\bibinfo {author} {\bibfnamefont {K.}~\bibnamefont
  {Adcox}} \emph {et~al.} (\bibinfo {collaboration} {PHENIX}),\ }\href
  {https://doi.org/10.1016/j.nuclphysa.2005.03.086} {\bibfield  {journal}
  {\bibinfo  {journal} {Nucl. Phys. A}\ }\textbf {\bibinfo {volume} {757}},\
  \bibinfo {pages} {184} (\bibinfo {year} {2005})},\ \Eprint
  {https://arxiv.org/abs/nucl-ex/0410003} {arXiv:nucl-ex/0410003} \BibitemShut
  {NoStop}%
\bibitem [{\citenamefont {Back}\ \emph {et~al.}(2005)\citenamefont {Back} \emph
  {et~al.}}]{Back:2004je}%
  \BibitemOpen
  \bibfield  {author} {\bibinfo {author} {\bibfnamefont {B.~B.}\ \bibnamefont
  {Back}} \emph {et~al.} (\bibinfo {collaboration} {PHOBOS}),\ }\href
  {https://doi.org/10.1016/j.nuclphysa.2005.03.084} {\bibfield  {journal}
  {\bibinfo  {journal} {Nucl. Phys. A}\ }\textbf {\bibinfo {volume} {757}},\
  \bibinfo {pages} {28} (\bibinfo {year} {2005})},\ \Eprint
  {https://arxiv.org/abs/nucl-ex/0410022} {arXiv:nucl-ex/0410022} \BibitemShut
  {NoStop}%
\bibitem [{\citenamefont {Adams}\ \emph {et~al.}(2005)\citenamefont {Adams}
  \emph {et~al.}}]{Adams:2005dq}%
  \BibitemOpen
  \bibfield  {author} {\bibinfo {author} {\bibfnamefont {J.}~\bibnamefont
  {Adams}} \emph {et~al.} (\bibinfo {collaboration} {STAR}),\ }\href
  {https://doi.org/10.1016/j.nuclphysa.2005.03.085} {\bibfield  {journal}
  {\bibinfo  {journal} {Nucl. Phys. A}\ }\textbf {\bibinfo {volume} {757}},\
  \bibinfo {pages} {102} (\bibinfo {year} {2005})},\ \Eprint
  {https://arxiv.org/abs/nucl-ex/0501009} {arXiv:nucl-ex/0501009} \BibitemShut
  {NoStop}%
\bibitem [{\citenamefont {Aad}\ \emph {et~al.}(2013)\citenamefont {Aad} \emph
  {et~al.}}]{Aad:2013xma}%
  \BibitemOpen
  \bibfield  {author} {\bibinfo {author} {\bibfnamefont {G.}~\bibnamefont
  {Aad}} \emph {et~al.} (\bibinfo {collaboration} {ATLAS}),\ }\href
  {https://doi.org/10.1007/JHEP11(2013)183} {\bibfield  {journal} {\bibinfo
  {journal} {JHEP}\ }\textbf {\bibinfo {volume} {11}},\ \bibinfo {pages}
  {183}},\ \Eprint {https://arxiv.org/abs/1305.2942} {arXiv:1305.2942 [hep-ex]}
  \BibitemShut {NoStop}%
\bibitem [{\citenamefont {de~Boer}\ \emph {et~al.}(2000)\citenamefont
  {de~Boer}, \citenamefont {Verlinde},\ and\ \citenamefont
  {Verlinde}}]{deBoer:1999tgo}%
  \BibitemOpen
  \bibfield  {author} {\bibinfo {author} {\bibfnamefont {J.}~\bibnamefont
  {de~Boer}}, \bibinfo {author} {\bibfnamefont {E.~P.}\ \bibnamefont
  {Verlinde}},\ and\ \bibinfo {author} {\bibfnamefont {H.~L.}\ \bibnamefont
  {Verlinde}},\ }\href {https://doi.org/10.1088/1126-6708/2000/08/003}
  {\bibfield  {journal} {\bibinfo  {journal} {JHEP}\ }\textbf {\bibinfo
  {volume} {08}},\ \bibinfo {pages} {003}},\ \Eprint
  {https://arxiv.org/abs/hep-th/9912012} {arXiv:hep-th/9912012} \BibitemShut
  {NoStop}%
\bibitem [{\citenamefont {York}(1972)}]{York:1972sj}%
  \BibitemOpen
  \bibfield  {author} {\bibinfo {author} {\bibfnamefont {J.~W.}\ \bibnamefont
  {York}, \bibfnamefont {Jr.}},\ }\href
  {https://doi.org/10.1103/PhysRevLett.28.1082} {\bibfield  {journal} {\bibinfo
   {journal} {Phys. Rev. Lett.}\ }\textbf {\bibinfo {volume} {28}},\ \bibinfo
  {pages} {1082} (\bibinfo {year} {1972})}\BibitemShut {NoStop}%
\bibitem [{\citenamefont {Gibbons}\ and\ \citenamefont
  {Hawking}(1977)}]{Gibbons:1976ue}%
  \BibitemOpen
  \bibfield  {author} {\bibinfo {author} {\bibfnamefont {G.~W.}\ \bibnamefont
  {Gibbons}}\ and\ \bibinfo {author} {\bibfnamefont {S.~W.}\ \bibnamefont
  {Hawking}},\ }\href {https://doi.org/10.1103/PhysRevD.15.2752} {\bibfield
  {journal} {\bibinfo  {journal} {Phys. Rev. D}\ }\textbf {\bibinfo {volume}
  {15}},\ \bibinfo {pages} {2752} (\bibinfo {year} {1977})}\BibitemShut
  {NoStop}%
\bibitem [{\citenamefont {de~Haro}\ \emph {et~al.}(2001)\citenamefont
  {de~Haro}, \citenamefont {Solodukhin},\ and\ \citenamefont
  {Skenderis}}]{deHaro:2000vlm}%
  \BibitemOpen
  \bibfield  {author} {\bibinfo {author} {\bibfnamefont {S.}~\bibnamefont
  {de~Haro}}, \bibinfo {author} {\bibfnamefont {S.~N.}\ \bibnamefont
  {Solodukhin}},\ and\ \bibinfo {author} {\bibfnamefont {K.}~\bibnamefont
  {Skenderis}},\ }\href {https://doi.org/10.1007/s002200100381} {\bibfield
  {journal} {\bibinfo  {journal} {Commun. Math. Phys.}\ }\textbf {\bibinfo
  {volume} {217}},\ \bibinfo {pages} {595} (\bibinfo {year} {2001})},\ \Eprint
  {https://arxiv.org/abs/hep-th/0002230} {arXiv:hep-th/0002230} \BibitemShut
  {NoStop}%
\bibitem [{\citenamefont {Bianchi}\ \emph {et~al.}(2002)\citenamefont
  {Bianchi}, \citenamefont {Freedman},\ and\ \citenamefont
  {Skenderis}}]{Bianchi:2001kw}%
  \BibitemOpen
  \bibfield  {author} {\bibinfo {author} {\bibfnamefont {M.}~\bibnamefont
  {Bianchi}}, \bibinfo {author} {\bibfnamefont {D.~Z.}\ \bibnamefont
  {Freedman}},\ and\ \bibinfo {author} {\bibfnamefont {K.}~\bibnamefont
  {Skenderis}},\ }\href {https://doi.org/10.1016/S0550-3213(02)00179-7}
  {\bibfield  {journal} {\bibinfo  {journal} {Nucl. Phys. B}\ }\textbf
  {\bibinfo {volume} {631}},\ \bibinfo {pages} {159} (\bibinfo {year}
  {2002})},\ \Eprint {https://arxiv.org/abs/hep-th/0112119}
  {arXiv:hep-th/0112119} \BibitemShut {NoStop}%
\bibitem [{\citenamefont {Skenderis}(2002)}]{Skenderis:2002wp}%
  \BibitemOpen
  \bibfield  {author} {\bibinfo {author} {\bibfnamefont {K.}~\bibnamefont
  {Skenderis}},\ }\href {https://doi.org/10.1088/0264-9381/19/22/306}
  {\bibfield  {journal} {\bibinfo  {journal} {Class. Quant. Grav.}\ }\textbf
  {\bibinfo {volume} {19}},\ \bibinfo {pages} {5849} (\bibinfo {year}
  {2002})},\ \Eprint {https://arxiv.org/abs/hep-th/0209067}
  {arXiv:hep-th/0209067} \BibitemShut {NoStop}%
\bibitem [{\citenamefont {Lindgren}\ \emph {et~al.}(2015)\citenamefont
  {Lindgren}, \citenamefont {Papadimitriou}, \citenamefont {Taliotis},\ and\
  \citenamefont {Vanhoof}}]{Lindgren:2015lia}%
  \BibitemOpen
  \bibfield  {author} {\bibinfo {author} {\bibfnamefont {J.}~\bibnamefont
  {Lindgren}}, \bibinfo {author} {\bibfnamefont {I.}~\bibnamefont
  {Papadimitriou}}, \bibinfo {author} {\bibfnamefont {A.}~\bibnamefont
  {Taliotis}},\ and\ \bibinfo {author} {\bibfnamefont {J.}~\bibnamefont
  {Vanhoof}},\ }\href {https://doi.org/10.1007/JHEP07(2015)094} {\bibfield
  {journal} {\bibinfo  {journal} {JHEP}\ }\textbf {\bibinfo {volume} {07}},\
  \bibinfo {pages} {094}},\ \Eprint {https://arxiv.org/abs/1505.04131}
  {arXiv:1505.04131 [hep-th]} \BibitemShut {NoStop}%
\bibitem [{\citenamefont {Elvang}\ and\ \citenamefont
  {Hadjiantonis}(2016)}]{Elvang:2016tzz}%
  \BibitemOpen
  \bibfield  {author} {\bibinfo {author} {\bibfnamefont {H.}~\bibnamefont
  {Elvang}}\ and\ \bibinfo {author} {\bibfnamefont {M.}~\bibnamefont
  {Hadjiantonis}},\ }\href {https://doi.org/10.1007/JHEP06(2016)046} {\bibfield
   {journal} {\bibinfo  {journal} {JHEP}\ }\textbf {\bibinfo {volume} {06}},\
  \bibinfo {pages} {046}},\ \Eprint {https://arxiv.org/abs/1603.04485}
  {arXiv:1603.04485 [hep-th]} \BibitemShut {NoStop}%
\bibitem [{\citenamefont {Bondi}\ \emph {et~al.}(1962)\citenamefont {Bondi},
  \citenamefont {van~der Burg},\ and\ \citenamefont {Metzner}}]{Bondi:1962px}%
  \BibitemOpen
  \bibfield  {author} {\bibinfo {author} {\bibfnamefont {H.}~\bibnamefont
  {Bondi}}, \bibinfo {author} {\bibfnamefont {M.~G.~J.}\ \bibnamefont {van~der
  Burg}},\ and\ \bibinfo {author} {\bibfnamefont {A.~W.~K.}\ \bibnamefont
  {Metzner}},\ }\href {https://doi.org/10.1098/rspa.1962.0161} {\bibfield
  {journal} {\bibinfo  {journal} {Proc. Roy. Soc. Lond. A}\ }\textbf {\bibinfo
  {volume} {269}},\ \bibinfo {pages} {21} (\bibinfo {year} {1962})}\BibitemShut
  {NoStop}%
\bibitem [{\citenamefont {Sachs}(1962)}]{Sachs:1962wk}%
  \BibitemOpen
  \bibfield  {author} {\bibinfo {author} {\bibfnamefont {R.~K.}\ \bibnamefont
  {Sachs}},\ }\href {https://doi.org/10.1098/rspa.1962.0206} {\bibfield
  {journal} {\bibinfo  {journal} {Proc. Roy. Soc. Lond. A}\ }\textbf {\bibinfo
  {volume} {270}},\ \bibinfo {pages} {103} (\bibinfo {year}
  {1962})}\BibitemShut {NoStop}%
\bibitem [{\citenamefont {Boyd}(2001)}]{boyd01}%
  \BibitemOpen
  \bibfield  {author} {\bibinfo {author} {\bibfnamefont {J.~P.}\ \bibnamefont
  {Boyd}},\ }\href@noop {} {\emph {\bibinfo {title} {{Chebyshev} and {Fourier}
  Spectral Methods}}},\ \bibinfo {edition} {2nd}\ ed.,\ Dover Books on
  Mathematics\ (\bibinfo  {publisher} {Dover Publications},\ \bibinfo {address}
  {Mineola, NY},\ \bibinfo {year} {2001})\BibitemShut {NoStop}%
\bibitem [{\citenamefont {van~der Schee}(2015)}]{wilkaodamassa}%
  \BibitemOpen
  \bibfield  {author} {\bibinfo {author} {\bibfnamefont {W.}~\bibnamefont
  {van~der Schee}},\ }\href
  {https://sites.google.com/site/wilkevanderschee/ads-numerics} {\emph
  {\bibinfo {title} {Lectures on numerics in dynamical AdS spacetimes}}}\
  (\bibinfo  {publisher} {Personal Website},\ \bibinfo {year}
  {2015})\BibitemShut {NoStop}%
\bibitem [{\citenamefont {Hawking}\ and\ \citenamefont
  {Ellis}(1975)}]{HawkingEllisBook}%
  \BibitemOpen
  \bibfield  {author} {\bibinfo {author} {\bibfnamefont {S.~W.}\ \bibnamefont
  {Hawking}}\ and\ \bibinfo {author} {\bibfnamefont {G.~F.~R.}\ \bibnamefont
  {Ellis}},\ }\href {https://doi.org/10.1017/CBO9780511524646} {\emph {\bibinfo
  {title} {The Large Scale Structure of Space-Time (Cambridge Monographs on
  Mathematical Physics)}}}\ (\bibinfo  {publisher} {Cambridge University
  Press},\ \bibinfo {year} {1975})\ p.\ \bibinfo {pages} {404}\BibitemShut
  {NoStop}%
\bibitem [{\citenamefont {Wald}(2010)}]{WaldBookGR1984}%
  \BibitemOpen
  \bibfield  {author} {\bibinfo {author} {\bibfnamefont {R.~M.}\ \bibnamefont
  {Wald}},\ }\href@noop {} {\emph {\bibinfo {title} {General relativity}}}\
  (\bibinfo  {publisher} {University of Chicago press},\ \bibinfo {year}
  {2010})\BibitemShut {NoStop}%
\bibitem [{\citenamefont {Visser}\ and\ \citenamefont
  {Barcelo}(1999)}]{Visser:1999de}%
  \BibitemOpen
  \bibfield  {author} {\bibinfo {author} {\bibfnamefont {M.}~\bibnamefont
  {Visser}}\ and\ \bibinfo {author} {\bibfnamefont {C.}~\bibnamefont
  {Barcelo}},\ }in\ \href {https://doi.org/10.1142/9789812792129_0014} {\emph
  {\bibinfo {booktitle} {{3rd International Conference on Particle Physics and
  the Early Universe}}}}\ (\bibinfo {year} {1999})\ \Eprint
  {https://arxiv.org/abs/gr-qc/0001099} {arXiv:gr-qc/0001099} \BibitemShut
  {NoStop}%
\bibitem [{\citenamefont {Janik}\ and\ \citenamefont
  {Peschanski}(2006)}]{Janik:2005zt}%
  \BibitemOpen
  \bibfield  {author} {\bibinfo {author} {\bibfnamefont {R.~A.}\ \bibnamefont
  {Janik}}\ and\ \bibinfo {author} {\bibfnamefont {R.~B.}\ \bibnamefont
  {Peschanski}},\ }\href {https://doi.org/10.1103/PhysRevD.73.045013}
  {\bibfield  {journal} {\bibinfo  {journal} {Phys. Rev. D}\ }\textbf {\bibinfo
  {volume} {73}},\ \bibinfo {pages} {045013} (\bibinfo {year} {2006})},\
  \Eprint {https://arxiv.org/abs/hep-th/0512162} {arXiv:hep-th/0512162}
  \BibitemShut {NoStop}%
\bibitem [{\citenamefont {Arnold}\ \emph {et~al.}(2014)\citenamefont {Arnold},
  \citenamefont {Romatschke},\ and\ \citenamefont {van~der
  Schee}}]{Arnold:2014jva}%
  \BibitemOpen
  \bibfield  {author} {\bibinfo {author} {\bibfnamefont {P.}~\bibnamefont
  {Arnold}}, \bibinfo {author} {\bibfnamefont {P.}~\bibnamefont {Romatschke}},\
  and\ \bibinfo {author} {\bibfnamefont {W.}~\bibnamefont {van~der Schee}},\
  }\href {https://doi.org/10.1007/JHEP10(2014)110} {\bibfield  {journal}
  {\bibinfo  {journal} {JHEP}\ }\textbf {\bibinfo {volume} {10}},\ \bibinfo
  {pages} {110}},\ \Eprint {https://arxiv.org/abs/1408.2518} {arXiv:1408.2518
  [hep-th]} \BibitemShut {NoStop}%
\bibitem [{\citenamefont {Groot}(1980)}]{degroot}%
  \BibitemOpen
  \bibfield  {author} {\bibinfo {author} {\bibfnamefont {S.~R.~D.}\
  \bibnamefont {Groot}},\ }\href@noop {} {\emph {\bibinfo {title} {Relativistic
  Kinetic Theory. Principles and Applications}}}\ (\bibinfo  {publisher}
  {Amsterdam, Netherlands: North-holland ( 1980) 417p},\ \bibinfo {year}
  {1980})\BibitemShut {NoStop}%
\bibitem [{\citenamefont {Bekenstein}(1973)}]{Bekenstein:1973ur}%
  \BibitemOpen
  \bibfield  {author} {\bibinfo {author} {\bibfnamefont {J.~D.}\ \bibnamefont
  {Bekenstein}},\ }\href {https://doi.org/10.1103/PhysRevD.7.2333} {\bibfield
  {journal} {\bibinfo  {journal} {Phys. Rev. D}\ }\textbf {\bibinfo {volume}
  {7}},\ \bibinfo {pages} {2333} (\bibinfo {year} {1973})}\BibitemShut
  {NoStop}%
\bibitem [{\citenamefont {Hawking}(1975)}]{Hawking:1974sw}%
  \BibitemOpen
  \bibfield  {author} {\bibinfo {author} {\bibfnamefont {S.~W.}\ \bibnamefont
  {Hawking}},\ }\href {https://doi.org/10.1007/BF02345020} {\bibfield
  {journal} {\bibinfo  {journal} {Commun. Math. Phys.}\ }\textbf {\bibinfo
  {volume} {43}},\ \bibinfo {pages} {199} (\bibinfo {year} {1975})},\ \bibinfo
  {note} {[Erratum: Commun.Math.Phys. 46, 206 (1976)]}\BibitemShut {NoStop}%
\bibitem [{\citenamefont {Friess}\ \emph {et~al.}(2007)\citenamefont {Friess},
  \citenamefont {Gubser}, \citenamefont {Michalogiorgakis},\ and\ \citenamefont
  {Pufu}}]{Friess:2006kw}%
  \BibitemOpen
  \bibfield  {author} {\bibinfo {author} {\bibfnamefont {J.~J.}\ \bibnamefont
  {Friess}}, \bibinfo {author} {\bibfnamefont {S.~S.}\ \bibnamefont {Gubser}},
  \bibinfo {author} {\bibfnamefont {G.}~\bibnamefont {Michalogiorgakis}},\ and\
  \bibinfo {author} {\bibfnamefont {S.~S.}\ \bibnamefont {Pufu}},\ }\href
  {https://doi.org/10.1088/1126-6708/2007/04/080} {\bibfield  {journal}
  {\bibinfo  {journal} {JHEP}\ }\textbf {\bibinfo {volume} {04}},\ \bibinfo
  {pages} {080}},\ \Eprint {https://arxiv.org/abs/hep-th/0611005}
  {arXiv:hep-th/0611005} \BibitemShut {NoStop}%
\bibitem [{\citenamefont {M\"uller}\ \emph {et~al.}(2020)\citenamefont
  {M\"uller}, \citenamefont {Rabenstein}, \citenamefont {Sch\"afer},
  \citenamefont {Waeber},\ and\ \citenamefont {Yaffe}}]{Muller:2020ziz}%
  \BibitemOpen
  \bibfield  {author} {\bibinfo {author} {\bibfnamefont {B.}~\bibnamefont
  {M\"uller}}, \bibinfo {author} {\bibfnamefont {A.}~\bibnamefont
  {Rabenstein}}, \bibinfo {author} {\bibfnamefont {A.}~\bibnamefont
  {Sch\"afer}}, \bibinfo {author} {\bibfnamefont {S.}~\bibnamefont {Waeber}},\
  and\ \bibinfo {author} {\bibfnamefont {L.~G.}\ \bibnamefont {Yaffe}},\ }\href
  {https://doi.org/10.1103/PhysRevD.101.076008} {\bibfield  {journal} {\bibinfo
   {journal} {Phys. Rev. D}\ }\textbf {\bibinfo {volume} {101}},\ \bibinfo
  {pages} {076008} (\bibinfo {year} {2020})},\ \Eprint
  {https://arxiv.org/abs/2001.07161} {arXiv:2001.07161 [hep-ph]} \BibitemShut
  {NoStop}%
\bibitem [{\citenamefont {Engelhardt}\ and\ \citenamefont
  {Wall}(2018)}]{Engelhardt:2017aux}%
  \BibitemOpen
  \bibfield  {author} {\bibinfo {author} {\bibfnamefont {N.}~\bibnamefont
  {Engelhardt}}\ and\ \bibinfo {author} {\bibfnamefont {A.~C.}\ \bibnamefont
  {Wall}},\ }\href {https://doi.org/10.1103/PhysRevLett.121.211301} {\bibfield
  {journal} {\bibinfo  {journal} {Phys. Rev. Lett.}\ }\textbf {\bibinfo
  {volume} {121}},\ \bibinfo {pages} {211301} (\bibinfo {year} {2018})},\
  \Eprint {https://arxiv.org/abs/1706.02038} {arXiv:1706.02038 [hep-th]}
  \BibitemShut {NoStop}%
\bibitem [{\citenamefont {Booth}\ \emph {et~al.}(2009)\citenamefont {Booth},
  \citenamefont {Heller},\ and\ \citenamefont {Spalinski}}]{Booth:2009ct}%
  \BibitemOpen
  \bibfield  {author} {\bibinfo {author} {\bibfnamefont {I.}~\bibnamefont
  {Booth}}, \bibinfo {author} {\bibfnamefont {M.~P.}\ \bibnamefont {Heller}},\
  and\ \bibinfo {author} {\bibfnamefont {M.}~\bibnamefont {Spalinski}},\ }\href
  {https://doi.org/10.1103/PhysRevD.80.126013} {\bibfield  {journal} {\bibinfo
  {journal} {Phys. Rev. D}\ }\textbf {\bibinfo {volume} {80}},\ \bibinfo
  {pages} {126013} (\bibinfo {year} {2009})},\ \Eprint
  {https://arxiv.org/abs/0910.0748} {arXiv:0910.0748 [hep-th]} \BibitemShut
  {NoStop}%
\bibitem [{\citenamefont {Kinoshita}\ \emph {et~al.}(2009)\citenamefont
  {Kinoshita}, \citenamefont {Mukohyama}, \citenamefont {Nakamura},\ and\
  \citenamefont {Oda}}]{Kinoshita:2008dq}%
  \BibitemOpen
  \bibfield  {author} {\bibinfo {author} {\bibfnamefont {S.}~\bibnamefont
  {Kinoshita}}, \bibinfo {author} {\bibfnamefont {S.}~\bibnamefont
  {Mukohyama}}, \bibinfo {author} {\bibfnamefont {S.}~\bibnamefont
  {Nakamura}},\ and\ \bibinfo {author} {\bibfnamefont {K.-y.}\ \bibnamefont
  {Oda}},\ }\href {https://doi.org/10.1143/PTP.121.121} {\bibfield  {journal}
  {\bibinfo  {journal} {Prog. Theor. Phys.}\ }\textbf {\bibinfo {volume}
  {121}},\ \bibinfo {pages} {121} (\bibinfo {year} {2009})},\ \Eprint
  {https://arxiv.org/abs/0807.3797} {arXiv:0807.3797 [hep-th]} \BibitemShut
  {NoStop}%
\bibitem [{\citenamefont {Nakamura}\ and\ \citenamefont
  {Sin}(2006)}]{Nakamura:2006ih}%
  \BibitemOpen
  \bibfield  {author} {\bibinfo {author} {\bibfnamefont {S.}~\bibnamefont
  {Nakamura}}\ and\ \bibinfo {author} {\bibfnamefont {S.-J.}\ \bibnamefont
  {Sin}},\ }\href {https://doi.org/10.1088/1126-6708/2006/09/020} {\bibfield
  {journal} {\bibinfo  {journal} {JHEP}\ }\textbf {\bibinfo {volume} {09}},\
  \bibinfo {pages} {020}},\ \Eprint {https://arxiv.org/abs/hep-th/0607123}
  {arXiv:hep-th/0607123} \BibitemShut {NoStop}%
\bibitem [{\citenamefont {Finazzo}\ \emph {et~al.}(2015)\citenamefont
  {Finazzo}, \citenamefont {Rougemont}, \citenamefont {Marrochio},\ and\
  \citenamefont {Noronha}}]{Finazzo:2014cna}%
  \BibitemOpen
  \bibfield  {author} {\bibinfo {author} {\bibfnamefont {S.~I.}\ \bibnamefont
  {Finazzo}}, \bibinfo {author} {\bibfnamefont {R.}~\bibnamefont {Rougemont}},
  \bibinfo {author} {\bibfnamefont {H.}~\bibnamefont {Marrochio}},\ and\
  \bibinfo {author} {\bibfnamefont {J.}~\bibnamefont {Noronha}},\ }\href
  {https://doi.org/10.1007/JHEP02(2015)051} {\bibfield  {journal} {\bibinfo
  {journal} {JHEP}\ }\textbf {\bibinfo {volume} {02}},\ \bibinfo {pages}
  {051}},\ \Eprint {https://arxiv.org/abs/1412.2968} {arXiv:1412.2968 [hep-ph]}
  \BibitemShut {NoStop}%
\bibitem [{\citenamefont {Finazzo}\ \emph {et~al.}(2016)\citenamefont
  {Finazzo}, \citenamefont {Critelli}, \citenamefont {Rougemont},\ and\
  \citenamefont {Noronha}}]{Finazzo:2016mhm}%
  \BibitemOpen
  \bibfield  {author} {\bibinfo {author} {\bibfnamefont {S.~I.}\ \bibnamefont
  {Finazzo}}, \bibinfo {author} {\bibfnamefont {R.}~\bibnamefont {Critelli}},
  \bibinfo {author} {\bibfnamefont {R.}~\bibnamefont {Rougemont}},\ and\
  \bibinfo {author} {\bibfnamefont {J.}~\bibnamefont {Noronha}},\ }\href
  {https://doi.org/10.1103/PhysRevD.94.054020} {\bibfield  {journal} {\bibinfo
  {journal} {Phys. Rev. D}\ }\textbf {\bibinfo {volume} {94}},\ \bibinfo
  {pages} {054020} (\bibinfo {year} {2016})},\ \bibinfo {note} {[Erratum:
  Phys.Rev.D 96, 019903 (2017)]},\ \Eprint {https://arxiv.org/abs/1605.06061}
  {arXiv:1605.06061 [hep-ph]} \BibitemShut {NoStop}%
\bibitem [{\citenamefont {Critelli}\ \emph
  {et~al.}(2017{\natexlab{b}})\citenamefont {Critelli}, \citenamefont
  {Noronha}, \citenamefont {Noronha-Hostler}, \citenamefont {Portillo},
  \citenamefont {Ratti},\ and\ \citenamefont {Rougemont}}]{Critelli:2017oub}%
  \BibitemOpen
  \bibfield  {author} {\bibinfo {author} {\bibfnamefont {R.}~\bibnamefont
  {Critelli}}, \bibinfo {author} {\bibfnamefont {J.}~\bibnamefont {Noronha}},
  \bibinfo {author} {\bibfnamefont {J.}~\bibnamefont {Noronha-Hostler}},
  \bibinfo {author} {\bibfnamefont {I.}~\bibnamefont {Portillo}}, \bibinfo
  {author} {\bibfnamefont {C.}~\bibnamefont {Ratti}},\ and\ \bibinfo {author}
  {\bibfnamefont {R.}~\bibnamefont {Rougemont}},\ }\href
  {https://doi.org/10.1103/PhysRevD.96.096026} {\bibfield  {journal} {\bibinfo
  {journal} {Phys. Rev. D}\ }\textbf {\bibinfo {volume} {96}},\ \bibinfo
  {pages} {096026} (\bibinfo {year} {2017}{\natexlab{b}})},\ \Eprint
  {https://arxiv.org/abs/1706.00455} {arXiv:1706.00455 [nucl-th]} \BibitemShut
  {NoStop}%
\bibitem [{\citenamefont {Grefa}\ \emph {et~al.}(2021)\citenamefont {Grefa},
  \citenamefont {Noronha}, \citenamefont {Noronha-Hostler}, \citenamefont
  {Portillo}, \citenamefont {Ratti},\ and\ \citenamefont
  {Rougemont}}]{Grefa:2021qvt}%
  \BibitemOpen
  \bibfield  {author} {\bibinfo {author} {\bibfnamefont {J.}~\bibnamefont
  {Grefa}}, \bibinfo {author} {\bibfnamefont {J.}~\bibnamefont {Noronha}},
  \bibinfo {author} {\bibfnamefont {J.}~\bibnamefont {Noronha-Hostler}},
  \bibinfo {author} {\bibfnamefont {I.}~\bibnamefont {Portillo}}, \bibinfo
  {author} {\bibfnamefont {C.}~\bibnamefont {Ratti}},\ and\ \bibinfo {author}
  {\bibfnamefont {R.}~\bibnamefont {Rougemont}},\ }\href
  {https://doi.org/10.1103/PhysRevD.104.034002} {\bibfield  {journal} {\bibinfo
   {journal} {Phys. Rev. D}\ }\textbf {\bibinfo {volume} {104}},\ \bibinfo
  {pages} {034002} (\bibinfo {year} {2021})},\ \Eprint
  {https://arxiv.org/abs/2102.12042} {arXiv:2102.12042 [nucl-th]} \BibitemShut
  {NoStop}%
\end{thebibliography}%
% name of the bibtex file (in the same directory as the main tex file)
% use @unpublished instead of @article to refer to still unpublished preprints in the bibtex file (this modification should be done manually for bibtex files generated with the Inspire's Bibliography Generator [https://inspirehep.net/bibliography-generator])

\end{document}